\documentclass[prb,aps,twocolumn,eqsecnum,showpacs,preprintnumbers,amsmath,amssymb]{revtex4}

\usepackage{graphicx}
\usepackage{dcolumn}
\usepackage{bm}

\begin{document}

\author{Peter B. Weichman$^1$ and Ranjan Mukhopadhyay$^2$}

\affiliation{$^1$BAE Systems, Advanced Information Technologies, 6
New England Executive Park, Burlington, MA 01803
\\
$^2$Department of Physics, Clark University, Worcester, MA 01610}

\title{Particle-hole symmetry and the dirty boson problem}

\date{\today}

\begin{abstract}

We study the role of particle-hole symmetry on the universality
class of various quantum phase transitions corresponding to the
onset of superfluidity at zero temperature of bosons in a quenched
random medium.  To obtain a model with an exact particle-hole
symmetry it is necessary to use the Josephson junction array, or
quantum rotor, Hamiltonian, which may include disorder in both the
site energies and the Josephson couplings between wavefunction phase
operators at different sites. The functional integral formulation of
this problem in $d$ spatial dimensions yields a $(d+1)$-dimensional
classical $XY$-model with extended disorder, constant along the
extra imaginary time dimension---the so-called \emph{random rod
problem}. Particle-hole symmetry may then be broken by adding
nonzero site energies, which may be uniform or site-dependent.  We
may distinguish three cases: (i) exact particle-hole symmetry, in
which the site energies all vanish, (ii) \emph{statistical}
particle-hole symmetry in which the site energy distribution is
symmetric about zero, vanishing on average, and (iii) complete
absence of particle-hole symmetry in which the distribution is
generic.  We explore in each case the nature of the excitations in
the non-superfluid Mott insulating and Bose glass phases. We show,
in particular, that, since the boundary of the Mott phase can be
derived exactly in terms of that for the pure, non-disordered
system, that there can be no direct Mott-superfluid transition.
Recent Monte Carlo data to the contrary can be explained in terms of
rare region effects that are inaccessible to finite systems. We find
also that the Bose glass compressibility, which has the
interpretation of a \emph{temporal} spin stiffness or superfluid
density, is positive in cases (ii) and (iii), but that it vanishes
with an essential singularity as full particle-hole symmetry is
restored. We then focus on the critical point and discuss the
relevance of type (ii) particle-hole symmetry breaking perturbations
to the random rod critical behavior, identifying a nontrivial
crossover exponent. This exponent cannot be calculated exactly but
is argued to be positive and the perturbation therefore relevant. We
argue next that a perturbation of type (iii) is irrelevant to the
resulting type (ii) critical behavior:  the statistical symmetry is
\emph{restored} on large scales close to the critical point, and
case (ii) therefore describes the dirty boson fixed point.  Using
various duality transformations we verify all of these ideas in one
dimension.  To study higher dimensions we attempt, with partial
success, to generalize the Dorogovtsev-Cardy-Boyanovsky double
epsilon expansion technique to this problem.  We find that when the
dimension of time $\epsilon_\tau < \epsilon_\tau^c \simeq
\frac{8}{29}$ is sufficiently small a type (ii) symmetry breaking
perturbation is \emph{irrelevant}, but that for sufficiently large
$\epsilon_\tau > \epsilon_\tau^c$ particle-hole asymmetry is a
\emph{relevant} perturbation and a new stable fixed point appears.
Furthermore, for $\epsilon_\tau > \epsilon_\tau^{c2} \approx
\frac{2}{3}$, this fixed point is stable also to perturbations of
type (iii):  at $\epsilon = \epsilon_\tau^{c2}$ the generic type
(iii) fixed point \emph{merges} with the new fixed point.  We
speculate, therefore, that this new fixed point becomes the dirty
boson fixed point when $\epsilon_\tau = 1$. We point out, however,
that $\epsilon_\tau = 1$ may be quite special. Thus, although the
qualitative renormalization group flow picture the double epsilon
expansion technique provides is quite compelling, one should remain
wary of applying it quantitatively to the dirty boson problem.

\end{abstract}


\maketitle

\section{Introduction}
\label{sec:intro}

Quantum phase transitions at zero temperature are driven entirely by
quantum fluctuations in the ground state wavefunction.  In many
cases a crucial requirement is the presence of quenched disorder.
Examples include random magnets with various kinds of site,
bond,\cite{DQM} or field\cite{RFM} disorder; the transitions between
plateaux in the two-dimensional quantized Hall effects;\cite{QHE}
metal-insulator,\cite{AL} metal-superconductor, and
superconductor-insulator\cite{SI} transitions in disordered
electronic systems; and the onset of superfluidity of $^4$He in
porous media.\cite{FWGF}

In this article, we will study the quantum transitions between the
superfluid (SF), the localized Bose glass (BG), and Mott insulating
(MI) phases, motivated mainly by the problem of superfluidity of
$^4$He in porous media. However, as is often the case in the study
of critical phenomena, the universality class of this transition, or
a straightforward generalization of it, also includes other physical
phenomena, such as many aspects of quantum magnetism and
superconductivity.

There have been a number of approaches to studying the SF--BG
transition:  mean field theories,\cite{mft} strong
coupling,\cite{strong} large $N$,\cite{largeN} real space
methods,\cite{RG} quantum Monte Carlo
calculations,\cite{QMC1,QMC2,LeeCha,Baranger06} double
dimensionality epsilon-expansions,\cite{DBC,WK89,MW96}, real space
renormalization group in the limit of strong disorder in one
dimension,\cite{Refael} renormalization group in fixed
dimension,\cite{herbut1} and in $1+\epsilon$
dimensions.\cite{herbut2} However, none of the above methods provide
a controlled analytical approach to the critical point for $d > 1$.

A satisfactory dimensionality expansion about the upper critical
dimension for the superfluid to Bose glass transition (analogous to
the epsilon expansion about $d=4$ for classical spin problems) does
not appear to exist. In particular, we previously found\cite{MW96}
that the approach based on a simultaneous expansion in both the
dimension, $\epsilon_\tau$, of imaginary time (physically equal to
unity), and the deviation, $\epsilon = 4 - D$, of the total
space-time dimensionality, $D = d + \epsilon_\tau$, from
four\cite{DBC} does not yield a perturbatively accessible
renormalization group fixed point, and therefore does not produce a
systematic expansion for the dirty boson problem.  Nevertheless it
does give one a fairly detailed picture of the renormalization group
fixed point and the flow structures, and produce an uncontrolled
expansion for the dirty boson problem, though further work needs to
be done in order to understand the analytic structure of the theory
as a function of $\epsilon_\tau$. Therefore, in spite of its poor
convergence, the double dimensionality expansion still appears to
provide the most flexible analytic approach to the study of the
SF--BG transition, including insight into the symmetries of the
fixed point. As detailed below, attention to the role of
particle-hole symmetry, especially, is essential to generating
correct RG flow structures. In three (and higher) dimensions, the
double dimensionality expansion is the only method that has been
able, with partial success, to access the critical fixed point and
obtain critical exponents.\cite{MW96}

\subsection{Particle-hole symmetry}
\label{sec:phsym}

As stated, it will transpire that an essential ingredient that is
necessary in order to correctly understand the physics of the SF--BG
transition is an extra ``hidden'' symmetry, which we call
\emph{particle-hole} symmetry, that is present at the critical
point, but not necessarily away from it. To make this notion
precise, we compare the following two lattice models of
superfluidity: the first is the usual lattice boson Hamiltonian,
\begin{eqnarray}
{\cal H}_B &=& -\frac{1}{2} \sum_{i,j} J_{ij}
[a_i^\dagger a_j + a_j^\dagger a_i]
+ \sum_i (\varepsilon_i-\mu) \hat n_i
\nonumber \\
&&+\ \frac{1}{2} \sum_{i,j} V_{ij}
\hat n_i (\hat n_j - \delta_{ij}),
\label{1.1}
\end{eqnarray}
where $J_{ij} = J_{ji}$ is the hopping matrix element between sites
$i$ and $j$, which we will allow to have a random component; $\mu$
is the chemical potential whose zero we fix by choosing the diagonal
components, $J_{ii}$, of $J_{ij}$ in such a way that $\sum_j J_{ij}
= 0$ for each $i$; $\varepsilon_i$ is a random site energy with
\emph{mean zero}; $V_{ij} = V_{ji}$ is the pair interaction
potential, assumed for simplicity to be \emph{non}random and
translation invariant; the only nonzero commutation relations are
$[a_i,a_j^\dagger] = \delta_{ij}$, and $\hat n_i \equiv a_i^\dagger
a_i$ is the number operator at site $i$.

The second model is the Josephson junction array Hamiltonian,
\begin{eqnarray}
{\cal H}_J &=& -\sum_{i,j} \tilde J_{ij}
\cos(\hat \phi_i - \hat \phi_j)
+ \sum_i (\tilde \varepsilon_i - \tilde \mu) \tilde n_i
\nonumber \\
&+& \frac{1}{2} \sum_{i,j} U_{ij} \tilde n_i \tilde n_j,
\label{1.2}
\end{eqnarray}
with analogous parameters, but now the commutation relations $[\hat
\phi_i,\tilde n_j] = i \delta_{ij}$.  These two Hamiltonians are, in
fact, very closely related.  It is easy to check that if $N_0$ is
any positive integer then
\begin{eqnarray}
a_i^\dagger = (N_0 + \tilde n_i)^\frac{1}{2} e^{i\hat \phi_i}
\nonumber \\
a_i = e^{-i \hat \phi_i} (N_0 + \tilde n_i)^\frac{1}{2}
\label{1.3}
\end{eqnarray}
satisfy the correct Bose commutation relations, and we identify
$\hat n_i = N_0 + \tilde n_i$.  Note, however, that the commutation
relations between $\hat \phi_i$ and $\tilde n_i$ permit $\tilde n_i$
to have any integer eigenvalue, positive or negative, whereas the
eigenvalues of $\hat n_i$ must be non-negative.  Therefore it is
only when $N_0$ is large, and the fluctuations in $\tilde n_i$ are
small compared to $N_0$, that ${\cal H}_J$ and ${\cal H}_B$ may be
compared quantitatively: inside the hopping term we may approximate
$a_i^\dagger \approx N_0^\frac{1}{2} e^{i\hat \phi_i}$, $a_i \approx
N_0^\frac{1}{2} e^{-i\hat \phi_i}$ and make the identifications
\begin{equation}
\tilde J_{ij} = N_0 J_{ij};\ U_{ij} = V_{ij};\
\tilde \varepsilon_i = \varepsilon_i;\
\tilde \mu = \mu - N_0 \hat V_0 + \frac{1}{2} V_0,
\label{1.4}
\end{equation}
where $V_0 = V_{ii}$ and $\hat V_0 = \sum_j V_{ij}$, and there
exists an overall additive constant term $E_0 N = (\frac{1}{2} N_0
\hat V_0 - \frac{1}{2} V_0 - \mu)N_0 N$, where $N$ is the number of
lattice sites.

Despite this asymptotic equivalence at large $N_0$, the Josephson
Hamiltonian (\ref{1.2}) has an exact discrete symmetry which the
boson Hamiltonian lacks.  Thus the constant shift, $\tilde
n_i^\prime = \tilde n_i + n_0$, where $n_0$ is any integer, has no
effect on the commutation relations or the eigenvalue spectrum of
the $\tilde n_i$. The Hamiltonian correspondingly transforms as
\begin{equation}
{\cal H}_J\{\tilde n_i^\prime\} = {\cal H}_J\{\tilde n_i\}
+ n_0 \hat U_0 \sum_i \tilde n_i + N \varepsilon^0(n_0,\tilde\mu),
\label{1.5}
\end{equation}
where $\varepsilon^0(n_0,\tilde\mu) = n_0(\frac{1}{2} n_0 \hat U_0 -
\tilde \mu)$, and $\hat U_0 = \sum_j U_{ij}$.  The free energy
density, $f_J = -\frac{1}{\beta N} \ln \left[{\rm
tr}\,e^{-\beta{\cal H}_J} \right]$, where $\beta = 1/k_BT$,
transforms as
\begin{equation}
f_J(\tilde \mu) = f_J(\tilde \mu - n_0 \hat U_0)
+ \varepsilon^0(n_0,\tilde \mu)
\label{1.6}
\end{equation}
independent of the $\tilde J_{ij}$ and $\tilde \varepsilon_i$.  This
implies that the only effect of a shift $n_0 \hat U_0$ in the
chemical potential is a trivial additive term in the free energy
which is linear in $\tilde \mu$.  This term serves only to increase
the overall density, $n = -\frac{\partial f_J}{\partial \tilde \mu}$,
by $n_0$ but otherwise has no effect whatsoever on the phase
diagram, which therefore will be precisely periodic in $\tilde \mu$,
with period $\hat U_0$.

Consider next the transformation $\tilde n_i^\prime = -\tilde n_i$,
$\hat \phi_i^\prime = -\hat \phi_i$.  The Hamiltonian transforms as
\begin{equation}
{\cal H}_J[\tilde n_i^\prime,\hat \phi_i^\prime,
\tilde \varepsilon_i - \tilde\mu]
= {\cal H}_J[\tilde n_i,\hat \phi_i,
-(\tilde\varepsilon_i-\tilde\mu)],
\label{1.7}
\end{equation}
so that
\begin{equation}
f_J(\tilde \mu,\{\tilde \varepsilon_i\}) =
f_J(-\tilde\mu,\{-\tilde \varepsilon_i\}).
\label{1.8}
\end{equation}
Combining the two symmetries (\ref{1.6}) and (\ref{1.8}) we see that
if all $\tilde \varepsilon_i = 0$, then for $\tilde \mu = \tilde \mu_k
\equiv \frac{1}{2}k \hat U_0$, where $k$ is any integer, the
Hamiltonian possesses a special \emph{particle-hole symmetry}, namely
invariance under the transformation $\tilde n_i^\prime = k - \tilde
n_i$, $\hat\phi^\prime = -\hat\phi$.  At $\tilde \mu = \tilde \mu_k$
the density is precisely $\frac{1}{2} k$ per site, and the
thermodynamics is symmetric under addition and removal of particles
(the removal of particles being synonymous with the addition of
holes).  If the $\varepsilon_i$ are nonzero, but have a symmetric
probability distribution, $p\{\tilde \varepsilon_i\} =
p\{-\tilde\varepsilon_i\}$, then the exact particle-hole symmetry is
lost, but there still exists a \emph{statistical} particle-hole
symmetry at the same special values $\tilde\mu_k$ of $\tilde \mu$:
self averaging will ensure that $f_J(\tilde
\mu_k,\{\tilde\varepsilon_i\}) =
f_J(\tilde\mu_k,\{-\tilde\varepsilon_i\})$.  The lattice boson
Hamiltonian (\ref{1.1}) clearly can never possess either form of
particle-hole symmetry since the hopping term mixes the number and
phase in an inextricable fashion.

\begin{figure*}

\includegraphics[scale=0.5]{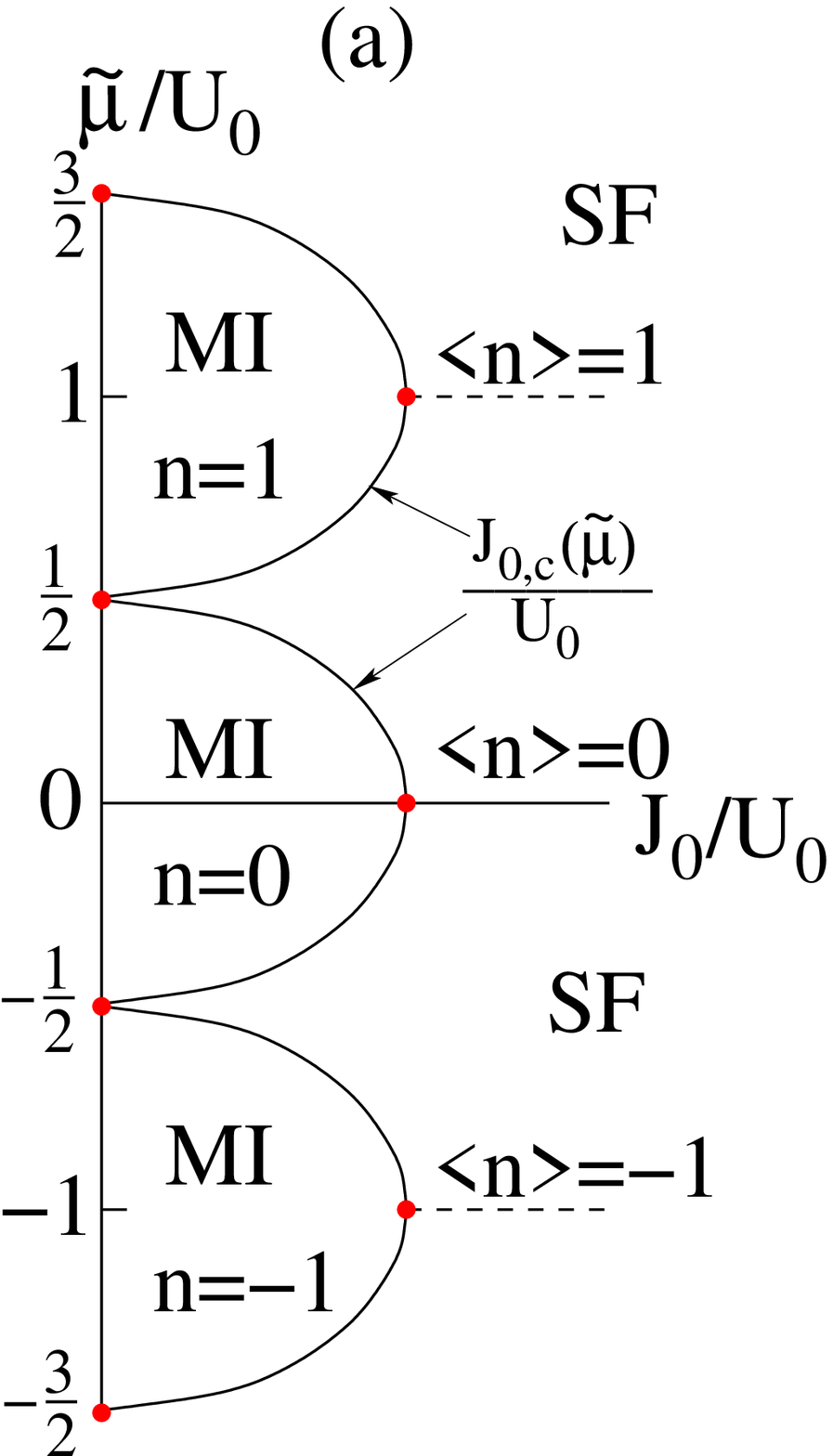}
\qquad
\includegraphics[scale=0.5]{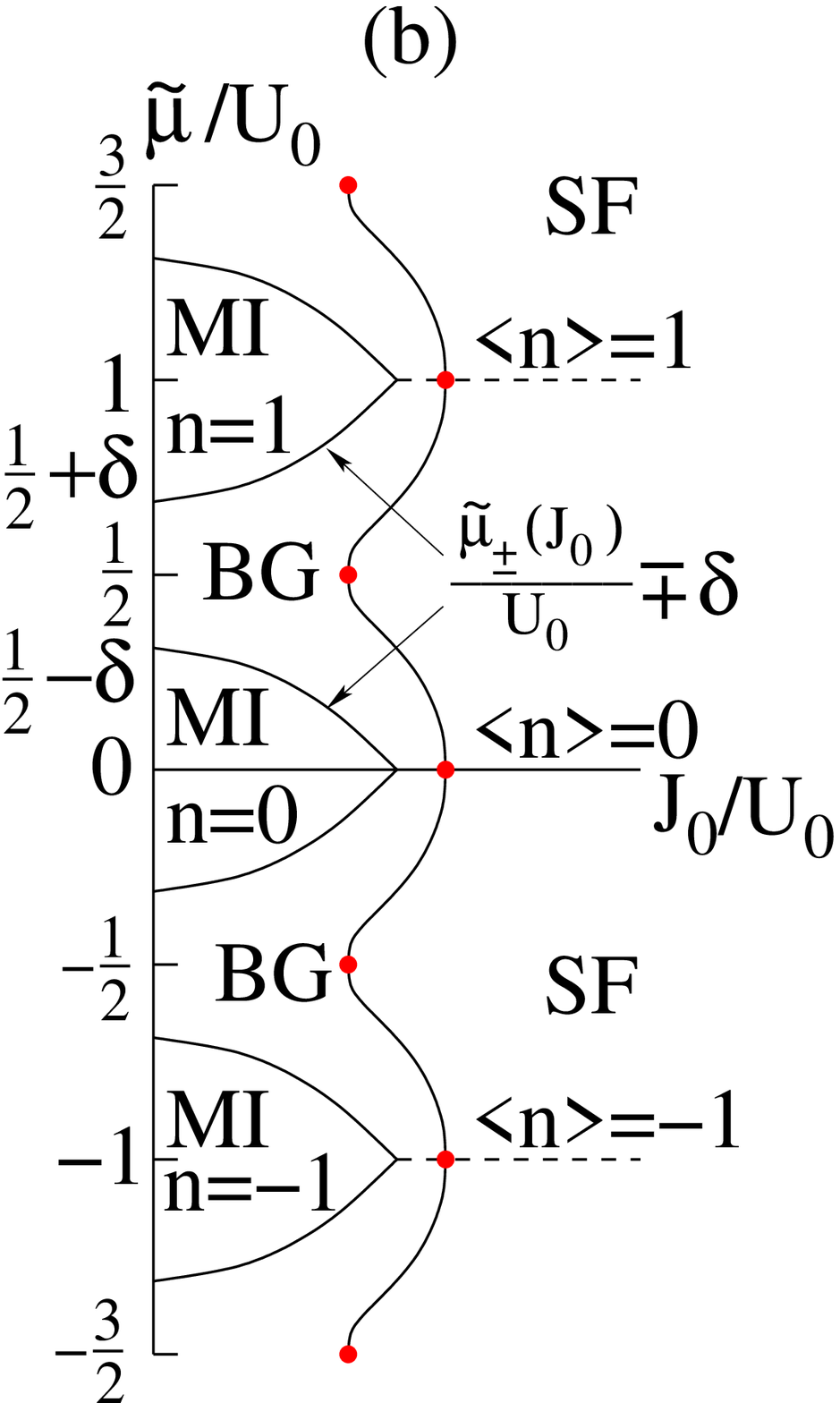}
\qquad
\includegraphics[scale=0.5]{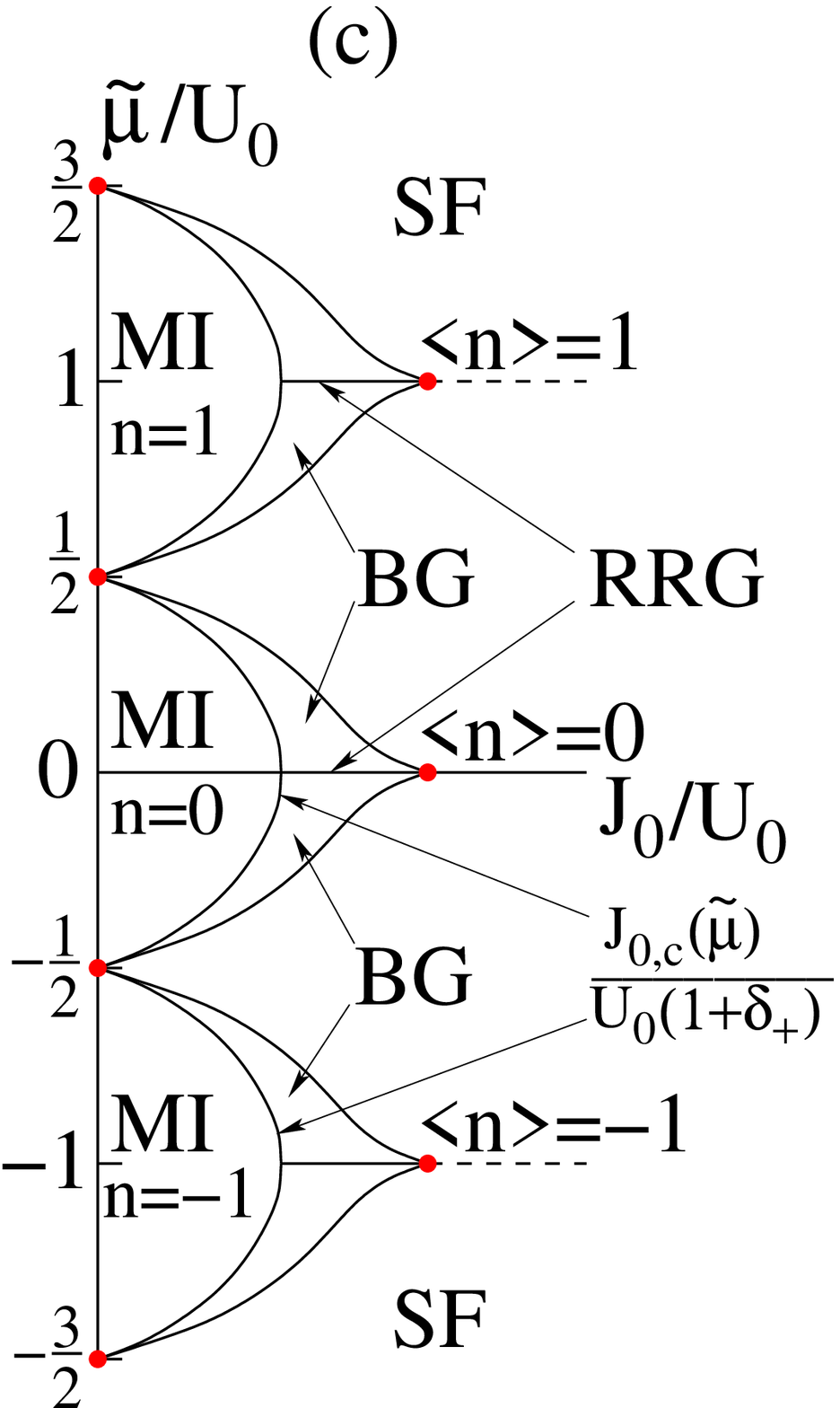}

\caption{(Color online) Schematic zero temperature phase diagram for
the Josephson junction model. (a) The model without disorder,
showing the periodic sequence of Mott lobes (MI) and the superfluid
phase (SF).  The transitions to superfluidity at the tips of the
Mott lobes, $J = J_c^0$, are special and are in the universality
class of the $(d+1)$-dimensional $XY$-model.  The points $\mu_k/U_0
= k/2$ have an exact particle-hole symmetry. As described in the
text, there is a corresponding singularity, determined by the
correlation length exponent $\nu_\mathrm{d+1}^\mathrm{XY}$, in the
shape of the lobe near its tip. (b) The model with site energy
disorder whose distribution is supported on the interval
$-\Delta_\varepsilon \leq \tilde \varepsilon_i \leq
\Delta_\varepsilon$, showing the now shrunken (or even absent, if
the disorder is sufficiently strong, $\delta \equiv
\Delta_\varepsilon/U_0 \geq \frac{1}{2}$) Mott lobes, and the new
Bose glass phase that now intervenes between them and the superfluid
phase. As indicated, the boundaries of the Mott lobes are determined
entirely by $\delta$ and their pure system counterparts. The
transition to superfluidity now takes place only from the Bose glass
phase (BG), and it is believed that the nature of this transition is
independent of where it occurs, even at the special points $\mu_k$
which now have only a statistical particle-hole symmetry. (c) Phase
diagram for the model with only nearest neighbor bond disorder,
$\tilde J_{ij} = J_0(1 + \Delta J_{ij})$ whose distribution has
support $-\delta_- \leq \Delta J_{ij} \leq \delta_+$, with $\delta_-
\leq 1$, showing again shrunken Mott lobes (which will vanish
entirely if $\Delta J_{ij}$ is unbounded from above). The boundaries
of the Mott lobes are now determined entirely by $\delta_+$ and
their pure system counterparts. At integer filling the Bose Glass
phase turns into an incompressible random rod glass (RRG). The
transition is in the universality class of the $(d+1)$-dimensional
classical random rod problem.  Away from integer filling the
compressibility turns on continuously, and the Bose glass phase
reappears. The transition is once more in the universality class of
the generic Bose glass to superfluid transition. The random rod
temporal correlation length exponent $\nu_{\tau,0} = z_0\nu_0$
(expected to be greater than unit in $d=3$\cite{DBC}) now describes
the shape of critical line near the commensurate point. In a
neighborhood of half filling (half-integer $\mu/U_0$), whose size
must depend on the precise distribution of $\Delta J_{ij}$, for $d >
1$,\cite{foot:1dsinglet} the system remains superfluid all the way
down to $J_0 = 0$: in the absence of site disorder the the exact
particle-hole symmetry that is restored at half filling guarantees
delocalized excitations.}

\label{fig:phases}
\end{figure*}

\subsection{Phase diagrams}
\label{sec:phases}

In Fig.\ \ref{fig:phases} we sketch zero temperature phase diagrams,
with and without various types of disorder, in the $\tilde
\mu$-$J_0$ plane, where $J_0$ is a measure of the overall strength
of the hopping matrix (e.g., $J_0 = \frac{1}{N} \sum_{i \neq j}
\tilde J_{ij}$), in the simplest case of onsite repulsion
only:\cite{foot1} $U_{ij} = U_0 \delta_{ij}$. This phase diagram has
been discussed in detail previously\cite{FWGF,QMC1} for the lattice
boson Hamiltonian, ${\cal H}_B$.  Here we emphasize the features
unique to ${\cal H}_J$, namely the periodicity in $\tilde \mu$, and
the special points $\tilde \mu_k$ corresponding to local extrema in
the phase boundaries.

\subsubsection{Phase diagram for the pure system}
\label{subsec:phases_pure}

\begin{figure}

\includegraphics[width=0.9\columnwidth]{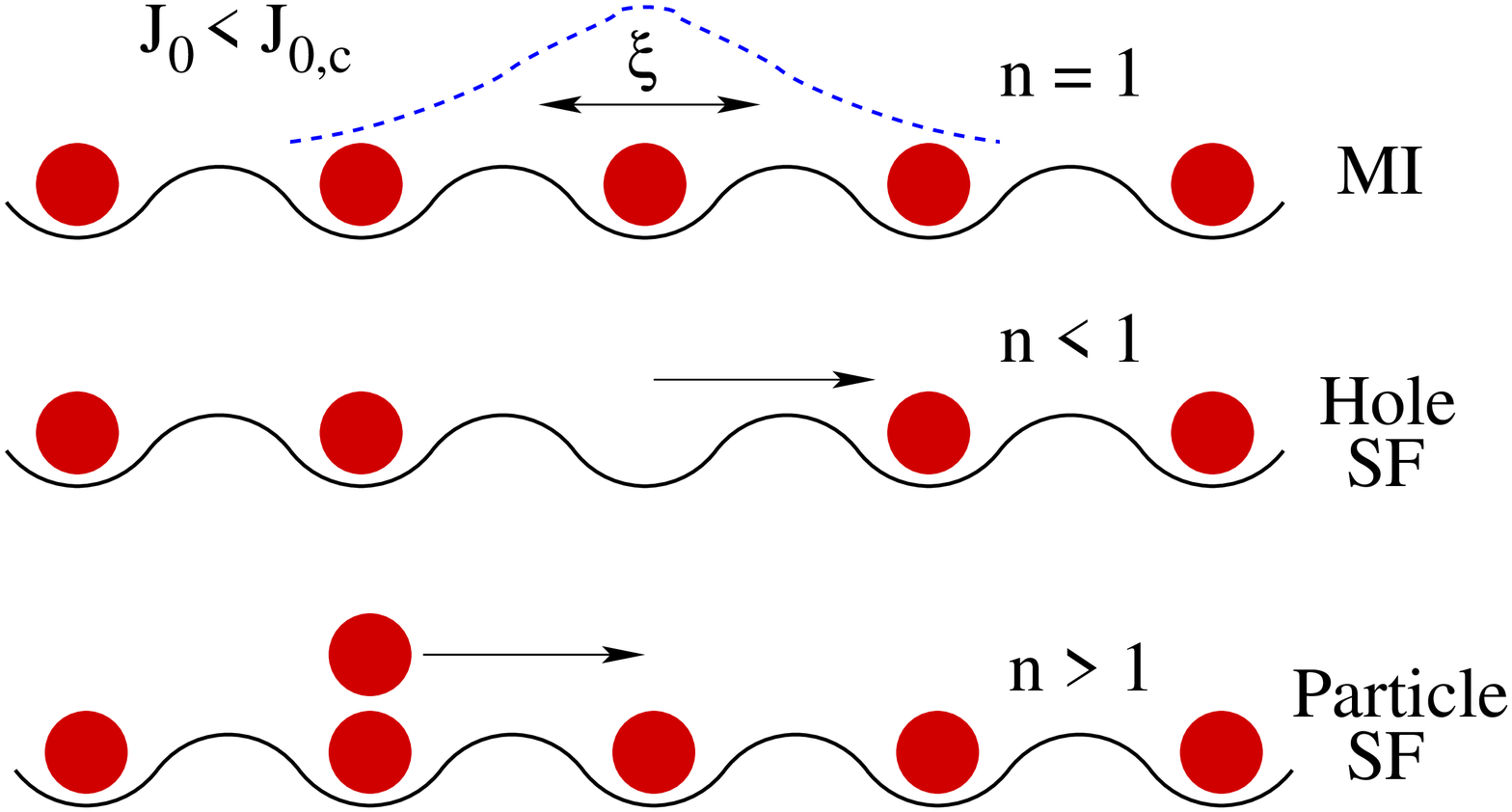}

\caption{(Color online) Schematic illustration of lattice bosons
near unit filling $n=1$ at some value of the hopping amplitude in
the interval $0 < J_0 < J_{0,c}$.  In the Mott insulating phase
($n=1$), the effective wavefunction of each particle spreads only a
finite distance $\xi(J_0)$, and the state is insulating.  For $n >
1$ ($n < 1$), the extra particles (holes) travel freely within the
essentially inert background, and the state is superfluid for
arbitrarily small $|n-1|$.}

\label{fig:hopping}
\end{figure}

In Fig.\ \ref{fig:phases}(a) we show the phase diagram in the
absence of all disorder.  For $J_{ij} \equiv 0$ the site occupancies
are good quantum numbers and each site has precisely $n_0$ particles
for $n_0-\frac{1}{2} < \tilde \mu/U_0 < n_0 + \frac{1}{2}$. The
points $\tilde\mu/U_0 = k + \frac{1}{2}$ for integer $k$ are $2^N$
fold degenerate with either $k$ or $k+1$ particles placed
independently on each site. For $J_{ij} > 0$ communication between
sites occurs and the effective wavefunction for each particle
spreads to neighboring sites (see Fig.\ \ref{fig:hopping}).  We
denote by $\xi(J_0)$ (to be defined carefully later) the range of
this spread. One can show within perturbation theory,\cite{FWGF}
however, that for sufficiently small, short ranged $J_{ij}$, there
is a finite energy gap for addition of particles, and the overall
density remains \emph{fixed} at $n_0$ for a finite range of $\tilde
\mu$.

Consider first $\tilde \mu \neq 0$.  Then only at a critical value
$J_{0,c}(\tilde\mu)$ of $J_0$ does the system favor adding extra
particles (or holes).  Equivalently, for a given $J_0$ there is an
interval $\tilde \mu_-(J_0) < \tilde \mu < \tilde \mu_+(J_0)$ of
fixed density $n_0$.  These extra particles may be thought of as a
dilute Bose fluid moving atop the essentially inert background
density, $n_0$ (see Fig.\ \ref{fig:hopping}).  The physics is
identical to that of a dilute Bose gas in the continuum, and is well
described by the Bogoliubov model.\cite{FW} From this one concludes
that the system immediately becomes superfluid (recall that we
assume $T=0$) with a superfluid density $\rho_s \sim n-n_0 \sim J_0
- J_{0,c}$, and an order parameter $\psi_0 \equiv \langle e^{i\hat
\phi_i} \rangle \sim (n-n_0)^\frac{1}{2} \sim
(J-J_{0,c})^\frac{1}{2}$.  The characteristic length in this phase
is $\xi_0 = J_0^\frac{1}{2}/[\tilde\mu -
\tilde\mu_\pm(J_0)]^\frac{1}{2} \sim (n-n_0)^{-\frac{1}{2}} \sim
(J_0-J_{0,c})^{-\frac{1}{2}}$ and represents the distance between
``uncondensed'' particles: $n-n_0-|\psi_0|^2 \sim \xi_0^{-d}$.  This
zero-temperature superfluid onset transition is therefore trivial,
in the sense that all exponents are mean-field-like.  In fact,
historically this onset was never really viewed as an example of a
phase transition.

Furthermore, although all quantities vary continuously as $J_0$
decreases toward $J_{0,c}$, the actual onset point is entirely
noncritical.  Thus, for a given value of $J_0$ within a Mott lobe,
the interval $\tilde \mu_-(J_0) < \tilde \mu < \tilde \mu_+(J_0)$
represents a \emph{single} unique (incompressible) thermodynamic
state. Incompressibility implies that for the given (integer)
density, the value of the chemical potential is ambiguous.  One
might just as well set $\tilde \mu = n_0 U_0$, its value at the
center of the lobe. The correlation length, $\xi(\tilde\mu,J_0)$, is
\emph{independent} of $\tilde \mu$, and remains perfectly finite in
the Mott phase at $J_{0,c}(\tilde\mu)$. In this sense the transition
has some elements of a first order phase transition.

The more important transition is the one occurring at fixed density,
$n = n_0$, at $\tilde\mu = n_0 U_0$ through the tip of the Mott lobe
at $J_{0,c}(0)$.  At this transition $\xi(J_0) \sim
(J_{0,c}-J_0)^{-\nu_\mathrm{pure}}$ diverges continuously with a
characteristic exponent, $\nu_\mathrm{pure}$.  One may show (see
Ref.\ \onlinecite{FWGF} and below) that the transition is precisely
in the universality class of the \emph{classical}
$(d+1)$-dimensional $XY$-model.  What distinguishes this transition
from the previous ones is precisely particle-hole symmetry:
superfluidity is achieved not by adding a small density of particles
or holes atop the inert background, but by the buildup of superfluid
fluctuations \emph{within} the background, to the point where
particles and holes \emph{simultaneously} overcome the potential
barrier $U_0$ and hop coherently without resistance.  The exponent
$\nu_\mathrm{pure}$ exhibits itself in the phase diagram as well: a
scaling argument shows that the shape of the Mott lobe is singular
near its tip,\cite{FWGF} $\mu_\pm(J_0) \sim \pm
(J_{0,c}-J_0)^{\nu_\mathrm{pure}}$. In $d = 2$, $\nu_\mathrm{pure}
\simeq \frac{2}{3}$, while in $d \geq 3$, $\nu_\mathrm{pure} =
\frac{1}{2}$. This will be discussed in a more general scaling
context in Sec.\ \ref{sec:xover}.

We have already observed that the lattice boson Hamiltonian
(\ref{1.1}) never has an exact particle-hole symmetry. Nevertheless,
one still has Mott lobes (now asymmetric and decreasing in size with
increasing $n_0$) for each integer density, and a unique extremal
point, $[J_{0,c}(n_0),\mu_c(n_0)]$, at which one exits the Mott lobe
at fixed density $n=n_0$.  One may show\cite{FWGF} that the
transition through these extremal points is still in the
$(d+1)$-dimensional $XY$ universality class, and that particle-hole
symmetry must therefore be asymptotically restored \emph{at} the
critical point. The difference now is that there is a nontrivial
balance between the densities of particle and hole excitations, and
the interactions between them. The position of the critical point is
no longer fixed by an explicit symmetry, but must be located by
carefully tuning both the hopping parameter and the chemical
potential.

This phenomenon of ``asymptotic symmetry restoration'' at a critical
point is actually fairly common (and we shall encounter it again
below). For example, though the usual Ising model of magnetism has
an up-down spin symmetry, the usual liquid--vapor or binary liquid
critical points do not.  However the Ising model correctly describes
the universality class of the transition, and one concludes that the
up-down symmetry must be restored near the critical point.
Similarly, the $p$-state clock model with Hamiltonian
\begin{equation}
{\cal H} = -J \sum_{\langle ij \rangle}
\cos\left[\frac{2\pi}{p}(q_i-q_j)\right],\ \
q_i = 1,2, \ldots, p,
\label{1.9}
\end{equation}
which may be thought of as a kind of discrete $XY$-model, has for
sufficiently large $p$ (specifically, $p > 4$ in $d = 2$; clearly $p
= 2$ corresponds to the Ising model and $p = 3$ to the three-state
Potts model) a transition precisely in the $XY$-model universality
class. Note, however, that in the \emph{ordered} phase,
corresponding to the zero temperature fixed point, the order
parameter will (for $d > 2$) spontaneously align along one of the
$p$ equivalent directions, $q_i$, breaking the $XY$-symmetry and
generating a mass for the spin-wave spectrum (more interestingly, in
$d = 2$ a power-law ordered Kosterlitz-Thouless phase exists for a
finite temperature interval below the transition, with a second
transition to a long-range ordered phase taking place only at a
lower temperature \cite{JKKN}). This latter property is not relevant
in the present case since breaking particle-hole symmetry does not
break the symmetry of the order parameter: the nature of the
superfluid phase is unaffected.

\begin{figure}

\includegraphics[width=0.9\columnwidth]{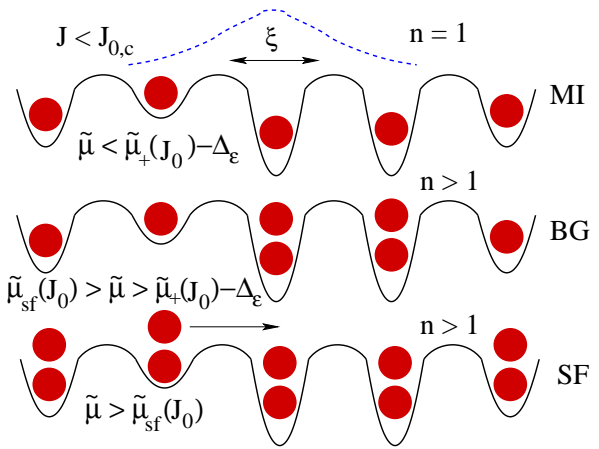}

\caption{(Color online) Schematic illustration of lattice bosons
near unit filling in the presence of bounded disorder.  If the
disorder is not too strong ($\Delta_\varepsilon < U_0/2$), there is
still a Mott phase with a finite energy gap for adding or removing
particles. Unlike in the pure case (Fig.\ \ref{fig:hopping}),
superfluidity is not generated immediately with the addition of
particles. For sufficiently small $n-1$, the additional particles
are Anderson localized by the residual random background potential
of the effectively inert layer. The finite compressibility
distinguishes this Bose glass phase from the Mott phase. The
superfluid critical point $\tilde \mu_\mathrm{sf}(J_0)$ occurs only
once the added particles have sufficiently smoothed the background
potential that its lowest lying states become extended.}

\label{fig:rand_hop}
\end{figure}

\subsubsection{Phase diagrams with disorder}
\label{subsec:phases_disorder}

We will consider two types of disorder: (i) onsite disorder in the
$\tilde \varepsilon_i$, and (ii) disorder in the hopping parameters,
$\tilde J_{ij}$. If $\tilde \mu \neq \tilde \mu_k$ for any $k$, that
is if particle-hole symmetry is broken, we expect the two types of
disorder to yield the same type of phase transition. In
renormalization group language, each in isolation will generate the
other under renormalization.  If $\tilde \mu = \tilde \mu_k$ and the
$\tilde \varepsilon_i$ have a symmetric distribution about zero so
that the Hamiltonian possesses a statistical particle-hole symmetry,
the obvious question is whether or not the transition in this case
is different from the one in the presence of generic nonsymmetric
disorder.  We shall argue below that it is \emph{not}, i.e., that
breaking particle-hole symmetry \emph{locally} is not substantially
different from breaking it \emph{globally}, and that in fact
statistical particle-hole symmetry is asymptotically restored at the
critical point.\cite{PHS} Only if $\tilde \mu = \tilde \mu_k$ and
$\tilde \varepsilon_i \equiv 0$ does the disorder fully respect
particle-hole symmetry.  We shall see that in this case the
transition is entirely different, lying in the same universality
class as the classical $(d+1)$-dimensional $XY$-model with
\emph{columnar} bond disorder, precisely the kind of system
addressed in Ref.\ \onlinecite{DBC}.

In Fig.\ \ref{fig:phases}(b) we sketch the phase diagram in the
presence of site disorder, whose distribution is supported on the
finite interval $-\Delta_\varepsilon \leq \varepsilon_i \leq
\Delta_\varepsilon$. We see that the Mott lobes have shrunk (and may
in fact disappear altogether for sufficiently strong disorder), and
a new \emph{Bose glass} phase separates these lobes from the
superfluid phase.\cite{FWGF} In this new phase the compressibility
is finite, but the particles do not hop large distances due to
localization effects: particles on top of the inert background still
see a residual random potential, whose lowest energy states will be
localized (Fig.\ \ref{fig:rand_hop}).  As particles are added to the
system, bosons will tend to fill these states until the residual
random potential has been smoothed out sufficiently that extended
states can form, finally producing superfluidity.\cite{FWGF} As
argued above, the nature of the superfluid transition is the same
everywhere along transition line.

As indicated in the figure, the boundaries of the Mott lobes are
determined entirely by the pure system, together with
$\Delta_\varepsilon$.\cite{strong} The upper half of the boundary,
$\tilde \mu_+(J_0,\Delta_\varepsilon) = \tilde \mu_+(J_0,0) -
\Delta_\varepsilon$, is pushed down by $\Delta_\varepsilon$, while
the lower half, $\tilde \mu_+(J_0,\Delta_\varepsilon) = \tilde
\mu_-(J_0,0) + \Delta_\varepsilon$, is pushed up by
$\Delta_\varepsilon$. This result, which relies on the existence of
large, rare regions of nearly uniform superfluid, will be derived in
Sec.\ \ref{sec:excite}.

Finally, in Fig.\ \ref{fig:phases}(c) we sketch the phase diagram in
the presence of bond disorder. For simplicity we consider here only
nearest neighbor hopping parameterized in the form $\tilde J_{ij} =
J_0(1+\Delta J_{ij})$, with $\left[\Delta J_{ij} \right]_\mathrm{av}
= 0$, whose distribution is supported on a finite interval
$-\delta_- \leq \Delta J_{ij} \leq \delta_+$, with $\delta_- \leq
1$. The Mott lobes have again shrunk (and may in fact disappear
altogether for unbounded disorder, $\delta_+ \to \infty$), and
glassy phases again separate these lobes from the superfluid phase.
At non-integer density, the glassy phase is the compressible Bose
glass. However, the existence of an exact particle-hole symmetry at
integer density changes the nature of the glassy phase there,
turning it into an \emph{incompressible} random rod glass (RRG). In
both cases, localization effects destroy superfluidity over a finite
interval.\cite{FWGF} As argued above, at non-integer density, the
universality class of the BG--SF transition is of the same as for
the site disorder model, while at integer density the RRG--SF
transition is in the special random rod universality class.

The random rod \emph{temporal} correlation length exponent
$\nu_{\tau,0} = z_0 \nu_0$ exhibits itself in the phase diagram: as
also discussed in Sec.\ \ref{sec:xover}, the superfluid transition
line has a singularity, $\mu_\pm(J_0) \sim \pm
(J^\mathrm{RR}_c-J_0)^{z_0 \nu_0}$, in the neighborhood of the
particle-hole symmetric points (note that $z_\mathrm{pure} = 1$).
One expects $\nu_{\tau,0} > 1$ in $d=3$, yielding the pictured
cusps.

The boundaries of the Mott lobes are again determined entirely by
the pure system, together with $\delta_+$. The boundary,
$J_{0,c}(\tilde \mu,\delta_+) = J_{0,c}(\tilde \mu,0)/(1+
\delta_+)$, is this time scaled to the left by $\Delta_J^+$. The
result again relies on the existence of large, rare regions of
nearly uniform superfluid, and will also be derived in Sec.\
\ref{sec:excite}.

At half filling, where $\tilde \mu = \tilde \mu_k = (k +
\frac{1}{2}) U_0$ is a half-integer, an exact particle-hole symmetry
is restored, and, for $d > 1$,\cite{foot:1dsinglet} the superfluid
phase can penetrate all the way to zero hopping. In general the
superfluid phase will survive on a finite interval in density (which
maps into the single point $\tilde \mu = \tilde \mu_k$ at $J_0=0$)
around half filling, whose size depends on the precise distribution
of $\tilde J_{ij}$.\cite{foot:percolate}

\begin{table*}

{\large
\begin{tabular}{l}
Pure PH-sym [$(d+1)$-dimensional XY model]:
\\
\ \ ${\cal L}_0 = -\int d^dx \int d\tau
\left\{\frac{1}{2} |\nabla \psi|^2
+ \frac{1}{2} |\partial_\tau \psi|^2
+ \frac{1}{2} r_0 |\psi|^2
+ \frac{1}{4} u_0 |\psi|^4 \right\}$
\\
\\
Pure PH-asym [$d$-dimensional dilute Bose gas]:
\\
\ \ ${\cal L}_1 = -\int d^dx \int d\tau
\left\{\frac{1}{2} |\nabla \psi|^2
- \frac{1}{2} \psi^*(\partial_\tau - g_0)^2 \psi
+ \frac{1}{2} r_0 |\psi|^2
+ \frac{1}{4} u_0 |\psi|^4 \right\}$
\\
\\
PH-sym RR [$(d+1)$-dimensional classical random rod model]:
\\
\ \ ${\cal L}_2 = -\int d^dx \int d\tau
\left\{\frac{1}{2} |\nabla \psi|^2
+ \frac{1}{2} |\partial_\tau \psi|^2
+ \frac{1}{2} [r_0 + \delta r({\bf x})] |\psi|^2
+ \frac{1}{4} u_0 |\psi|^4 \right\}$
\\
\\
PH-asym RR [$(d+1)$-dimensional incommensurate random rod model]:
\\
\ \ ${\cal L}_3 = -\int d^dx \int d\tau
\left\{\frac{1}{2} |\nabla \psi|^2
- \frac{1}{2} \psi^*(\partial_\tau - g_0)^2 \psi
+ \frac{1}{2} [r_0 + \delta r({\bf x})] |\psi|^2
+ \frac{1}{4} u_0 |\psi|^4 \right\}$
\\
\\
Statistical PH-sym [commensurate dirty boson problem]:
\\
\ \ ${\cal L}_4 = -\int d^dx \int d\tau
\left\{\frac{1}{2} |\nabla \psi|^2
- \frac{1}{2} \psi^* [\partial_\tau - \delta g({\bf x})]^2 \psi
+ \frac{1}{2} [r_0 + \delta r({\bf x})] |\psi|^2
+ \frac{1}{4} u_0 |\psi|^4 \right\}$
\\
\\
Generic PH-asym [incommensurate dirty boson problem]:
\\
\ \ ${\cal L}_5 = -\int d^dx \int d\tau
\left\{\frac{1}{2} |\nabla \psi|^2
- \frac{1}{2} \psi^* [\partial_\tau - g_0 - \delta g({\bf x})]^2 \psi
+ \frac{1}{2} [r_0 + \delta r({\bf x})] |\psi|^2
+ \frac{1}{4} u_0 |\psi|^4 \right\}$
\end{tabular}}

\caption{$\psi^4$ representation of models with various types of
disorder and various degrees of particle-hole symmetry. The
coefficients of $|\nabla \psi|^2$ and $|\partial_\tau \psi|^2$ have
been normalized to $\frac{1}{2}$. The control parameters $r_0$ and
$g_0$ are analogous to $J_0$ and $\tilde \mu$, respectively.
Disorder in the hopping strengths is represented by $\delta r$,
while that in the site energies is represented by $\delta g$. Both
are independent of $\tau$. In field theoretic treatments, both are
taken as quenched Gaussian random fields with zero mean and
delta-function correlations characterized by variances $\Delta_r$
and $\Delta_g$, respectively. Disorder in the other parameters
(including the unit gradient-squared coefficients) may also be
introduced, but produces no new critical behavior.}

\label{table1}
\end{table*}

\subsection{Criticality and restoration of statistical particle-hole
symmetry}
\label{sec:phsym_restore}

\begin{figure}

\includegraphics[width=0.9\columnwidth]{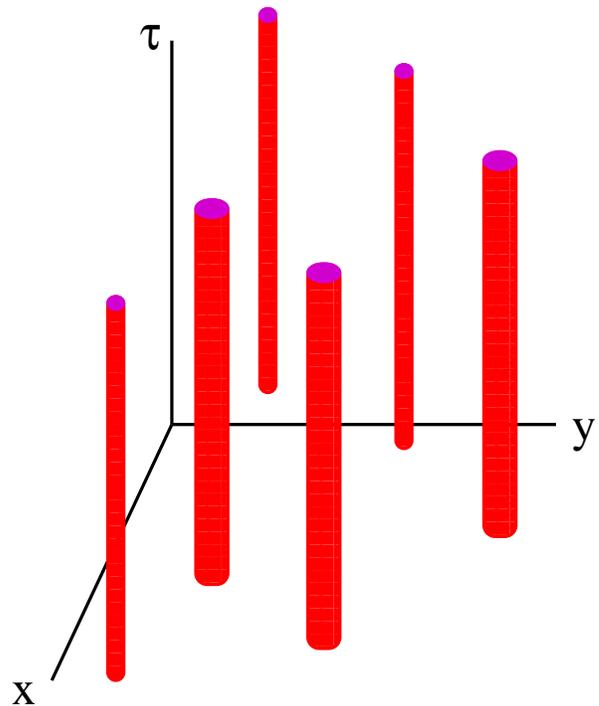}

\caption{(Color online) Schematic illustration of random rod-like
structure generated by quenched disorder in quantum models. The
cylinders represent, for example, a picture of random regions of
enhanced or suppressed hopping strength.}

\label{fig:rods}
\end{figure}

In Sec.\ \ref{sec:funcint} various functional integral
representations of the Hamiltonians (\ref{1.1}) and (\ref{1.2}) will
be introduced. In order to discuss, in the most transparent fashion,
the role of various symmetries at the superfluid transition, we
summarize in Table \ref{table1} the basic classical continuum
$\psi^4$ models that may be abstracted from these representations.
The phase of $\psi = |\psi| e^{i\phi}$ represents the Josephson
phases in (\ref{1.2}). The control parameter $r_0$ represents the
hopping strength $J_0$, while $g_0$ represents the chemical
potential $\tilde \mu$. Quenched hopping and site energy disorder
are correspondingly represented, respectively, by the random fields
$\delta r({\bf x})$ and $\delta g({\bf x})$.  The fact that they are
$\tau$-independent generates rod-like disorder (Fig.\
\ref{fig:rods}). These models are intended to be used in the
vicinity of the $n=0$ Mott lobe, hence $g_0$ in the neighborhood of
zero, since fluctuations in the amplitude $|\psi|$ destroy any
translation symmetry in $g_0$, analogous to (\ref{1.6}), and the
nonzero lobes are no longer symmetric.

The Lagrangian ${\cal L}_0 = {\cal L}_1(g_0=0)$ generates the
classical $(d+1)$-dimensional XY critical behavior along the line
$\tilde \mu = 0$ in Fig.\ \ref{fig:phases}(a): the superfluid
transition occurs for decreasing $r_0$ at a critical value
$r_{0,c}$. Nonzero $g_0$ in ${\cal L}_1$ generates the remainder of
the Mott lobe, $r_{0,c}(g_0)$, corresponding to the remainder of the
phase diagram in Fig.\ \ref{fig:phases}(a). The transition at
nonzero $g_0$ is in the universality class of the dilute Bose gas
superfluid onset transition (described by the Bogoliubov model) in
$d$ dimensions.\cite{W88} The usual coherent state representation of
the (continuum version of the) boson Hamiltonian (\ref{1.1}) is
obtained by dropping the $|\partial_\tau \psi|^2$ term in ${\cal
L}_1$ and setting $g_0 = 1$.  This demonstrates explicitly the lack
of a simple interpolation between the particle-hole symmetric and
asymmetric models in the original boson Hamiltonian.

The Lagrangian ${\cal L}_2 = {\cal L}_3(g_0=0)$ incorporates hopping
disorder in the form of $\tau$-independent disorder $\delta r$ in
$r_0$, and represents the previously analyzed (particle-hole
symmetric) random rod model.\cite{DBC} The model describes the
transitions along the line $g_0 = 0$ in Fig.\ \ref{fig:phases}(c),
including the incompressible random rod glass (RRG) separating the
tip of the Mott lobe from the superfluid phase. Nonzero $g_0$ in
${\cal L}_3$ generates the remainder of the phase diagram in Fig.\
\ref{fig:phases}(c).  We will show in Sec.\ \ref{sec:excite} that
the breaking of particle-hole symmetry implies that the RRG becomes
the compressible Bose glass exists at all nonzero $g_0$---in the
renormalization group sense, nonzero $g_0$, together with nonzero
$\delta r$, generates a finite $\delta g$ even if it vanishes to
begin with.

Under the condition that the distribution of $\delta g$ is
symmetric, he Lagrangian ${\cal L}_4 = {\cal L}_5(g_0=0)$ maintains
a statistical particle-hole symmetry. It is believed, however, that
the superfluid transition at zero and nonzero $g_0$ are in the same
universality class.  In renormalization group language, $g_0$ must
flow to zero at large scales, so that ${\cal L}_4$, not ${\cal
L}_5$, describes the critical fixed point. This is the technical
definition of the restoration of statistical particle-hole symmetry.
We will confirm this picture within the double-epsilon expansion in
Sec.\ \ref{sec:epstau}.

The significance of this symmetry restoration at the critical point
is the following.  In Ref.\ \onlinecite{WK89} only the boson
Hamiltonian (\ref{1.1}), and corresponding coherent state
Lagrangian, was considered, and a fixed point sought only in the
space of parameters accessible to this Hamiltonian. Such a fixed
point can therefore never possess a statistical particle-hole
symmetry. However if the true fixed point does possess this
symmetry, it is clear that it must then lie outside the space of
boson Hamiltonians of the form (\ref{1.1}) (or Lagrangians without a
$|\partial_\tau \psi|^2$ term). Accessing this fixed point requires
an enlargement of the parameter space so that both particle-hole
symmetric and asymmetric problems may be treated within the same
model.  The Josephson junction array Hamiltonian, (\ref{1.2}), from
which all of the Lagrangians listed in Table \ref{table1} may be
derived, satisfies this requirement.

Using ${\cal L}_4$ and ${\cal L}_5$, we shall see in Sec.\
\ref{sec:epstau} that, indeed, a new statistically particle-hole
symmetric dirty boson fixed point can be identified, and that all
technical problems encountered in Ref.\ \onlinecite{WK89} may then
be avoided. Specifically, certain diagrams that were ignored in
Ref.\ \onlinecite{WK89} are in fact important and accomplish the
required enlargement of the parameter space. There is, however, a
price: this new fixed point is not perturbatively accessible within
the double epsilon expansion. Thus, the random rod problem may be
analyzed for small $\epsilon$ in $d = 4-\epsilon-\epsilon_\tau$
dimensions if the dimension, $\epsilon_\tau$, of time is also
small.\cite{DBC}  We find, however, that for small $\epsilon$ and
$\epsilon_\tau$, site disorder is \emph{irrelevant} at the random
rod fixed point, and that \emph{full} particle-hole symmetry is
therefore restored on large scales close to criticality.  This
remains true for sufficiently small $\epsilon_\tau <
\epsilon^c_\tau(D)$, with $\epsilon_\tau^c(D) = O(1)$.  Only for
$\epsilon_\tau > \epsilon^c_\tau(D)$ does a new fixed point appear
which breaks full particle-hole symmetry. Extrapolation of the
critical behavior associated with this new fixed point, though
uncontrolled, shows significant similarities to some features of the
known behavior at $\epsilon_\tau = 1$.  To lowest nontrivial order
in $\epsilon_\tau$ we find $\epsilon_\tau^c(D=4) \simeq
\frac{8}{29}$ ($D=4$, hence $\epsilon = 0$, corresponding to $d = 3$
at $\epsilon_\tau = 1$). This value of $\epsilon_\tau^c$ is
significantly less than unity, and leads us to hope that estimates
based on these extrapolations from small $\epsilon_\tau$ are not too
unreasonable.

\subsection{Outline}
\label{sec:outline}

The outline of the remainder of this paper is as follows.  In Sec.\
\ref{sec:funcint} we introduce various useful functional integral
formulations for the thermodynamics of the Hamiltonians (\ref{1.1})
and (\ref{1.2}). We begin in Sec.\ \ref{sec:excite} by considering
the role of particle-hole symmetry in the nature of the excitation
spectra of the glassy phases.  Using a phenomenological model in
which we view the structure of the random rod and Bose glass phases
as a set of random sized, randomly placed, isolated superfluid
droplets, we focus on the density and compressibility and examine
how they vanish as full particle-hole symmetry is restored.  The
droplet model also confirms the relation between the boundaries of
the pure and disordered Mott lobes in Figs.\ \ref{fig:phases}. In
Sec.\ \ref{sec:scaling} we begin focusing on the critical point
through various phenomenological scaling arguments. In particular we
identify a new crossover exponent that describes the relevance of
particle-hole symmetry breaking perturbations to the random rod
critical behavior.  We revisit the original
arguments\cite{FF88,FWGF} for the dynamical exponent scaling
equality $z=d$, showing that they can be violated, and hence that
$z$ may remain an independent exponent.\cite{WM07} This is
consistent with recent quantum Monte Carlo simulations in $d=2$ that
find $z = 1.40 \pm 0.02$.\cite{Baranger06} We also discuss the
asymptotic restoration of statistical particle-hole symmetry at the
dirty boson critical point. In Sec.\ \ref{sec:1d} we illustrate all
of these ideas using an exactly soluble one-dimensional model. The
analysis is very similar to that of the Kosterlitz-Thouless
transition in the classical two-dimensional $XY$-model.  In Sec.\
\ref{sec:epstau} we generalize the previous analyses to general
$\epsilon_\tau \neq 1$, observing along the way some apparent
pathologies that make $\epsilon_\tau = 1$ very special, leading one
to question how smooth the limit $\epsilon_\tau \rightarrow 1$ might
be.  For example, the Bose glass phase has finite compressibility
for $\epsilon_\tau = 1$, but is incompressible for all
$\epsilon_\tau < 1$. It is distinguished from the Mott phase only by
having a divergent order parameter susceptibility, $\chi_s$. We then
introduce the Dorogovtsev-Cardy-Boyanovsky double epsilon expansion
formalism\cite{DBC} and derive the results outlined in the previous
subsection. Finally, two appendices outline the derivations of
various path integral formulations and duality transformations used
in the body of the paper.

\section{Functional Integral Formulations}
\label{sec:funcint}

In order to obtain a formulation of the problem more amenable to
analytic treatment, we turn to functional integral representations
of the partition function.  It will turn out to be important to have
an \emph{exact} representation.  Representations which involve
dividing the Hamiltonian into two pieces, ${\cal H} = {\cal H}_0 +
{\cal H}_1$, then using the Kac-Hubbard-Stratanovich transformation
to decouple ${\cal H}_1$, generate effective classical actions with
an infinite number of terms, which must be truncated at some finite
order.\cite{FWGF}  In addition, such representations work only when
$\tilde \mu$ lies within a Mott phase when $J_0=0$, and hence break
down for unbounded, e.g., Gaussian, distributions of site energies.
We turn instead to representations obtained from the Trotter
decomposition (see App.\ \ref{app:a}).

\subsection{Lattice boson model}
\label{sec:latboson}

For the lattice boson model, the coherent state representation is
most appropriate, and yields a classical Lagrangian
\begin{equation}
{\cal L}_B = \int_0^{\beta} d\tau
\left[\sum_i \psi_i^*(\tau) \partial_\tau \psi_i(\tau)
- {\cal H}_B\{\psi_i^*(\tau),\psi_i(\tau)\} \right],
\label{2.1}
\end{equation}
where ${\cal H}\{\psi_i^*(\tau),\psi_i(\tau)\}$ is obtained by
substituting the classical complex variable $\psi_i(\tau)$ for the
boson site annihilation operator $a_i$ [and $\psi_i^*(\tau)$ for the
creation operator $a_i^{\dagger}$] wherever it appears in the
(normally ordered form of the) quantum Hamiltonian. The partition
function is given by $Z = \mathrm{tr}^{\psi}[e^{\cal L}]$, where
$\mathrm{tr}^{\psi}[\cdot]$ is an unrestricted integral over all
complex fields, $\psi_i(\tau)$.  Notice that the only term that
couples different time slices is the ``Berry phase''
$\psi^*\partial_\tau \psi$ term which arises from the overlap of two
coherent states at neighboring times.  This should be contrasted
with the spatial coupling, $\frac{1}{2} \sum_{i,j} J_{ij}\psi_i^*
\psi_j$ (essentially a discrete version of $\psi^* \nabla^2 \psi$),
which appears in ${\cal H}_B$.  The imaginary time dimension is
therefore highly anisotropic. This anisotropy is increased further
if disorder is present since the $\varepsilon_i$ and $J_{ij}$ are
$\tau$-independent:  the disorder appears in perfectly correlated
\emph{columns}, rather than as point-like defects, in
$(d+1)$-dimensional space-time.

If the $\psi^*\partial_\tau \psi$ term were replaced by
$\psi^*\partial_\tau^2 \psi$, only the disorder would contribute to
the anisotropy (the fact that the coefficients of
$\psi^*\partial_\tau^2 \psi$ and $\psi^* \nabla^2 \psi$ are
different is not important, and may be cured by a simple rescaling).
The model becomes precisely a special case in the family of
classical models with rod-like disorder treated in Ref.\
\onlinecite{DBC}. The linear time derivative term in (\ref{2.1}) is
more singular than a term with a second derivative in time.  It
should therefore not be too surprising that its presence leads to
new critical behavior.

Yet another crucial property of the $\psi^* \partial_\tau \psi$ term
is that it is purely imaginary:
\begin{equation}
\left[\int_0^\beta d\tau \psi^* \partial_\tau \psi \right]^*
= -\int_0^\beta d\tau \psi^* \partial_\tau \psi,
\label{2.2}
\end{equation}
where integration by parts and periodic boundary conditions have
been used. Therefore the statistical factor, $e^{{\cal L}_B}$, used
to compute the thermodynamics is in general a complex number, and
leads to interference between different configurations of the
$\psi_i(\tau)$.  Unlike that with coupling $\psi^* \partial_\tau^2
\psi$, the resulting model therefore does not correspond to any
classical model with a well defined real Hamiltonian in one higher
dimension.  In fact, it is precisely this property that reflects the
particle-hole asymmetry in the model.  Interchanging particles and
holes is equivalent to interchanging $\psi_i^*(\tau)$ and
$\psi_i(\tau)$. The $\psi^* \partial_\tau \psi$ term changes sign
under this operation, while ${\cal H}_B[\psi^*,\psi]$ is unaffected.
Thus although ${\cal L}_B$ is invariant under the combination (known
as \emph{time reversal}) of complex conjugation and $\tau
\rightarrow -\tau$, the boson model always violates each separately.

\subsection{Josephson model}
\label{sec:josephson}

Consider now the canonical coordinate Lagrangian for the Josephson
array model (see App.\ \ref{app:a} for a derivation):
\begin{eqnarray}
&&{\cal L}_J = \int_0^\beta d\tau \bigg\{
\sum_{i,j} \tilde J_{ij} \cos[\phi_i(\tau)-\phi_j(\tau)]
\nonumber \\
&&+\ \frac{1}{2} \sum_{i,j} (U^{-1})_{ij}
\left[i\dot \phi_i(\tau) + \tilde \mu - \tilde \epsilon_i \right]
\left[i\dot \phi_j(\tau) + \tilde \mu - \tilde \epsilon_j \right]
\bigg\},
\nonumber \\
\label{2.3}
\end{eqnarray}
with partition function $Z = {\rm tr}^\phi e^{{\cal L}_J}$.  Notice
that the linear time derivative, $\dot \phi_i$, now appears in a much
more symmetric looking fashion.  If $\tilde \mu - \tilde \epsilon_i
\equiv 0$, ${\cal L}_J$ is particle-hole symmetric and real:
\begin{eqnarray}
{\cal L}_J[\tilde\mu+\tilde\epsilon_i \equiv 0] &=& \int_0^\beta
d\tau \bigg[ \sum_{i,j} \tilde J_{ij}
\cos[\phi_i(\tau)-\phi_j(\tau)]
\nonumber \\
&&-\ \frac{1}{2} \sum_{i,j} (U^{-1})_{ij} \dot \phi_i(\tau) \dot
\phi_j(\tau)\bigg],
\label{2.4}
\end{eqnarray}
and takes precisely the form of a classical $XY$-model in $(d+1)$
dimensions.

The periodicity, (\ref{1.6}), of the phase diagram is a consequence
of the periodicity of the $\phi_i$ in $\tau$: substituting $\tilde
\mu - n_0 \hat U_0$ for $\tilde \mu$ and multiplying out the
$(U^{-1})_{ij}$ term yields
\begin{eqnarray}
{\cal L}_J[\tilde\mu-n_0\hat U_0] &=& {\cal L}_J[\tilde \mu]
- i n_0 \int_0^\beta d\tau \sum_i \dot \phi_i(\tau)
\nonumber \\
&&+\ \beta N \varepsilon^0(n_0,\tilde \mu).
\label{2.5}
\end{eqnarray}
However $\int_0^\beta d\tau \dot \phi_i(\tau) = 2\pi m_i$ and $e^{i
2\pi m_i n_0} = 1$, so the second term simply drops out of the
statistical factor, $e^{{\cal L}_J}$, and we recover the free energy
identity (\ref{1.6}). Notice that if $\tilde \mu - \tilde \epsilon_i
\equiv \frac{1}{2} \hat U_0$, we obtain a statistical factor
\begin{equation}
e^{{\cal L}_J[\tilde \mu - \tilde \epsilon_i
= \frac{1}{2}\hat U_0]} = (-1)^{\sum_i m_i}
e^{{\cal L}_J[\tilde\mu - \tilde\epsilon_i = 0]
+ \beta N \varepsilon^0(\frac{1}{2},0)},
\label{2.6}
\end{equation}
which, though real, is not always positive.  Although this
Lagrangian is also particle-hole symmetric, it too does not
correspond to the Hamiltonian of any classical model.  This model is
very different from that with $\tilde \mu - \tilde \epsilon_i \equiv
0$.  For example, as shown in Fig.\ \ref{fig:phases}(c), it always
has superfluid order at $T=0$, for arbitrarily small $J_{ij}$, as
opposed to (\ref{2.4}) which orders only for sufficiently large
$J_{ij}$.\cite{FWGF}

\subsection{Josephson model for general $\epsilon_\tau$}
\label{sec:genepstau}

For later reference, note that, in contrast to (\ref{2.1}), the
Lagrangian (\ref{2.3}) has an obvious generalization to noninteger
dimensions of time.\cite{foot:fractau}  If $\epsilon_\tau$ is the
dimension of time, we simply write
\begin{widetext}
\begin{equation}
{\cal L}_J^{(\epsilon_\tau)} =
\int_0^\beta d^{\epsilon_\tau}\tau
\left\{\sum_{i,j} \tilde J_{ij}
\cos[\phi_i({\bm \tau})-\phi_j({\bm \tau})]
+\ \frac{1}{2} \sum_{i,j} (U^{-1})_{ij}
\left[i\nabla_\tau \phi_i({\bm \tau}) + {\bm \mu}
- {\bm \epsilon}_i \right] \cdot
\left[i\nabla_\tau \phi_j({\bm \tau}) + {\bm \mu}
- {\bm \epsilon}_j \right] \right\},
\label{2.7}
\end{equation}
where ${\bm \tau}$, ${\bm \mu}$ and ${\bm \epsilon}_i$ are
$\epsilon_\tau$-dimensional vectors: ${\bm \tau} = (\tau_1, \ldots,
\tau_{\epsilon_\tau})$, etc.

Renormalization group calculations are performed most conveniently
on Lagrangians, such as (\ref{2.1}), which are polynomials in
unbounded, continuous fields and their gradients.  Therefore we
would like to convert (\ref{2.7}) to such a model, while retaining
the essential physics.  If we write $\psi_i({\bm \tau}) = e^{i
\phi_i({\bm \tau})}$, then (\ref{2.7}) may be written
\begin{equation}
{\cal L}_J^{(\epsilon_\tau)} = \int_0^\beta d^{\epsilon_\tau}\tau
\left\{\sum_{i,j} \tilde J_{ij}
[\psi_i^*({\bm \tau})\psi_j({\bm \tau}) + \mathrm{c.c.}]
+ \frac{1}{2} \sum_{i,j} (U^{-1})_{ij} \psi_i^*({\bm \tau})
(\nabla_\tau + {\bm \mu}- {\bm \epsilon}_i) \psi_i({\bm \tau})
\cdot \psi^*_j({\bm \tau}) (\nabla_\tau + {\bm \mu} - {\bm \epsilon}_j)
\psi_j({\bm \tau})
\right\}.
\label{2.8}
\end{equation}
\end{widetext}
For onsite interactions only, $U_{ij} = U_0\delta_{ij}$, the second
term simplifies to
\begin{equation}
\int d^{\epsilon_\tau} \tau \frac{1}{2 U_0}
\sum_i \psi^*_i({\bm \tau})
(\nabla_\tau + {\bm \mu} - {\bm \epsilon}_i)^2 \psi_i({\bm \tau}),
\label{2.9}
\end{equation}
which is conveniently quadratic in $\psi$.  We now relax the
assumption $|\psi_i|=1$, employing instead the usual $r|\psi|^2 +
v|\psi|^4$ Landau-Ginzburg-Wilson weighting factor, obtaining
finally
\begin{eqnarray}
{\cal L}_\psi^{(\epsilon_\tau)} &=& \int d^{\epsilon_\tau}\tau
\bigg\{\frac{1}{2} \sum_{i,j} \tilde J_{ij}
[\psi_i^*({\bm \tau}) \psi_j({\bm \tau}) + c.c.]
\nonumber \\
&&+\ \frac{1}{2U_0} \sum_i \psi_i^*({\bm \tau})
(\nabla_\tau + {\bm \mu} - {\bm \epsilon}_i)^2 \psi_i({\bm \tau})
\nonumber \\
&&-\ \sum_i\left[r|\psi_i({\bm \tau})|^2 + v |\psi_i({\bm \tau})|^4
\right] \bigg\}.
\label{2.10}
\end{eqnarray}
This model retains the exact particle-hole symmetry at ${\bm \mu} +
{\bm \epsilon}_i \equiv 0$, but loses the precise periodicity of the
phase diagram when $\epsilon_\tau = 1$:  thus the second term in
(\ref{2.5}) now becomes
\begin{equation}
n_0 \int_0^\beta d\tau \sum_i \psi_i^* \partial_\tau \psi_i,
\label{2.11}
\end{equation}
[compare the first term in (\ref{2.1})] which reduces to the
previous form if $|\psi_i| = 1$.  However if $|\psi_i|$ fluctuates,
as in (\ref{2.10}), this term is no longer a perfect time derivative
and will not integrate to a simple integer result.  We will
therefore only use (\ref{2.10}) near $\tilde \mu = 0$ when we study
the role of particle-hole symmetry near the phase transition.

\subsection{Continuum models}
\label{sec:continuum}

The field theoretic LGW-type Lagrangians listed in Table
\ref{table1} follow (for $\epsilon_\tau = 1$, but with obvious
generalizations to general $\epsilon_\tau$) from the continuum limit
of (\ref{2.10}) (and its various special cases), in which the
nearest neighbor hopping term maps to $|\nabla \psi|^2$.  Space and
time are also rescaled to produce unit coefficients of the
$\partial_\tau \psi|^2$ and $|\nabla \psi|^2$ terms. As a result,
the hopping disorder now maps to disorder in the $|\psi|^2$
coefficient. As is standard, the mapping is not exact, but is
intended only to produce a minimal model that preserves the basic
symmetries of the original, so that its phase transitions lie in the
correct universality class. The control parameters $r_0$ and $g_0$
have only a rough correspondence with $J_0$ and $\tilde \mu$, but
nevertheless generate phase diagrams with the same topology.

\section{Particle-hole symmetry and the excitation spectrum of
the glassy phases}
\label{sec:excite}

In this section we will consider the nature of the non-superfluid
phases in the presence of the two types of disorder, $\epsilon_i$
and $J_{ij}$ (to simplify the notation we henceforth drop the tildes
on the Josephson junction model parameters).  Recall that for the
boson problem the $\epsilon_i$ produce a Bose glass
phase,\cite{FWGF} with a finite compressibility and a finite density
of excitation states at zero energy.  We shall contrast this with
the case of the particle-hole symmetric, random $J_{ij}$ model,
which we will show has a vanishing compressibility and an excitation
spectrum with an exponentially small density of states,
$\rho(\varepsilon) \sim e^{-{\varepsilon_0/\varepsilon}}$, i.e. a
``soft gap.''  We will show that the compressibility is precisely
the spin-wave stiffness in the time direction, which therefore
vanishes in the ``symmetric glass,'' but is finite in the Bose
glass.  This yields an upper bound $z_0 \leq d$ for the dynamical
exponent at the particle-hole symmetric transition.  An effective
\emph{lower bound} on $z_0$ may be obtained by demanding that
particle-hole asymmetry be a \emph{relevant} operator at the
symmetric transition.  This is a necessary condition in order that
the particle-hole asymmetric transition be in a different
universality class from the symmetric one.  We will obtain estimates
for this lower bound within the double $\epsilon$-expansion in Sec.\
\ref{sec:epstau}.

\subsection{Superfluid densities or helicity modulii}
\label{sec:sfdensity}

We begin by defining the superfluid density---the ``helicity
modulus,'' or ``spin wave stiffness,'' in classical spin models.
This quantity is computed from the change in free energy under a
change in boundary conditions.  Consider a box-shaped system with
sides $L_\alpha$, $\alpha=1,\ldots,D$.  We are interested in $D =
d+1$ and $L_D = \beta$.  We say that $\psi$ obeys
$\theta_\alpha$-boundary conditions if
\begin{eqnarray}
\psi(x_1,\ldots,x_\alpha+L_\alpha,\ldots,x_D)
&=& e^{i\theta} \psi(x_1,\ldots,x_\alpha,\ldots,x_D)
\nonumber \\
\psi(x_1,\ldots,x_\beta+L_\beta,\ldots,x_D)
&=& \psi(x_1,\ldots,x_\beta,\ldots,x_D),
\nonumber \\
&&\ \ \ \ \ \ \ \ \beta \neq \alpha
\label{3.1}
\end{eqnarray}
i.e., a twist angle $\theta$ is imposed in the $\alpha$ direction,
while periodic boundary conditions are maintained in all other
directions.  Let
\begin{equation}
f^{\theta_\alpha} = - \frac{1}{V_D} \ln \mathrm{tr}
\left(e^{{\cal L}^{\theta_\alpha}}\right),\
V_D \equiv \prod_{\beta=1}^D L_\beta,
\label{3.2}
\end{equation}
where ${\cal L}^{\theta_\alpha}$ is the Lagrangian, be the free
energy obtained using $\theta_\alpha$-boundary conditions.  We may
define
\begin{equation}
\tilde \psi(x_1,\ldots,x_D) = e^{-i \theta x_\alpha/L_\alpha}
\psi(x_1,\ldots,x_D),
\label{3.3}
\end{equation}
which obeys periodic boundary conditions in all directions.  In all
cases of interest one may write
\begin{equation}
{\cal L}^{\theta_\alpha}[\psi] = {\cal L}^0[\tilde \psi]
+ \delta {\cal L}[\tilde \psi;\theta/L_\alpha]
\label{3.4}
\end{equation}
where $\delta {\cal L}$ may be expanded as a Taylor series in powers
of $\theta/L_\alpha$, and superscript ``0'' denotes periodic
boundary conditions.  Thus
\begin{equation}
\delta f^{\theta_\alpha} \equiv f^{\theta_\alpha} - f^0
= -\frac{1}{V_D} \left[\langle \delta{\cal L} \rangle
+ \frac{1}{2} \langle \delta{\cal L}^2 \rangle_c + \ldots \right],
\label{3.5}
\end{equation}
where $\langle \delta{\cal L}^2 \rangle_c = \langle \delta{\cal L}^2
\rangle - \langle \delta{\cal L} \rangle^2$, and the averages are
with respect to ${\cal L}^0[\tilde \psi]$.  Equation (\ref{3.5})
yields a series of terms in powers of $\theta/L_\alpha$, and we use
the notation
\begin{equation}
\delta f^{\theta_\alpha} = -\frac{i\theta}{L_\alpha} \rho_\alpha
+ \frac{1}{2} \left(\frac{\theta}{L_\alpha} \right)^2 \Upsilon_\alpha
+ \ldots
\label{3.6}
\end{equation}
to define the coefficients in this series.  This only makes sense
for $|\theta| \leq \pi$, since it is clear from the definition that
$f^{\theta_\alpha}$ is periodic in $\theta_\alpha$ with period
$2\pi$. We shall see that the first term, which in most previous
cases was completely {\it absent}, arises from particle-hole
asymmetry.  The second term defines the \emph{helicity modulus},
$\Upsilon_\alpha$, in the direction $\alpha$.

\subsubsection{Boson model helicity modulus}
\label{subsec:bhelicity}

Let us now turn to specific cases with Lagrangians defined by
(\ref{2.1}) and (\ref{2.3}).  In both cases, when $\alpha$ is a
spatial coordinate ($\alpha = 1,\ldots,d$) the sensitivity to
boundary conditions comes only from the hopping term so that for the
boson model [see (\ref{1.1}) and (\ref{2.1})],
\begin{eqnarray}
\delta{\cal L}_B &=& \frac{1}{2} \int_0^\beta d\tau
\sum_{i,j} J_{ij} \left[ \tilde \psi^*_i
(e^{i \theta (x_i^\alpha - x_j^\alpha)/L_\alpha} - 1)
\tilde \psi_j + {\rm c.c.} \right]
\nonumber \\
&=& \frac{i\theta}{2 L_\alpha} \int_0^\beta d\tau
\sum_{i,j} J_{ij} (x_i^\alpha - x_j^\alpha)
[\tilde\psi^*_i \tilde\psi_j - {\rm c.c.}]
\nonumber \\
&&-\ \frac{\theta^2}{4L_\alpha^2} \int_0^\beta d\tau
\sum_{i,j} J_{ij} (x_i^\alpha - x_j^\alpha)^2
[\tilde\psi^*_i \tilde\psi_j + {\rm c.c.}]
\nonumber \\
&&+\ O(\theta^3/L_\alpha^3),
\label{3.7}
\end{eqnarray}
which, due to the assumed vanishing of the net current under
periodic boundary conditions, yields $\rho_\alpha=0$,
$\alpha=1,\ldots,d$, and
\begin{eqnarray}
\Upsilon_\alpha &=& \frac{1}{4} \int_0^\beta d\tau \sum_{i,j,k}
\Big[ J_{ij} (x_i^\alpha-x_j^\alpha) J_{k0} x_k^\alpha
\nonumber \\
&&\ \ \times\ \langle [\psi_i^*(\tau) \psi_j(\tau) - {\rm c.c.}]
[\psi_k^*(0) \psi_0(0) - {\rm c.c.}] \rangle \Big]_{\rm av}
\nonumber \\
&&+\ \frac{1}{2} \sum_i \Big[ J_{i0}(x_i^\alpha)^2
\langle \psi^*_i(0) \psi_0(0) + {\rm c.c.}
\rangle \Big]_{\rm av},
\nonumber \\
&&\ \ \ \ \ \ \ \ \ \alpha=1,\ldots,d,
\label{3.8}
\end{eqnarray}
where $[ \cdot ]_{\rm av}$ denotes an average over the disorder (we
assume self averaging).  For nonrandom $J_{ij}$, with nearest
neighbor hopping $J$ only, (\ref{3.8}) reduces to
\begin{eqnarray}
\Upsilon_\alpha &=& J^2 a^2 \int_0^\beta d\tau
\sum_i \Big[\langle [\psi_i^*(\tau) \partial_\alpha \psi_i(\tau)
- \psi_i(\tau) \partial_\alpha \psi_i^*(\tau)]
\nonumber \\
&&\ \ \ \ \ \ \times\ [\psi_0^*(0) \partial_\alpha \psi_0(0)
- \psi_i(0) \partial_\alpha \psi_i^*(0)]\rangle \Big]_{\rm av}
\nonumber \\
&&+\ Ja^2 \Big[\langle \psi_{{\bf \hat x}_\alpha}^* \psi_0
+ \psi_0^* \psi_{{\bf \hat x}_\alpha} \rangle \Big]_{\rm av},\
\alpha=1,\ldots,d,
\label{3.9}
\end{eqnarray}
where $\partial_\alpha \psi_i \equiv \psi_{i+{\bf \hat x}_\alpha} -
\psi_i$, $a$ is the lattice spacing, and, in an obvious notation, $i
+ {\bf \hat x}_\alpha$ labels the nearest neighbor lattice site in
direction $\alpha$.  We recognize this as the discrete version of
the usual definition of $\Upsilon_\alpha$ in terms of the
current-current correlation function.

\subsubsection{Josephson model helicity modulus}
\label{subsec:jhelicity}

The Josephson Lagrangian yields precisely the same expressions if one
identifies $\psi_i(\tau) = e^{i\phi_i(\tau)}$.  Thus, (3.8) becomes
\begin{eqnarray}
\Upsilon_\alpha &=& -\int_0^\beta d\tau \sum_{i,j,k}
\Big[ J_{ij}(x_i^\alpha - x_j^\alpha) J_{k0} x_k^\alpha
\nonumber \\
&&\ \ \ \ \times\ \langle \sin[\phi_i(\tau)-\phi_j(\tau)]
\sin[\phi_k(0)-\phi_0(0)] \rangle \Big]_{\rm av}
\nonumber \\
&&+\ \sum_i J_{i0} (x_i^\alpha)^2
\Big[ \langle \cos[\phi_i(0)-\phi_0(0)] \rangle \Big]_{\rm av},
\nonumber \\
&&\ \ \ \ \ \ \ \ \ \ \ \alpha=1,\ldots,d
\label{3.10}
\end{eqnarray}
and (\ref{3.9}) becomes
\begin{eqnarray}
\Upsilon_\alpha &=& -4 J^2 a^2 \int_0^\beta d\tau \sum_i
\Big[\langle \sin[\phi_{i+{\bf \hat x}_\alpha}(\tau)-\phi_i(\tau)]
\nonumber \\
&&\ \ \ \ \
\times\ \sin[\phi_{{\bf \hat x}_\alpha}(0)-\phi_0(0)]
\rangle \Big]_{\rm av}
\nonumber \\
&&+\ 2 J a^2 \Big[\langle
\cos[\phi_{{\bf \hat x}_\alpha}(0)-\phi_0(0)]
\rangle \Big]_{\rm av},
\nonumber \\
&&\ \ \ \ \ \ \ \ \ \ \alpha=1,\ldots,d
\label{3.11}
\end{eqnarray}

\subsubsection{Temporal helicity modulus and compressibility}
\label{subsec:thelicity}

Consider next the stiffness in the time direction.  We will show
that it is precisely the compressibility, $\kappa \equiv -
\frac{\partial^2 f}{\partial \mu^2}$.  To see this, note that only
terms with time derivatives are sensitive to $\theta_\tau$-boundary
conditions.  In the Bose case, Eq.\ (\ref{2.1}), we obtain
\begin{equation}
\delta {\cal L} = \frac{i\theta}{\beta}
\int_0^\beta d\tau \tilde \psi_i^*(\tau) \tilde \psi_i(\tau)
\label{3.12}
\end{equation}
which corresponds precisely to an imaginary shift, $\mu^\prime = \mu
+ i\frac{\theta}{\beta}$, in the chemical potential.  Similarly, for
the Josephson case, Eq.\ (\ref{2.3}), we define $\tilde \phi_i(\tau)
= \phi_i(\tau) - \frac{\theta}{\beta}\tau$, leading to exactly the
same chemical potential shift.  Thus in both ${\cal L}_B$ and ${\cal
L}_J$ the time derivatives appear with the chemical potential in
just the right way to give rise to what amounts to the Josephson
relation between the time derivative of the phase and changes in the
chemical potential. We immediately conclude that the series
(\ref{3.6}) takes the form
\begin{eqnarray}
\delta f^{\theta_\tau} &=& \frac{i\theta}{\beta}
\frac{\partial f^0}{\partial \mu}
+ \frac{1}{2} \left(\frac{i\theta}{\beta} \right)^2
\frac{\partial^2 f^0}{\partial \mu^2} + \ldots
\nonumber \\
&=& -\frac{i\theta}{\beta} \rho
+ \frac{1}{2} \left(\frac{\theta}{\beta} \right)^2 \kappa
+ \ldots
\label{3.13}
\end{eqnarray}
where $\rho = -\frac{\partial f^0}{\partial \mu}$ is the number
density, and, as promised, we identify $\Upsilon_\tau = \kappa$.

Classical intuition tells us that $\Upsilon_\alpha$ should be
nonzero only when the model has long range order in the phase of the
order parameter, i.e. only in the superfluid phase.  Although this
statement is true for the spatial directions, $\alpha=1,\ldots,d$,
this is not necessarily true for $\alpha = \tau$.  Our intuition
regarding $^4$He in porous media would lead us to be very surprised
if the system were incompressible, $\kappa=0$, throughout the
nonsuperfluid phase.  Thus there should be no barrier to the
continuous addition of particles to the system, even when it is
completely localized (only Mott phases, in which disorder is
unimportant, are incompressible because the density is pinned at
special values commensurate with the lattice\cite{FWGF}).  The Bose
glass phase is therefore rather special in that the order parameter
has a \emph{temporal stiffness}, $\Upsilon_\tau = \kappa > 0$, even
when there is no spatial stiffness, $\Upsilon_x \propto \rho_s = 0$.
Our classical intuition breaks down because the Lagrangian is
typically not real and, as discussed earlier, does not have a proper
classical interpretation.  The droplet model picture developed below
will make clear the origin of this special partial order.

The particle-hole symmetric model described by (\ref{2.4}), however,
\emph{does} have a classical interpretation, and despite the fact
that the $J_{ij}$ are random and the model anisotropic (the disorder
being fixed in time) it would be surprising if the disordered phase
possessed long range order in time.  Thus we expect $\kappa$ to
vanish when $\rho_s$ does, so that the glassy disordered phase is
incompressible.  This is permitted because the particle-hole
symmetry now dictates that the density be an integer.  What
distinguishes this glassy phase from the Mott phase, however, is
that $\kappa$ is not zero for an entire interval of $\mu$, but
vanishes only for the special value $\mu = 0$ where particle-hole
symmetry holds.

\subsubsection{Continuum model helicity moduli}
\label{subsec:chelicity}

For the continuum models listed in Table \ref{table1}, the simple
squared gradient term produces in all cases the isotropic result
\begin{eqnarray}
\Upsilon_\alpha &=&  \left[\langle |\psi|^2 \rangle
\right]_\mathrm{av}
\label{3.14} \\
&&+\ \int d^dx \int_0^\beta d\tau
\left[\langle \psi^* \partial_\alpha \psi({\bf x},\tau)
\psi^* \partial_\alpha \psi({\bf 0},0)\rangle \right]_\mathrm{av}
\nonumber
\end{eqnarray}
for the spatial helicity moduli. The temporal phase twist response
is given by (\ref{3.14}) with $g_0$ replacing $\mu$:  $\rho =
-\frac{\partial f^0}{\partial g_0}$, $\Upsilon_\tau = \kappa =
-\frac{\partial^2 f^0}{\partial g_0^2}$.  Clearly, for ${\cal L}_0$,
${\cal L}_2$, and ${\cal L}_4$, one should set $g_0 = 0$ after
taking the derivatives.

\subsection{Droplet model of the glassy phases}
\label{sec:dropletmodel}

Let us now understand in detail how these two different behaviors
merge with each other in the full phase diagram, Fig.\
\ref{fig:phases}, and in particular confirm the positions of the
Mott phase boundaries. Consider therefore the particle-hole
symmetric model (\ref{2.4}) with, for concreteness, $J_{ij} = J_0(1
+ \Delta J_{ij}) > 0$, nonzero on nearest neighbor bonds only, with
all $\Delta J_{ij}$ \emph{independent} random variables with zero
mean. Let $J_c$ be its critical point, and let $J_c^0$ be the
critical point when all $\delta J_{ij} = 0$ (note that it is
entirely possible that $J_c < J_c^0$, since random fluctuations can
sometimes \emph{compete} with Mott phase commensuration effects and
thereby enhance superfluid order\cite{QMC1}). In the latter,
nonrandom case, the transition is from a Mott insulating phase for
$J_0 < J_c^0$ to a superfluid phase for $J_0 > J_c^0$. Suppose now
that $0 < \Delta J_{ij} \leq \delta_+$ is bounded from above (as
well as, trivially, from below) with $\delta_+$ the essential
supremum (i.e., the largest value of $\Delta J_{ij}$ achievable with
finite probability density).  Then for $J_0$ such that $J_0 +
\delta_+ < J_c^0$, all $J_{ij}$ are smaller than $J_c^0$, and the
system must have a Mott gap---reducing any set of the $J_{ij}$ from
an initially uniform value in the Mott phase can only enhance its
stability. However, for $J_c > J_0 > J_c^0/(1+\delta_+)$ one will
form, via probabilistic fluctuations, exponentially rare, but
arbitrarily large regions of bonds in which all $J_{ij} > J_c^0$.
These regions therefore represent finite droplets of
superfluid---increasing any set of the $J_{ij}$ from an initially
uniform value within the superfluid phase can only enhance the
stability of the superfluid phase phase.  It is here that the
$\tau$-independence of the $J_{ij}$ becomes critical---in the
classical interpretation these droplets are one-dimensional
cylinders (see Fig.\ \ref{fig:rods}) with arbitrarily large
cross-section, made of material that would be ferromagnetically
ordered in the bulk. The fact that these regions are already
infinite along one dimension clearly enhances magnetic ordering more
than would finite (zero-dimensional) pieces of ferromagnet. We shall
see now that these droplets generate Griffiths
singularities\cite{G69} that close the Mott gap.  It then follows
that the Mott phase boundary must lie precisely at
$J_c^0/(1+\delta_+)$, as shown in Fig.\ \ref{fig:phases}(c).

\subsection{Correlations and excitations in the random bond model}
\label{sec:symglass}

Consider a superfluid droplet with (spatial) volume $V$, which will
occur roughly with density $e^{-p_0V}$, for some constant $p_0$. The
behavior of $V \times \infty$ cylinders of magnet has been discussed
in detail by Fisher and Privman,\cite{FP85} who were concerned with
finite size scaling theory of ferromagnets with a continuous $O(n)$
symmetry below their bulk critical points.  Their main result (which
will be rederived in a more general context below) was that the
correlation length, $\xi_\parallel$, along the cylinder is governed
by the bulk helicity modulus along the same direction:
\begin{equation}
\xi_\parallel = \frac{2 \Upsilon(T) V}{(n-1)k_B T}.
\label{3.15}
\end{equation}
In our case, $k_B T = 1$, $n=2$ and $\Upsilon(T) \equiv
\Upsilon_\tau(J_0)$.  The correlation function, $G_0(\tau)$, along
the cylinder then varies as
\begin{equation}
G_0(\tau) \sim e^{-|\tau|/\xi_\parallel},\ \ |\tau| \gg \xi_\parallel.
\label{3.16}
\end{equation}
There is some ambiguity in what we should take for $V$ and
$\Upsilon_\tau(J)$ in (\ref{3.15}): the droplets are neither
perfectly spherical, nor is $J_{ij}$ uniform throughout the droplet.
Thus $V$ should be some effective volume, while $\Upsilon_\tau(J_0)$
should be the bulk temporal helicity modulus associated with some
effective uniform $J_0 > J_c^0$, say roughly the average of $J_{ij}$
over the droplet.  None of these ambiguities change the order of
magnitude estimates made below.

\subsubsection{Stretched exponential correlations in the symmetric
glass}
\label{subsec:stretchedexp}

The full temporal correlation function, $G(\tau)$, is obtained by
averaging $G_0(\tau)$ over all droplets (considered independent in
the present picture).  We therefore estimate
\begin{equation}
G(\tau) \approx \int dV \int d\Upsilon_\tau
p(V,\Upsilon_\tau) G_0(\tau),
\label{3.17}
\end{equation}
where $p(V,\Upsilon_\tau)$ is the probability density for droplets
of volume $V$ and bulk helicity modulus $\Upsilon_\tau$,
\begin{equation}
p(V,\Upsilon_\tau) \sim e^{-V/V_0(\Upsilon_\tau)}.
\label{3.18}
\end{equation}
The coefficient $V_0(\Upsilon_\tau)$, which we interpret as the
``typical'' droplet size for a given $\Upsilon_\tau$, will depend on
the detailed shape of the tail of the probability distribution for
$J_{ij} > J_c^0$.  Using (\ref{3.15}), for large $\tau$ we may
perform the integral over $V$ using the saddle point method.  The
integration will be dominated by $V$ near the solution of
$\frac{d}{dV}(V/V_0 + \tau/2\Upsilon_\tau V) = 0$.  This yields
\begin{equation}
G(\tau) \sim \int d\Upsilon_\tau e^{-\sqrt{2\tau/\Upsilon_\tau
V_0(\Upsilon_\tau})},\ \ \tau \rightarrow \infty.
\label{3.19}
\end{equation}
The coefficient $\Upsilon_\tau V_0(\Upsilon_\tau)$ will have a
minimum at some value, $\bar \Upsilon_\tau(J_0)$, corresponding to
the most probable large droplets, and this will govern the
asymptotic behavior of the integral (\ref{3.19}) to yield finally,
\begin{equation}
G(\tau) \sim e^{-\sqrt{\tau/\tau_0(J_0)}},\ \ \tau_0(J_0)
= \frac{1}{2} \bar \Upsilon_\tau V_0(\bar\Upsilon_\tau).
\label{3.20}
\end{equation}
The droplets therefore yield a \emph{stretched} exponential
behavior, to be contrasted with the purely exponential behavior in
the Mott phase.\cite{foot:sfsusc}  From (\ref{3.20}) we may derive
the quantum mechanical single-particle density of
states,\cite{FWGF,FW} $\rho_1(\epsilon)$, defined as the inverse
Laplace transform of $G(\tau)$:
\begin{equation}
G(\tau) = \int_0^\infty d\epsilon \rho_1(\epsilon)
e^{-\epsilon |\tau|}.
\label{3.21}
\end{equation}
It is easy to see that exponential decay in $G(\tau)$ requires a
\emph{gap} in $\rho_1(\epsilon)$,
\begin{equation}
\rho_1(\epsilon) = 0,\ \epsilon < \epsilon_c\ \ \Leftrightarrow \ \
G(\tau) \sim e^{-\epsilon_c |\tau|},
\label{3.22}
\end{equation}
while slower than exponential decay permits $\rho_1(\epsilon) > 0$
for all $\epsilon > 0$.  The form (\ref{3.20}) yields
\begin{equation}
\rho_1(\epsilon) \sim e^{-\frac{1}{4 \tau_0(J) \epsilon}},\
\epsilon \to 0^+,
\label{3.23}
\end{equation}
a ``soft gap.'' The non-exponential form of (\ref{3.20}) and the gap
free form of (\ref{3.23}) are known as Griffiths
singularities,\cite{G69} and define the RRG phase for $J_0$ within
some interval above $J_0^c < J_c^0 - \Delta_J^+ < J_c$.

\subsubsection{Lack of a direct Mott--superfluid transition}
\label{subsec:nodirectMSF}

The fact that there cannot be a direct Mott--SF transition (i.e.,
$J_c = J_0^c$) follows from the fact that the correlation lengths in
both the superfluid droplets and in the ``background'' Mott phase
(defined, for example, as the region between droplets in which all
$J_{ij}$ are some specified finite distance below $J_c^0$) are
finite.

The superfluid transition can occur only if the droplets grow to be
large enough, and/or close enough together, that they begin to
coalesce.  Thus, we have treated the droplets as independent, but
there will actually be exponentially small interactions $\sim
e^{-d(J_0)/\xi(J_0)}$ between them due to the finite background
correlation length, where $d(J_0)$ is the typical droplet separation
[which diverges exponentially as $J_0 \to J_c^0/(1+\delta_+)$] and
$\xi(J_0)$ is the effective correlation length in the surrounding
Mott phase (which remains finite in this same limit).  These
couplings will increase the correlation length slightly, but only
when $J_0$ is sufficiently large [a finite distance above
$J_c^0/(1+\delta_+)$], and $d_0(J_0)$ sufficiently small, will the
superfluid droplets effectively overlap, and the correlation length
diverge. This defines $J_c$, which, as noted earlier, could lie
below $J_c^0$ for certain disorder distributions.

\subsubsection{Correlations, excitations and compressibility at
small nonzero $\mu$: Bose glass onset}
\label{subsec:excbf}

Now consider the compressibility.  Its computation requires the
addition of a small uniform chemical potential, $\mu$.  As alluded
to earlier, we expect $\rho_1(\epsilon)$ to be \emph{finite} at
$\epsilon = 0$ in the presence of $\mu$, signifying a Bose glass,
and implying power law behavior for $G(\tau)$:\cite{foot:sfsusc}
\begin{equation}
G(\tau) \approx \rho_1(0;\mu)/\tau,\ \ \tau \rightarrow \infty,
\label{3.24}
\end{equation}
though we shall see that $\rho_1(0;\mu)$ is exponentially small in
$\frac{1}{\mu}$.

Such behavior lies far outside any classical intuition.  To see how
it comes about we must generalize the ideas of Ref.\
\onlinecite{FP85} to this case.  Fortunately this is relatively
straightforward: a compact statement of the Fisher-Privman result is
that long time correlations along $V \times \infty$ cylinders (for
$n=2$) are governed by an effective one dimensional classical action
\begin{equation}
S_{\rm eff}^0 = -\frac{1}{2} V \Upsilon_\tau
\int_0^\beta d\tau [\partial_\tau \phi(\tau)]^2,
\label{3.25}
\end{equation}
where $\phi(\tau)$ is a coarse-grained phase.  This immediately
yields
\begin{equation}
G_0(\tau) \equiv \langle e^{i[\phi(\tau)-\phi(0)]} \rangle
= e^{-\frac{1}{2} \langle [\phi(\tau)-\phi(0)]^2 \rangle},
\label{3.26}
\end{equation}
which, upon using
\begin{equation}
\frac{1}{2} \langle [\phi(\tau)-\phi(0)]^2 \rangle =
\int_{-\infty}^\infty \frac{1 - e^{i\omega \tau}}{\omega^2}
\frac{d\omega}{\Upsilon_\tau V}
= \frac{|\tau|}{2 \Upsilon_\tau V},
\label{3.27}
\end{equation}
yields (\ref{3.15}) and (\ref{3.16}).

Now we must generalize (\ref{3.25}) to finite $\mu$.  This is
accomplished using (\ref{3.6}): effective long wavelength, long time
``hydrodynamic'' fluctuations in the phase $\phi$ are governed by
precisely the same elastic moduli that govern equilibrium twists in
the phase.  Thus in (\ref{3.6}) one simply replaces
$\frac{\theta}{L_\alpha}$ by $\partial_\alpha \phi$ and integrates
over all space. If, as in the present case, the twists in different
directions, $\alpha$, linearly superimpose [this may be checked
directly from (\ref{3.7}), where now one defines $\tilde \psi =
e^{-i\sum_\alpha \theta_\alpha x_\alpha /L_\alpha} \psi$, with
$\psi$ obeying $\theta_\alpha$-boundary conditions simultaneously in
each direction], one simply sums over all directions $\alpha$ to
obtain the final result:
\begin{equation}
S_{\rm eff} = -\sum_{\alpha=1}^D \int d^dx d\tau
\left[-i \rho_\alpha \partial_\alpha \phi
+ \frac{1}{2} \Upsilon_\alpha (\partial_\alpha \phi)^2 \right].
\label{3.28}
\end{equation}
In the case where the interactions are spatially isotropic one has
$\rho_\alpha = 0$ and $\Upsilon_\alpha = \Upsilon$ for $\alpha =
1,\ldots,d$.  With the identifications (\ref{3.13}) for $\alpha =
\tau$ we have
\begin{equation}
S_{\rm eff} = -\int d^dx d\tau
\left[-i \rho \partial_\tau \phi
+ \frac{1}{2} \kappa (\partial_\tau \phi)^2
+ \frac{1}{2} \Upsilon |\nabla \phi|^2\right].
\label{3.29}
\end{equation}
The validity of this hydrodynamic form in the presence of disorder
relies on a hidden assumption that no new, unforseen low energy
excitations develop, e.g., in the amplitude, rather than just the
phase, of the order parameter. This appears unlikely, and there are
concrete proposals for such excitations (but see Ref.\
\onlinecite{QMC2} for some discussion on this point in the context
of interpreting quantum Monte Carlo data).

For $V\times\infty$ cylindrical geometries, the effective one
dimensional result, (\ref{3.15}) and (\ref{3.16}), is obtained by
assuming that for each $\tau$, $\phi({\bf x},\tau)$ is essentially
constant in space, and hence that only the temporal fluctuations are
important.  More formally, the finiteness of $V$ implies an energy
gap in the spatial spin-wave spectrum, between uniform $\phi({\bf
x})$ and the next excited state in which $\phi$ twists by $2\pi$
from one side of the system to the other, of order $V^{-2/d}$.  The
temporal spectrum has no such gap (the frequency, $\omega$, in
(\ref{3.27}) is continuous), and therefore the asymptotic long time,
large distance behavior may be obtained by assuming $\phi({\bf
x},\tau) = \phi(\tau)$ only.  The $|\nabla \phi|^2$ term in
(\ref{3.28}) may be treated as an additive constant that drops out
of any temporal average, and we obtain the proper generalization of
(\ref{3.24}):
\begin{equation}
S_\mathrm{eff}^{(1)} = -V \int_0^\beta d\tau
\left[\frac{1}{2} \kappa (\partial_\tau \phi)^2
- i\rho \partial_\tau \phi \right].
\label{3.30}
\end{equation}
All the effects of particle-hole asymmetry are in the $\rho$ term.

Let us now study the consequences of (\ref{3.30}).  First, when
$\rho V$ is an integer (i.e., the density in the bulk is
commensurate with the droplet volume, $V$) the $2\pi$-periodic
boundary conditions on $\phi$ imply that the $\rho$ term simply
drops out of the statistical factor, $e^{S_\mathrm{eff}^{(1)}}$.
This implies that only the fractional part, $\rho V \mod 1$,
matters in (\ref{3.30}). One must be careful to distinguish $\rho$
and $\kappa$ from the \emph{actual} density and compressibility of
the droplet of volume $V$.  The values of $\rho$ and $\kappa$ are
appropriate to a bulk superfluid system with some effective $J >
J_c^0$.  The bulk compressibility, $\kappa_0(J)$, of such a system
is finite and nonzero.  When $\mu = 0$ the density is $\rho = 0$
(or, more generally, some integer), so for small $\mu$,
\begin{eqnarray}
\rho &=& \kappa_0(J) \mu + O(\mu^2)
\nonumber \\
\kappa &=& \kappa_0(J) + O(\mu).
\label{3.31}
\end{eqnarray}

The actual values of $\rho$ and $\kappa$ in the droplet are now
computed from (\ref{3.30}) as follows: the free energy density is
given by
\begin{equation}
f = f_0 - \frac{1}{\beta V} \mathrm{tr}^\phi
\left[e^{-S_{\rm eff}^{(1)}} \right],
\label{3.32}
\end{equation}
in which $f_0$ is the {\it bulk} free energy density corresponding to
the input parameters, $\kappa$ and $\rho$.  Thus, for example, at a
given value of the chemical potential, $\mu = \mu_0$, we have
\begin{equation}
-\left(\frac{\partial f_0}{\partial \mu}\right)_{\mu=\mu_0} =
\rho(\mu_0) \equiv \rho^0,\ \
\left(\frac{\partial^2 f}{\partial \mu^2}\right)
= \kappa(\mu_0) \equiv \kappa^0,
\label{3.33}
\end{equation}
and hence, correct to quadratic order in $\mu-\mu_0$, we may take
\begin{equation}
f_0(\mu) = f_0(\mu_0) - \rho^0 (\mu-\mu_0) - \frac{1}{2} \kappa^0
(\mu-\mu_0)^2.
\label{3.34}
\end{equation}
The effective action must also be correct to quadratic order,
therefore for the purposes of computing the full free energy,
consistency requires that in $S_\mathrm{eff}^{(1)}$ we take $\kappa
\equiv \kappa^0$ and $\rho = \rho^0 + \kappa^0 (\mu-\mu_0)$. For the
purposes of computing derivatives with respect to $\mu$, the only
$\mu$-dependence in the fluctuation part of the free energy is now
in $\rho$.  We obtain
\begin{eqnarray}
f &=& f_0 - \frac{1}{\beta V}
\ln \left[\sum_{m=-\infty}^\infty \mathrm{tr}^{\phi^m}
\left\{e^{S_{\rm eff}^{(1)}}\right\} \right]
\label{3.35} \\
&=& f_0 - \frac{1}{\beta V} \ln \left[\sum_{m=-\infty}^\infty
e^{2\pi i m \rho V} \mathrm{tr}^{\phi^m}
\left\{e^{S_{\rm eff}^{(0)}}\right\} \right],
\nonumber
\end{eqnarray}
where ${\rm tr}^{\phi^m}$ means that we impose the temporal boundary
condition $\phi(\beta) = \phi(0) + 2\pi m$.  Now define $\tilde
\phi(\tau) = \phi(\tau) - 2\pi m \tau/\beta$, so that $\tilde
\phi(\beta) = \tilde \phi(0)$, to obtain
\begin{eqnarray}
f &=& f_0 - \frac{1}{\beta V} \ln \bigg[\sum_{m=-\infty}^\infty
e^{i2\pi m \rho V} e^{-2\pi^2 m^2 \kappa^0/\beta}
\nonumber \\
&&\ \ \ \ \ \ \ \ \ \
\times\ {\rm tr}^{\tilde\phi} \left\{
e^{-S_{\rm eff}^{(0)}[\tilde\phi]}\right\} \bigg]
\label{3.36} \\
&=& f_0 + f_{00}[\kappa^0] - \frac{1}{\beta V}
\ln\left[ \sum_{l=-\infty}^\infty
e^{-\frac{\beta}{2\kappa^0 V} (\rho V-l)^2} \right],
\nonumber
\end{eqnarray}
where we have used (see App.\ \ref{app:b})
\begin{equation}
\sum_{m=-\infty}^\infty e^{i2\pi m x}
\frac{e^{-m^2/2K}}{\sqrt{2\pi K}}
= \sum_{l=-\infty}^\infty e^{-2\pi^2 K (x-l)^2},
\label{3.37}
\end{equation}
with $K = \beta/4\pi^2\kappa^0 V$, and
\begin{equation}
f_{00}[\kappa^0] = \ln \left[\sqrt{\frac{\beta}{2\pi\kappa^0 V}}
\mathrm{tr}^{\tilde\phi}\left\{e^{S_{\rm eff}^{(0)}[\tilde\phi]}
\right\}\right]
\label{3.38}
\end{equation}
is independent of $\mu-\mu_0$.  In the limit $\beta
\rightarrow\infty$ only the term with minimal $(x-l)^2$, i.e.
$-\frac{1}{2} \leq x-l \leq \frac{1}{2}$, contributes (at the
boundaries, two neighboring terms are degenerate). Let
$l_0(\kappa^0,\mu)$ be this minimizing value of $l$. We then obtain
finally,
\begin{equation}
f = f_0 + f_{00} + \frac{1}{\kappa^0 V} (\rho V - l_0)^2,
\label{3.39}
\end{equation}
and the actual density in the droplet is
\begin{equation}
-\frac{\partial f}{\partial \mu}
= \rho - \frac{1}{V}(\rho V -l_0) = \frac{l_0}{V}.
\label{3.40}
\end{equation}
There are exactly $l_0$ particles in the droplet for the interval
of $\mu$ such that $|\rho(\mu) V - l_0| < \frac{1}{2}$, and we have
established the desired result that the droplet is incompressible
on this same interval.

Consider next the temporal correlation function, given by
\begin{eqnarray}
&&G^{(0)}_\rho(\tau-\tau^\prime)
= \langle e^{i[\phi(\tau)-\phi(\tau^\prime)]} \rangle
\nonumber \\
&&\ \ \ \ \ \ \ \ \
=\ \frac{{\rm tr}^\phi \left[e^{S_{\rm eff}^{(1)}}
e^{i[\phi(\tau)-\phi(\tau^\prime)]} \right]}
{{\rm tr}^\phi \left[ e^{S_{\rm eff}^{(1)}} \right]}
\label{3.41} \\
&=& \frac{\sum_{m=-\infty}^\infty
e^{i2\pi m \rho V} {\rm tr}^{\phi^m}
\left[ e^{S_{\rm eff}^{(0)}}
e^{i[\phi(\tau)-\phi(\tau^\prime)]} \right]}
{\sum_{m=-\infty}^\infty
e^{i2\pi m \rho V} {\rm tr}^{\phi^m}
\left[e^{S_{\rm eff}^{(0)}} \right]}.
\nonumber
\end{eqnarray}
Defining the same periodic field, $\tilde \phi(\tau)$, we obtain
\begin{eqnarray}
&&G^{(0)}_\rho(\tau-\tau^\prime)
= \langle e^{i[\phi(\tau)-\phi(\tau^\prime)]}
\rangle_{S_{\rm eff}^{(0)}}
\nonumber \\
&&\times\ \frac{\sum_{m=-\infty}^\infty e^{i2\pi m(\rho V +
\frac{\tau-\tau^\prime}{\beta})} e^{-2\pi^2 m^2 \kappa^0 V/\beta}}
{\sum_{m=-\infty}^\infty e^{i2\pi m \rho V} e^{-2\pi^2 m^2
\kappa^0 V / \beta}}
\nonumber \\
&=& e^{-|\tau-\tau^\prime|/2\kappa^0 V} e^{-(\tau-\tau^\prime) (\rho
V \mod 1)/\kappa^0 V},\ \ \beta \rightarrow \infty,
\nonumber \\
\label{3.42}
\end{eqnarray}
where we have used (\ref{3.37}).  Once again, in the limit $\beta
\rightarrow \infty$ only the term with $-\frac{1}{2} \leq \rho V-l
\equiv \rho V \mod 1 \leq \frac{1}{2}$, contributes.  One sees now
that $G_\rho(\tau)$ decays exponentially for both $\tau \rightarrow
\pm \infty$, but at different rates:
\begin{eqnarray}
G^{(0)}_\rho(\tau) &=& e^{-(1 \pm \gamma) |\tau|/2\kappa^0 V},\
\tau \rightarrow \pm \infty
\nonumber \\
-1 < \gamma &\equiv& 2(\rho V \mod 1) \leq 1.
\label{3.43}
\end{eqnarray}
This exponential decay signifies an energy gap, proportional to
$\frac{1}{\kappa^0 V}$, for adding a particle, and is equivalent to
the incompressibility result above.  However, for large $V$ this gap
is very small, and one need only increase $\mu$ (and hence $\rho$)
by a small amount to add a single particle to the droplet.

For given $\mu$ the number of particles in the droplet will be $l =
[\rho V]$, the greatest integer less than or equal to $\rho V$.
Since there exist arbitrarily large droplets, an arbitrarily small
change in $\mu$ will add particles to the system in precisely those
droplets with volume $V \geq \frac{1}{\rho} \approx
\frac{1}{\kappa^0 \mu}$. Focusing on $\mu$ near zero (where
$\kappa^0 = \kappa_0$), we may estimate the total density as
\begin{eqnarray}
\rho_\mathrm{tot} &\sim& \int dV d\kappa_0 p(V,\kappa_0)
\frac{[\kappa_0 \mu V]}{V}
\nonumber \\ &\sim& \bar \kappa_0 \mu
\int_{V > \frac{1}{\kappa_0 \mu}} dV e^{-V/V_0(\bar \kappa_0)}
\nonumber \\
&\sim& \bar \kappa_0 \mu e^{-1/ \bar \kappa_0 \mu V_0(\bar \kappa_0)},
\label{3.44}
\end{eqnarray}
where $\bar \kappa_0$ is defined analogously to $\bar \Upsilon_\tau$
in (\ref{3.20}).  In the derivation of this formula we have assumed
that $\mu > 0$, but the result is valid also for $\mu < 0$ if $\mu$
is replaced by $|\mu|$ in the exponent (only).  The total
compressibility may be estimated as
\begin{equation}
\kappa_\mathrm{tot} \sim
\left(\bar \kappa_0 + \frac{1}{V_0 |\mu|} \right)
e^{-1/\bar \kappa_0 V_0(\bar \kappa_0) |\mu|},
\label{3.45}
\end{equation}
which also vanishes exponentially as $|\mu| \to 0$.

\begin{figure*}

\includegraphics[width=0.9\columnwidth]{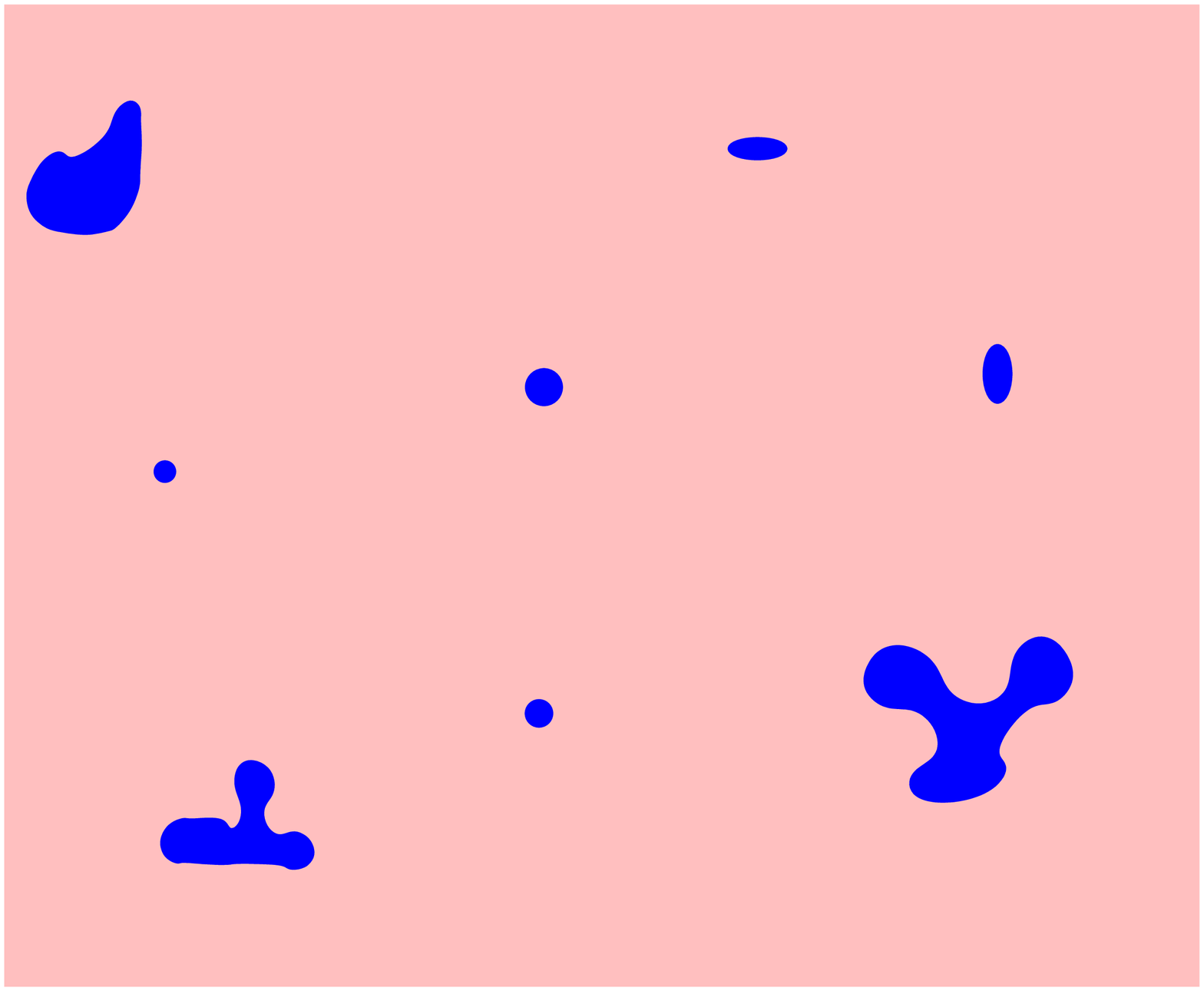}
\qquad
\includegraphics[width=0.9\columnwidth]{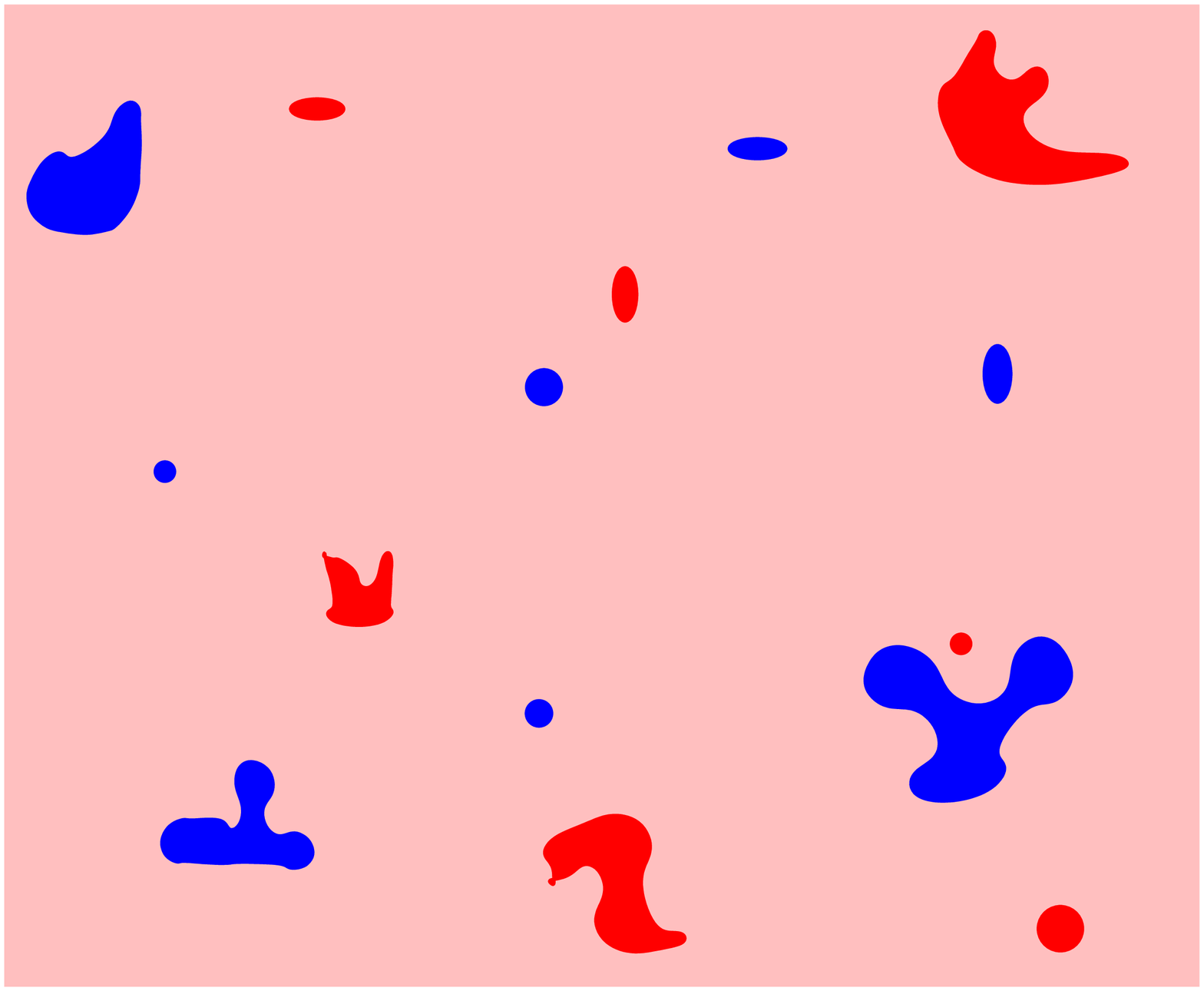}

\caption{(Color online) Schematic illustration of the droplet model
of the Bose glass phase for the random site energy model.
\textbf{Left:} Exiting the Mott lobe at fixed $J_0 < J_0^M$ below
its tip with increasing $\mu > \mu_+(J_0) - \Delta$ [or decreasing
$\mu < \mu_-(J_0,\Delta) = \mu_-(J_0) + \Delta$]. For arbitrarily
small $\epsilon = \mu - \mu_-(J_0,\Delta)$ there will be an
(exponentially small) density of droplets with volume large enough
that the energy gap to adding a particle is smaller than $\epsilon$.
\textbf{Right:} Exiting the Mott lobe from its tip at fixed $\mu=0$
with increasing $J_0 > J_0^M$, in the statistically particle-hole
symmetric model. For arbitrarily small $\Delta J = J_0 -
J_0^M(\Delta)$ there will similarly be arbitrarily large droplets
whose local chemical potential magnitude is above the energy gap
required to add a particle (previous droplets from the left
illustration), or remove a particle (additional droplets).}

\label{fig:droplet}
\end{figure*}

Finally we may use the above results to estimate the total temporal
correlation function and to exhibit the finite density of states,
(\ref{3.24}), at $\bar \epsilon = 0$.  Once again, the total
correlation function, $G_\rho(\tau)$ is the average of
$G_\rho^{(0)}(\tau)$ over all droplets:
\begin{equation}
G_\rho(\tau) = \int dV d\kappa_0 p(V,\kappa_0)
G_\rho^{(0)}(\tau;\kappa_0,V).
\label{3.46}
\end{equation}
For large $\tau$ and small $\rho > 0$, only large volumes contribute
to the integral.  It is clear from (\ref{3.43}) that
$G_\rho^{(0)}(\tau)$ decays most slowly when $\rho V$ is close to
half-integer, and those droplets with such ``resonant'' values of
$V$ will contribute the leading large $\tau$-dependence.  The
smallest resonant volume (into which a single particle will be
added) is precisely $V = \frac{1}{2\rho}$, and contributions from
higher order resonances, $V = \frac{3}{2\rho},
\frac{5}{2\rho},\ldots$, will be exponentially smaller in
$\frac{1}{\rho}$. Thus
\begin{eqnarray}
G_\rho(\tau) &\sim& \int dV d\kappa_0 p(V,\kappa_0)
e^{-|\tau|/2\kappa_0 V} e^{-\gamma\tau/2\kappa_0 V}
\nonumber \\
&\sim& e^{-1/ 2\bar \kappa_0 |\mu| V_0(\bar \kappa_0)}
\int_0^\delta \frac{dx}{\bar\kappa_0 |\mu|}
e^{-\frac{1}{4} x |\mu\tau|},
\label{3.47}
\end{eqnarray}
where $x = |\rho V - \frac{1}{2}|$ and $\delta < \frac{1}{2}$ is a
cutoff and we have replaced $V$ by its smallest resonant value,
$\frac{1}{2\rho} \approx \frac{1}{2\kappa_0 \mu}$, everywhere except
in $\gamma = 2(\rho V\ \mod 1)$.  The integration is now trivial,
and we obtain
\begin{equation}
G_\rho(\tau) \sim \frac{4}{\bar \kappa_0 \mu^2 |\tau|}
\left[1-e^{-\frac{1}{4} \delta |\mu \tau|} \right]
e^{-1/2 \bar \kappa_0 |\mu| V_0(\bar \kappa_0)}.
\label{3.48}
\end{equation}
This reproduces the $\frac{1}{\tau}$ behavior (\ref{3.24}),
predicted for the Bose glass phase with
\begin{equation}
\rho_1(\epsilon=0) \sim \frac{4}{\bar \kappa_0 \mu^2}
e^{-1/2 \bar \kappa_0 |\mu| V_0(\bar \kappa_0)}.
\label{3.49}
\end{equation}
Note that the power law prefactors (in $\mu$) of the exponential
must not be taken seriously because we have made a very crude
estimate for the probability function $p(V,\kappa_0)$.  Recall that
$\bar\kappa_0$ is the ``most probable'' compressibility for large
droplets.

One direct consequence of the slow power law decay of temporal
correlations, (\ref{3.24}) or (\ref{3.48}), is a divergent
superfluid susceptibility,\cite{FWGF}
\begin{eqnarray}
\chi_s &=& \int d^dx \int d\tau G({\bf x},\tau)
\nonumber \\
&\sim& V_0(\bar \kappa_0) \int d\tau G_\rho(\tau) \to \infty.
\label{3.50}
\end{eqnarray}
This provides another signature, in addition to the finite
compressibility, distinguishing the Bose glass from the Mott phases.

To summarize, we have seen that for the particle-hole symmetric
model the correlation function, $G(\tau)$, has stretched exponential
behavior coming from large rare regions in which $J > J_c^0$.  This
is known as a Griffiths singularity\cite{G69}, and this kind of
effect is ubiquitous in random systems.  Since $G(\tau)$ still
decays faster than any power law, the effects of these singularities
are obviously physically rather subtle.  In contrast, when $\mu \neq
0$ the model no longer has a classical interpretation, and the
behavior is far more singular: for given $\mu$, \emph{finite}
droplets of size $V \approx \frac{1}{2} \bar \kappa_0 |\mu|$ give
rise to power law decay of $G_\rho(\tau)$---no longer do the
singularities occur only in the limit $V \rightarrow \infty$.
Quantum mechanically, we understand this as being a consequence of
the existence of arbitrarily low energy single particle excitations,
arising from superfluid droplets with very small energy gaps for the
addition of an extra particle. It is interesting to see this derived
explicitly from the interference terms in the Lagrangian [see
(\ref{3.35})-(\ref{3.43})].

\begin{figure}

\includegraphics[width=0.95\columnwidth]{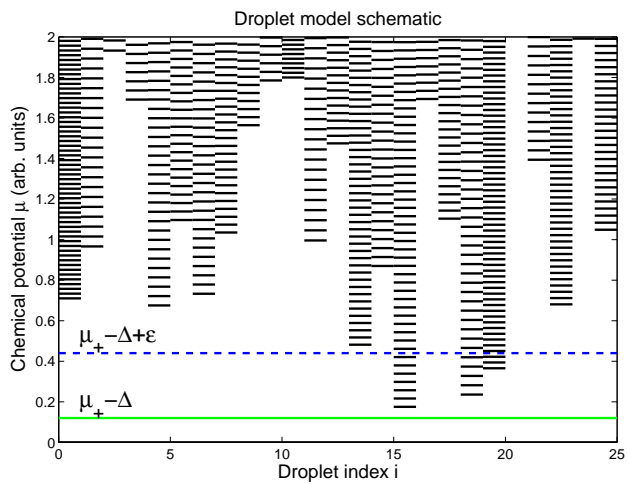}

\caption{(Color online) Schematic illustration of spectrum of
droplet excitations. Droplets are assumed to be well separated, and
independent (see Fig.\ \ref{fig:droplet}). The horizontal axis is
some convenient droplet index. The vertical axis shows the critical
chemical potential levels (in some arbitrary units) at which a new
particle is added for each droplet. The lower bound of the
excitation spectrum lies at the Mott phase boundary $\mu_+(J_0) -
\Delta$ (solid horizontal line).  For any $\epsilon \equiv \mu -
\mu_+(J_0) + \Delta > 0$ (dashed horizontal line), there will be a
finite (but exponentially small) density of large droplets with
local chemical potential lying sufficiently far above $\mu_+(J_0)$
that one or more extra particles are added. The total number of
particles added is given by counting the number of ``occupied''
levels below $\mu$. For a given droplet, while the lowest excitation
depends primarily on the local chemical potential, subsequent
particles are added in sequence, with gaps scaling as $1/\kappa_+ V$
where $V$ is the droplet volume, and $\kappa_+(J_0,\mu_+)$ is the
bulk compressibility of the pure (superfluid) phase just above Mott
phase boundary.  In an infinite system there are infinitely many
droplets, implying a continuous distribution of excitation energies,
and the bulk density will increase continuously with increasing
$\epsilon > 0$.}

\label{fig:mibg_excite}
\end{figure}

\subsection{Droplets in the random site energy model}
\label{sec:ranmudrops}

\subsubsection{MI--BG phase boundary}
\label{subsec:mibgbdy}

Consider now the random site energy model, with uniform (nonrandom)
hopping $J_{ij}$.  Above, we studied the crossover between the RRG
and BG phases with application of a small uniform $\mu$, with $J_0$
beyond the tip of the Mott lobe. Here we begin by considering $J_0 <
J_0^M(\Delta)$ below the tip of the Mott lobe (to be computed
below), and consider, for specificity, the transition to the BG
phase with increasing $\mu$ (identical arguments, with particles
replaced by holes, go through for the transition with decreasing
$\mu$).

For $\mu < \mu_+(J_0)-\Delta$, all local chemical potentials $\mu -
\varepsilon_i$ lie below the Mott gap $\mu_+(J_0)$, and no extra
particles can enter the system (if a uniform chemical potential
$\mu$ lies below the Mott gap, then reducing it on some sites cannot
reduce the gap\cite{foot:Mottgap}). Thus, $\mu_+(J_0)-\Delta$ is a
lower bound for the Mott phase boundary.

Another large rare region argument now shows that it is also an
upper bound.  Let $\epsilon = \mu - \mu_+(J_0) + \Delta > 0$ be
positive, but arbitrarily small.  Then there will be a finite
density $\sim e^{-V/V_0(\epsilon)}$ of arbitrarily large, but very
rare regions of volume $V$, in which all of the site energies are
larger than, say, $\Delta-\epsilon/2$---see the left panel of Fig.\
\ref{fig:droplet}. The scale $V_0$ will diverge as $\epsilon \to 0$
in a fashion determined by the shape of the tail of the
distribution. The region will then look like a finite size droplet
with local chemical potential larger than $\mu_+(J_0)$ by at least
$\epsilon/2$ [additional near-uniformity constraints may be imposed
on the site energies if necessary to make the droplet as homogeneous
as desired, but this will not be critical to the argument as it
entails only an adjustment of the definition of $V_0(\epsilon)$].

\begin{figure}

\includegraphics[width=0.95\columnwidth]{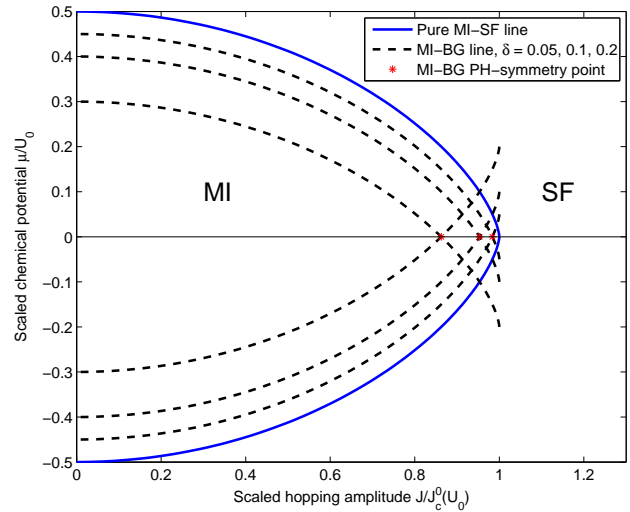}

\caption{(Color online) Mott lobe boundaries as a function of
disorder. The phase boundary is determined by that of the pure
system and the half-width $\Delta$ of the disorder distribution:
$\mu_\pm(J_0,\Delta) = \mu_\pm(J_0,0) \mp \Delta$.  The tip of the
Mott lobe, where a statistical particle-hole symmetry obtains,
occurs at $J_0 = J_0^M(\Delta)$ defined by $\mu_\pm(J_0) = \Delta$.}

\label{fig:mibg_bdy}
\end{figure}

Let $\kappa_+(J_0)$ be the (finite) compressibility of a homogeneous
system just above the Mott lobe. From the Bogoluibov theory of the
dilute Bose gas of quasiparticle excitations, one may estimate
$\kappa_+ \sim m/\hbar^2 \xi(J_0)^{d-2}$ (replaced by a logarithmic
form in $d=2$\cite{FH88}), where $\xi(J_0)$ estimates the diameter
of the quasiparticles which determines the s-wave scattering
length.\cite{FW} Then the added number of particles may be estimated
as $l = [\epsilon \kappa_+ V/2]$ [compare (\ref{3.43}) and below].
Thus, if $V > 2/\kappa_+ \epsilon$ the droplet will have at least
one extra particle, and additional particles are added each with
energy gap $2/\kappa_+ V$. Note that one should have as well $V \gg
\xi(J_0)^d$, the quasiparticle volume, for this dilute Bose gas
argument to make sense. The droplet excitation spectrum following
from these arguments is illustrated in Fig.\ \ref{fig:mibg_excite}.

A calculation identical in form to (\ref{3.44}) and (\ref{3.45})
(with $\mu$ replaced by $\epsilon/2$) can now be used to show that
the bulk compressibility is finite, though exponentially small in
$1/\epsilon$.  This finally demonstrates that $\mu = \mu_+(J_0) -
\Delta$ is also a lower bound on the Bose glass phase boundary, and
hence is, in fact, the phase boundary.  The resulting shrinking of
the pure system Mott lobe is illustrated in Fig.\ \ref{fig:mibg_bdy}
for various values of $\delta = \Delta/U_0$.  Note that the pure
system $\mu_\pm(J_0) \sim |J_0-J_c^0|^{\nu_0}$ critical
singularity\cite{FWGF} is replaced by a slope discontinuity at the
tip of the Mott lobe, $J_0 = J_0^M$ [defined by $\mu_\pm(J_0^M) =
\Delta$].

\subsubsection{Statistical particle-hole symmetry}
\label{subsec:statphsym}

Since we assume that the random site energies $\epsilon_i$ have a
symmetric distribution, the line $\mu = 0$, passing through the tip
of the Mott lobe, has a statistical particle-hole symmetry (see the
discussion in Sec.\ \ref{sec:funcint}). Does this change the nature
of the Bose-glass phase along this line, $J_0 > J_0^M$?

It is clear that the answer must be no: for any $J_0 > J_0^M$, hence
$\delta \mu \equiv \Delta - |\mu_\pm(J_0)| > 0$, we will be able to
find an interpenetrating, but independent, distribution of droplets
of arbitrarily large size in which all $\epsilon_i < -\mu_+(J_0) -
\delta\mu/2$, or all $\epsilon_i > -\mu_-(J_0) + \delta\mu/2$. The
former will have additional particles, while the latter will have a
particle deficit (additional holes)---right panel of Fig.\
\ref{fig:droplet}. Precisely at $\mu = 0$, symmetry implies that the
overall density is still fixed at the Mott value. The previous
analysis, generating finite compressibility and nonzero energy
density of excitation states, then goes through precisely as before,
but now with $\epsilon$ replaced by $\delta \mu$, and a different
volume scale, $V_0(\bar \kappa_\pm,\delta\mu)$, now depending on
$\delta \mu$.

Thus, at least at this qualitative level, statistical particle-hole
symmetry differs from generic particle-hole \emph{asymmetry} only in
that there are now two sets of droplets (particle droplets and hole
droplets) contributing independently to the excitation spectrum. It
seems very unlikely then that the nature of the superfluid
transition would be any different either. We shall address this
issue further in Sec.\ \ref{sec:scaling}.

\subsubsection{Lack of a direct Mott--superfluid transition}
\label{subsec:nomisf}

Using arguments similar to that presented in Sec.\
\ref{subsec:nodirectMSF}, we can rule out a direct MI--SF transition
in this model as well. The excited droplets, containing a dilute
superfluid of excitations, have been treated as independent.  There
will, however, be exponentially decaying interactions $\sim
e^{-d/\xi(J_0)}$ between them, where $d$ is the typical droplet
separation, and $\xi(J_0)$ is the correlation length in the
background insulating Mott phase. Even for very large $d$
(exponentially large in $1/\epsilon$), it is not immediately obvious
that there might be some (exponentially) small hopping of
quasiparticles between droplets, generating bulk superfluid
coherence. However, the excitation spectrum of each individual
droplet is discrete, and the usual Anderson localization arguments
imply that for small $\epsilon$ the competition between hopping
distance and energy level matching implies that the excited states
remain strongly localized to the neighborhood of their droplets.
Stated slightly differently, for $\epsilon$ not too large, the
excited particles see a residual random potential whose effective
low lying single particle states must be localized. Identical
arguments, applied to the two sets of droplets, imply that there can
be no direct transition through the statistical particle-hole
symmetric tip of the Mott lobe either.

In apparent violation of these, essentially rigorous, arguments,
there have been some recent Monte Carlo simulations claiming to see
a direct MI--SF transition, over a finite segment of the Mott lobe
around $\mu = 0$, for sufficiently weak disorder,\cite{LeeCha} and
even claiming evidence for new multicritical behavior at the
endpoints of this segment.  However, it is well known that rare
region effects are invisible to Monte Carlo
simulations,\cite{strong,QMC1} which are, by necessity, limited to
finite volumes with at most a few thousand sites.

Note, in addition, that since the boundary of the Mott phase is
known exactly, a direct transition would imply that the SF phase
moves in to take over \emph{all} of what used to be the pure system
Mott phase.  Although disorder can indeed increase the stability of
the superfluid over the MI,\cite{QMC1} that it would eat up the
putative intervening BG phase entirely is disproven by the rare
region arguments.

At the risk of belaboring the point, notice also how extraordinarily
sensitive to the type of disorder the direct MI--SF transition would
have to be if it existed. Consider a model in which the usual
top-hat disorder model of sufficiently small half-width $\Delta$ is
augmented by an additional Gaussian disorder with the same width
$\Delta$, but with a miniscule relative amplitude---say $10^{-9}$.
Since the onsite potentials are now unbounded (though one would have
to survey billions of sites to know it) the MI phase no longer
exists. However, by any conceivable measure, the total disorder is
still small, but a direct MI--SF transition would now require a SF
phase all the way down to $J_0 = 0$.  In other words, a
one-in-a-billion change must cause the phase boundary to move an
infinite distance (on the natural scale of $1/J_0$).  It is obvious,
however, that the simulations would see no change at all with the
infinitesimal added Gaussian.

Regarding the apparent multicritical scaling, it is very easy to
find apparent scaling of data over the limited ranges of system
sizes that are available, if one has a sufficient number of free
parameters available to fit (the transition point, and the exponents
$\nu$ and $z$ in this case). A more likely explanation for the
apparent scaling is a combination of finite size effects, and a
crossover from the pure system critical behavior to dirty boson
critical behavior at weak disorder.  The latter implies a crossover
scaling variable of the form $\Delta/|J_0-J_c^0|^\phi$, where $\phi$
is a crossover exponent quantifying the instability of the pure
MI--SF transition to small disorder.  At small $\mu$, this crossover
will also mix with the pure system particle-hole symmetry-breaking
scaling variable $\mu/|J_0-J_c^0|^{\nu_0}$. For small $\Delta$ and
limited system sizes, these scaling variables will saturate before
the asymptotic BG--SF dirty boson criticality can become visible
very close to $J_{0,c}(\mu)$. The corresponding multi-crossover
scaling form, which will be discussed in more detail in Sec.\
\ref{sec:xover} below, could easily mimic multicriticality.

\section{Particle-hole symmetry and scaling near criticality}
\label{sec:scaling}

In order to discuss scaling it is convenient (but by no means
necessary) to use the $\psi^4$ Lagrangian, (\ref{2.10}), further
simplified by taking the continuum limit and dropping all
unnecessary dimensionful coefficients. To begin we take
$\epsilon_\tau = 1$ only, and consider the Lagrangian,
\begin{eqnarray}
{\cal L}_c &=& -\int d^dx \int d\tau
\bigg[\frac{1}{2} |\nabla \psi|^2
- \frac{1}{2} \psi^* [\partial_\tau - g({\bf x})]^2 \psi
\nonumber \\
&+&\frac{1}{2} r({\bf x}) |\psi|^2
+ \frac{1}{4} u|\psi|^4 \bigg],
\label{4.1}
\end{eqnarray}
equivalent to ${\cal L}_5$ in Table \ref{table1} with $g({\bf x}) =
g_0 + \delta g({\bf x})$ and $r({\bf x}) = r_0 + \delta r({\bf x})$.
The phase transition occurs when the control parameter $r_0$ becomes
sufficiently negative. As described earlier, when $g \equiv 0$ the
particle-hole symmetric problem is recovered. If the $|\partial_\tau
\psi|^2$ term is dropped and we take $g \equiv 1$, we obtain the
closest approximation to the boson coherent state Lagrangian,
(\ref{2.1}), which was the starting point for the work in Ref.\
\onlinecite{WK89}.  We shall see that the $|\partial_\tau \psi|^2$
term, which was ignored in Ref.\ \onlinecite{WK89}, is actually
crucial for a correct understanding of the critical behavior.  When
$g \equiv 0$ one obtains precisely the random rod model studied in
Ref.\ \onlinecite{DBC}.

As part of our scaling discussion, we will revisit previous
arguments\cite{FWGF} for the scaling relation $z = d$ for the
dynamical exponent for the dirty boson problem. Recent quantum Monte
Carlo results contradict this relation, finding $z = 1.40 \pm 0.02$
in $d = 2$.\cite{Baranger06} We will show that all of the previous
arguments, when looked at more carefully, in fact \emph{place no
constraint} on the exponent $z$.\cite{WM07} Lacking deeper
arguments, it would appear that $z$ remains an independent exponent,
undetermined by any scaling relation.\cite{Dohm}

\subsection{Scaling of superfluid density and compressibility}
\label{sec:sfscaling}

Recall that, as described in Sec.\ \ref{sec:sfdensity}, the
superfluid density, $\rho_s$, and compressibility, $\kappa$, measure
the system response to twists, spatial and temporal respectively, in
the superfluid order parameter. As in (\ref{3.3}), we introduce
\begin{equation}
\tilde{\psi}({\bf x},\tau)
= e^{-i({\bf k}_0 \cdot {\bf x} + \omega_0 \tau)}
\psi({\bf x}, \tau),
\label{4.2}
\end{equation}
and impose periodic boundary conditions on $\tilde{\psi}({\bf
x},\tau)$, identifying $\omega_0 = \theta_0 /\beta$ and ${\bf k}_0 =
(\theta_1/L_1,\ldots, \theta_d/L_d)$.  Analogously to (\ref{3.4}),
one obtains
\begin{equation}
{\cal L}_c^{{\bf k}_0,\omega_0}[\psi]
= {\cal L}_c^0[\tilde \psi] + \delta {\cal L}_c[\tilde \psi]
\label{4.3}
\end{equation}
with
\begin{eqnarray}
\delta {\cal L}_c[\tilde \psi] &=& -\int d^dx \int d\tau
\Big\{\frac{1}{2} (k_0^2 + \omega_0^2) |\tilde \psi|^2
\label{4.4} \\
&&-\ i\omega_0 \tilde \psi^* [\partial_\tau - g({\bf x})] \tilde \psi
+ i{\bf k}_0 \cdot \tilde \psi^* \nabla \tilde \psi \Big\}.
\nonumber
\end{eqnarray}
The expansion of the free energy (\ref{3.5}) in powers of $k_0$ and
$\omega_0$ takes the form
\begin{equation}
\delta f(k_0,\omega_0) =  - i \rho \omega_0
+ \frac{1}{2} (\kappa \omega_0^2 + \Upsilon k_0^2)
+ O(\omega_0^3,\omega_0 k_0^2),
\label{4.5}
\end{equation}
In addition to the result (\ref{3.14}) for the helicity modulus, one
may then identify
\begin{eqnarray}
\kappa &\equiv& \Upsilon_\tau = \left[\langle |\psi|^2 \rangle
\right]_{\rm av} + \int d^dx \int d\tau
\nonumber \\
&\times& \left[\langle \psi^*(\partial_\tau -  g)\psi]({\bf x},\tau)
[\psi^*(\partial_\tau  - g)\psi]({\bf 0},0) \rangle_c \right]_{\rm av}
\nonumber \\
\rho &\equiv& \rho_\tau
= -\left[\langle \psi^* (\partial_\tau - g)
\psi \rangle \right]_{\rm av}.
\label{4.6}
\end{eqnarray}
The spatial isotropy of (\ref{4.1}) implies that $\Upsilon_\alpha
\equiv \Upsilon$ is independent of the direction $\alpha$.  The
subscript ``$c$'' on the right hand side of the second equation
indicates a cummulant average, i.e., that the product of the
averages, namely $\rho^2$, should be subtracted from the integrand.
We shall ultimately require these expressions only when $g \equiv
0$, where $\rho \equiv 0$ as well.

\subsubsection{Random rod critical point scaling}
\label{subsec:rrscaling}

Let us  first consider the scaling of $\rho_s$ and $\kappa$ for the
classical random rod problem, $g(x) \equiv 0$.  Note that ${\bf k}_0
\cdot \tilde \psi^* \nabla \tilde \psi$ and $\omega_0 \tilde \psi^*
\partial_\tau \tilde\psi $ are \emph{symmetry breaking
perturbations}: the first breaks the ${\bf x} \leftrightarrow -{\bf
x}$ spatial inversion symmetry, and the second breaks the time
inversion symmetry $\tau \leftrightarrow -\tau$ symmetries of random
rod Lagrangian.  The latter corresponds precisely to particle-hole
symmetry.

Critical universality classes are, by definition, insensitive to
most changes in the detailed parameters of the Hamiltonian, but
symmetry breaking perturbations are often an exception, leading to
changes in the asymptotic critical behavior. One therefore expects
$k_0,\omega_0$ to be \emph{relevant} perturbations to the random rod
problem, entering the thermodynamics through scaling combinations
that diverge as the critical point is approached. Since $k_0$ is an
inverse length and $\omega_0$ is an inverse time, one expects them
to be scaled by the corresponding divergent correlation length and
time, respectively. This motivates the following form for the
singular part of the free energy:
\begin{equation}
f_s(k_0,\omega_0) \approx A |\delta|^{2-\alpha}
\Phi(k_0 \xi,\omega_0 \xi_\tau),
\label{4.7}
\end{equation}
where $\xi \approx \xi_0 |\delta|^{-\nu_0}$ and $\xi_\tau \approx
\xi_{\tau,0} |\delta|^{-\nu_{\tau 0}}$ are the correlation lengths
in the spatial and temporal directions, respectively, $A$ is a
nonuniversal amplitude, and the \emph{dynamical exponent} is defined
by $z_0 = \nu_{\tau 0}/\nu_0$. The subscript 0 on the exponents
indicate that they are those appropriate to the classical random rod
problem, and the generic auxillary parameter, $\delta$ is $r_0 -
r_{0,c}$ in (\ref{4.1}), but more generally is any parameter such as
chemical potential, pressure, strength of disorder, film thickness,
or magnetic field, which moves the system through the phase
transition at $T=0$, defined to occur at $\delta = 0$.  We assume
that $\delta>0$ corresponds to the disordered phase and $\delta < 0$
to the ordered (superfluid or superconducting)
phase.\cite{foot:signdelta} The two scaling arguments $k_0 \xi$ and
$\omega_0 \xi_\tau$ indeed diverge as $\delta \to 0$ for arbitrarily
small $k_0$ and $\omega_0$, consistent with their expected
relevance.

Although (\ref{4.8}) motivates the correct result, the fact that
$\omega_0$ and $k_0$ are infinitesimal in the zero temperature
thermodynamic limit, $\beta, L_\alpha \to \infty$, means that a
little more care is required to construct a rigorous scaling
formulation. More properly, the boundary condition dependence
appears in a finite size scaling ansatz for the free
energy,\cite{KW91}
\begin{equation}
\delta f^{\theta} \approx \beta^{-1} L^{-d}
\Phi_0^{\theta} [\delta (L/\xi_0)^{1/\nu_0},
\delta (\beta/\xi_{\tau,0})^{1/\nu_{\tau 0}}].
\label{4.8}
\end{equation}
The existence of a nonzero stiffness (in the ordered phase), i.e.,
via (\ref{3.6}), a leading finite-size correction of order $L^{-2}$
or $\beta^{-2}$, now requires that the scaling function obey
$\Phi_0^\theta(x,y) \approx x^{d\nu_0} y^{z_0\nu_0} (\Phi_1^\theta
x^{-2\nu_0} + \Phi_2^\theta y^{-2z_0 \nu_0})$ for large $x,y$, (and
$\delta < 0$), yielding
\begin{eqnarray}
\Upsilon &\approx& \xi_0^{2-d} \xi_{\tau,0}^{-1}
(\Phi_1^\theta/2 \pi^2 \theta^2) \delta^{\upsilon_0}
\nonumber \\
\kappa &\approx& \xi_0^{-d} \xi_{\tau,0}
(\Phi^{\theta}_2/2\pi^2 \theta^2) \delta^{\upsilon_{\tau 0}}
\label{4.9}
\end{eqnarray}
with the \emph{Josephson scaling relations,}\cite{FWGF}
\begin{eqnarray}
\upsilon_0 &=& (d + z_0 - 2)\nu_0 = 2 - \alpha_0 - 2\nu_0
\nonumber \\
\upsilon_{\tau 0} &=& (d - z_0)\nu_0 = 2 - \alpha_0 - 2 z_0 \nu_0,
\label{4.10}
\end{eqnarray}
and requiring in addition that $\Phi_{1,2}^{\theta} \propto
\theta^2$. We emphasize that the crucial assumption is that the
leading boundary condition dependence is all in the singular, i.e.,
finite size scaling part, of the free energy.  We can make this
assumption only because $k_0$ and $\omega_0$ introduce relevant
perturbations which fundamentally alter the symmetry of the
Lagrangian.  Further support for this assumption is that we expect
all stiffnesses to vanish identically in the disordered phase of the
classical model: thus $\Upsilon$ and $\kappa$ can have no analytic
contributions at all.

\subsubsection{Dirty boson critical point scaling: revisiting $z=d$}
\label{subsec:revisit_zvsd}

Now, we can try to extend the above arguments for nonzero $g$,
following Ref.\ \onlinecite{FWGF}.  Thus, one may posit identical
forms (\ref{4.7}) or (\ref{4.8}) for the singular part of the free
energy, but with exponents and scaling functions appropriate to the
dirty boson critical point.  One obtains then
\begin{eqnarray}
\Upsilon &\sim& |\delta|^\upsilon,\
\upsilon = 2-\alpha-2\nu = (d+z-2)\nu
\nonumber \\
\kappa &\sim& |\delta|^{\upsilon_\tau},\
\upsilon_\tau = 2-\alpha-2z\nu = (d-z)\nu.
\label{4.11}
\end{eqnarray}
For $g \neq 0$, both the Bose glass and superfluid phases are
compressible, so it is expected that the compressibility remains
finite right through the transition. This leads to the prediction
$z=d$.

However, although the result for $\Upsilon$ is believed to be
correct, there are a number of questionable assumptions underlying
the argument above for $\kappa$, as we shall now discuss.  For $g
\neq 0$, the density of the Bose glass phase varies smoothly with
$g_0$ (the control parameter analogous to the chemical potential
$\mu$). A temporal twist only perturbs slightly the term, $g \psi^*
\partial_\tau \psi$, that is \emph{already present} in the
Lagrangian, and is therefore not expected to produce a new relevant
perturbation.  Thus, the scaling variable \emph{cannot} be $\omega_0
\xi_\tau$.  Rather $\omega_0$ produces only an infinitesimal shift
$g_0 \rightarrow g_0 - i \omega_0$, which, through the above
analyticity argument (including also the possibility of complex
shifts), alters the free energy in a completely predictable fashion
unrelated to scaling. By way of contrast, spatial twists still
represent a relevant perturbation, and we therefore predict a
scaling form at $\omega_0 = 0$:
\begin{equation}
\delta f^{\theta} = \beta^{-1} L^{-d}
\Phi^\theta[\delta (L/\xi_0)^{1/\nu},
\delta (\beta/\xi_{\tau,0})^{1/\nu_\tau}],
\label{4.12}
\end{equation}
with $\Phi(x,y) \approx \Phi_1 x^{(d-2)\nu} y^{z\nu}$ for large
$x,y$, yielding $\Upsilon \approx (\Phi_1^{\theta}/2\pi^2 \theta^2)
\delta^\upsilon$, with $\upsilon = (d + z - 2)\nu = 2 - \alpha - 2
\nu$ as before.  We expect only small subleading corrections in
$y^{z\nu}$: since no temporal twist has been imposed, there can be
no $O(\beta^{-1},\beta^{-2})$ corrections.

Now, if we include a finite $\omega_0$, the basic change in
(\ref{4.12}) is that $g_0 \rightarrow g_0 - i\omega_0$ everywhere,
and in addition one must include changes arising from boundary
condition dependence of the analytic part of the free energy. One
obtains
\begin{eqnarray}
\delta f^{\theta} &=& \beta^{-1} L^{-d}
\Phi^{\theta}[\delta_\theta (L/\xi_0)^{1/\nu},
\delta_\theta (\beta/\xi_{\tau,0})^{1/\nu_\tau}]
\nonumber \\
&&+\ f^0(r_0,g_0 - i\omega_0) - f^0(r_0,g_0),
\label{4.13}
\end{eqnarray}
where $f^0$ is the free energy density in the absence of all twists
(i.e., under fully periodic boundary conditions), including both
analytic and singular parts, and $\delta_\theta = r_0 - r_{0,c}(g_0 -
i\omega_0) \approx \delta + i\omega_0 r_{0,c}'(g_0)$ is the perturbed
deviation from the critical line $r_{c,0}(g_0)$.  Most importantly,
$\Phi^{\theta}$ is the \emph{same function} as that in (\ref{4.12}),
and therefore produces only small corrections in
$(\beta/{\xi_{\tau}})^{-1}$.  The scaling function itself therefore
contributes only vanishing corrections to $\kappa$ in the limit
$\beta \to \infty$.

All contributions to $\kappa$ are therefore contained in $f^0$, and
arise from (a) its analytic part, (b) the (linear) $\omega_0$
dependence of $\delta_\theta$ in its singular part.\cite{FBJ}
Regarding (b), the leading singular part of $f^0$ is of the form
$f_{\rm sing}^0 \sim |\delta_\theta|^{2-\alpha}$.  The
$g_0$-dependence of $r_{0,c}$ couples derivatives with respect to
$g_0$ to those with respect to $\delta$, and one obtains therefore
contributions $\rho_\mathrm{sing} \sim |\delta_\theta|^{1-\alpha}$
to the density and $\kappa_\mathrm{sing} \sim
|\delta_\theta|^{-\alpha}$ to the compressibility.  The hyperscaling
relation $\alpha = 2-(d+z)\nu$ implies that $\alpha < 0$ under the
rather weak condition that $\nu
> 2/(d+z)$ (which by all evidence to date appears to be strongly
satisfied,\cite{foot:CCFS}) and the singular parts of $\rho$ and
$\kappa$ are expected to vanish at criticality. Put slightly
differently, given that $\kappa$ is already nonzero in the Bose
glass phase, signifying the existence of long range temporal
correlations on both sides of the transition, it would not be
surprising to find that the critical singularity, signifying the
adjustment of the density of states $\rho_1(\epsilon = 0)$ as the
density of large droplets increases, leads to only small corrections
to $\kappa$.

Regarding (a), the analytic part of the free energy has an expansion
\begin{eqnarray}
f^0_a(r_0,g_0) &=& -\rho_c(r_0)[g_0 - g_{0,c}(r_0)]
\label{4.14} \\
&&-\ \frac{1}{2} \kappa_c(r_0)[g_0 - g_{0,c}(r_0)]^2 + \ldots,
\nonumber
\end{eqnarray}
about the transition line $g_{0,c}(r_0)$, where $\rho_c(r_0)$ and
$\kappa_c(r_0)$ are now recognized as the finite values of $\rho$
and $\kappa$ at the transition.  These finite values are predicted
irrespective of the value of the exponent $z$, now seen as
unconstrained by the present arguments. Note further that since the
leading $\omega_0$-dependence is linear, if the leading boundary
condition dependence were indeed through the scaling combination
$\omega_0 \xi_\tau$ (or, more properly, the finite size scaling
variable $\delta/\beta^{1/z\nu}$), the leading contribution to the
density must take the form $\rho \sim |\delta|^{d\nu}$,
contradicting the fact that the density must be finite at the
transition.

To end this discussion, it is worth emphasizing why the arguments
above fail for the classical random rod model, $g({\bf x}) \equiv
0$.  The key point is that $r_{0,c}(g_0)$ is singular at the special
value $g_0 = 0$, and $\delta_\theta$ is no longer an analytic
function of $\omega_0$. Essentially, the special symmetry at $g_0
\equiv 0$ implies that $\delta$ and $\mu$ are ``orthogonal''
thermodynamic coordinates and the derivatives with respect to $\mu$
that define $\kappa \equiv \kappa_s$ do not mix with derivatives
with respect to $\delta$.  We have already seen, therefore, that
$\kappa \equiv 0$ in the disordered phase. We therefore expect
$\kappa$ to rise continuously from zero for $\delta < 0$, with the
exponent $\upsilon_\tau > 0$. This implies that $z \leq d$ in this
case (equality is still permitted and would imply a discontinuity in
$\kappa$ at $\delta=0$, which indeed is the case in $d=1$---see
Sec.\ \ref{sec:1d}).  Note that for homogeneous classical disorder,
where the coefficient $r$ in (\ref{4.1}) depends on \emph{both}
${\bf x}$ and $\tau$, we will have isotropic scaling, $z=1$.  The
rod disorder should increase $z$.

\subsubsection{Scaling of correlations}
\label{subsec:corrscale}

Let us now consider the two-point correlation function,
\begin{equation}
G({\bf x},\tau) = \left[\langle \psi({\bf x},\tau)
\psi({\bf 0},0) \rangle_c \right]_\mathrm{av},
\label{4.15}
\end{equation}
whose Fourier transform is normally assumed to scale in the form,
\begin{equation}
\hat G({\bf k},\omega) \approx C|\delta|^{-\gamma}
\hat g(k\xi,\omega\xi_\tau),
\label{4.16}
\end{equation}
where $\gamma$ is the susceptibility exponent. At small $k,\omega$
in the superfluid phase, the dynamics is governed by the
hydrodynamic Lagrangian (\ref{3.29}).  Since this is now a bulk
Lagrangian, with $\rho$ the actual bulk density, with
$2\pi$-periodic boundary conditions on the phase $\phi$, the $i \rho
\partial_\tau \phi$ term integrates to $2\pi i n \rho V = 2\pi i N$,
and therefore drops out of $e^{S_\mathrm{eff}}$.  Recalling that at
hydrodynamic scales, $\psi = \psi_0 e^{i\phi}$ where $\psi_0$ is the
order parameter, the resulting Gaussian Lagrangian yields
\begin{eqnarray}
G({\bf x},\tau) &\approx& |\psi_0|^2
\left[e^{-\frac{1}{2} \langle
[\phi({\bf x},\tau) - \phi({\bf 0},0)]^2 \rangle}
- e^{-\langle [\phi({\bf 0},0)]^2 \rangle} \right]
\nonumber \\
&\approx& |\psi_0|^2
\langle \phi({\bf x},\tau) \phi({\bf 0},0) \rangle,
\label{4.17}
\end{eqnarray}
in which fluctuations about the ordered state are assumed small. In
Fourier space one therefore obtains
\begin{equation}
\hat G({\bf k},\omega) \approx
\frac{|\psi_0|^2}{\Upsilon k^2 + \kappa \omega^2}.
\label{4.18}
\end{equation}
If one naively matches (\ref{4.16}) and (\ref{4.17}), one concludes
that $\hat g(x,y) \approx (g_1 x^2 + g_2 y^2)$ for small $x,y$ and
hence that
\begin{equation}
\frac{\Upsilon}{|\psi_0|^2}
\approx \frac{g_1 \xi_0^2}{C|\delta|^{2\nu-\gamma}},\ \
\frac{\kappa}{|\psi_0|^2}
\approx \frac{g_2 \xi_{\tau,0}^2}{C|\delta|^{2z\nu-\gamma}}.
\label{4.19}
\end{equation}
Using the well known scaling relation $\alpha + 2\beta + \gamma =
2$, one recovers (\ref{4.11}).

However, (\ref{4.19}) is really just a disguised version of the free
energy argument. Thus, $S_\mathrm{eff}$ assumes that the energetics
of global phase twists also describes slowly varying local phase
twists.  Thus, locally we replace ${\bf k}_0$ by $\nabla \phi$ and
$\omega_0$ by $\partial_\tau \phi$, then integrate over space-time.
Once again, (\ref{4.16})--(\ref{4.19}) are expected to be valid for
the random rod problem. However, for the dirty boson problem, if
$\kappa$ arises from the nonscaling part of the free energy, it is
unlikely that it can now arise from the scaling part of the
two-point function. Thus, in place of (\ref{4.16}), we propose
instead the scaling form
\begin{equation}
\hat G({\bf k},\omega) \approx
\frac{|\psi_0|^2}{D |\delta|^{2-\alpha}
\hat \Gamma(k\xi,\omega\xi_\tau)
+ \hat \Gamma_a({\bf k},\omega)},
\label{4.20}
\end{equation}
where $\Gamma_a$ is analytic. Although we presently have no
theoretical support for this ``self-energy'' scaling form, we appeal
to the similarity of the denominator to (\ref{4.13}). If one matches
(\ref{4.16}) to (\ref{4.18}), one now assumes that $\Gamma(x,y)
\approx \gamma_1 x^2$ for small $x,y$ while $\Gamma_a({\bf
k},\omega) \approx \kappa \omega^2$ for small $k,\omega$. The $x^2$
term yields $\Upsilon \approx D \gamma_1 \xi_0^2 |\delta|^\upsilon$,
with $\upsilon$ given by (\ref{4.11}) as as required. From the
analytic term we recover a finite $\kappa$ without any scaling
constraint on $z$. Standard static scaling, without any unusual
analytic corrections, is recovered for $\omega = 0$.

To summarize key the results of this subsection, the leading
behavior of $\kappa$ is governed by the analytic part of the free
energy, and is unconstrained (by any argument so far presented) by
the dynamical exponent $z$.\cite{WM07} The critical behavior of
$\kappa$ is governed by the exponent $\alpha$, and yields only
subleading corrections to the analytic behavior.

\subsection{Crossover exponent associated with particle-hole
symmetry breaking}
\label{sec:xover}

Since we expect the presence of $g({\bf x})$ to change the
universality class of the the phase transition, there must be an
associated positive crossover exponent, $\phi_g$, which quantifies
the instability of the classical random rod fixed point with respect
to this term. What is the value of $\phi_g$, and what conditions
does it place on the values of the classical fixed point exponents?

To begin to answer this question, let us write ${\cal L}_c = {\cal
L}_0 + {\cal L}_g$, where
\begin{equation}
{\cal L}_g = \int d^dx \int d\tau \left[\frac{1}{2} g({\bf x})^2
|\psi|^2 - g({\bf x}) \psi^* \partial_\tau \psi \right].
\label{4.21}
\end{equation}
We assume (see below) that $[g({\bf x})]_{\rm av} = 0$, $[g({\bf x})
g({\bf x}^\prime)]_{\rm av} = \Delta_g \varphi({\bf x}-{\bf
x}^\prime)$, which implies a statistical particle-hole symmetry, and
that $g({\bf x})$ and $r({\bf x})$ are statistically independent.
The correlation function takes the form $\varphi({\bf x}) =
\delta({\bf x})$ for uncorrelated disorder, but for reasons that
will become evident below, we shall allow more general long-range
power-law correlated disorder with
\begin{equation}
\varphi({\bf x}) \sim |{\bf x}|^{-(d+a)}
\label{4.22}
\end{equation}
for large $|{\bf x}|$ and some exponent $a$.  The crossover
exponent, $\phi_g$, which in general will be a function of $a$, is
defined by the following scaling form, valid for small $\Delta_g$:
\begin{equation}
f_s \approx A |\delta_g|^{2-\alpha_0} \Phi_g
\left(\frac{B \Delta_g}{|\delta_g|^{\phi_g}} \right)
\label{4.23}
\end{equation}
where $\alpha_0$ is the random rod (quantum) specific heat exponent,
and the subscript on $\delta_g$ is to serve as a reminder that
$\Delta_g$ will also generate a shift in the position of the
critical point: $\delta_g = \delta + c_1 \Delta_g + \ldots$.  The
value of $\phi_g$ may now be inferred from the derivative
\begin{eqnarray}
\left(\frac{\partial f_s}{\partial \Delta_g} \right)_{\Delta_g=0}
&\approx& A |\delta|^{2-\alpha_0} [B|\delta|^{-\phi_g}
- (2-\alpha_0)c_1 |\delta|^{-1}],
\nonumber \\
&&\ \ \ \ \ \ \ \ \ \ |\delta| \rightarrow 0,
\label{4.24}
\end{eqnarray}
where we choose $A$ and $B$ so that $\Phi_g(0) = \Phi_g'(0) = 1$.
Note the very singular $|\delta|^{-1}$ term generated by the shift,
which may often dominate the $|\delta|^{-\phi_g}$ term of interest.
Now, this derivative may also be calculated directly within
perturbation theory:
\begin{equation}
f(\Delta_g)-f(0) = -\frac{1}{V_D}
\left[\langle {\cal L}_g \rangle_0
+ \frac{1}{2} [\langle {\cal L}_g^2 \rangle_0
- \langle {\cal L}_g \rangle_0^2] + O(g^3) \right]
\label{4.25}
\end{equation}
where the averages are with respect to ${\cal L}_0$.  Assuming that
$g({\bf x})$ and $r({\bf x})$ self average, we obtain
\begin{widetext}
\begin{eqnarray}
f(\Delta_g)-f(0) &=& -\frac{1}{V_D} \left[
[\langle {\cal L}_g \rangle_0]_{\rm av}
+ \frac{1}{2}[\langle {\cal L}_g^2 \rangle_0]_{\rm av}
+ O(g^4) \right]
\label{4.26} \\
&=& -\frac{1}{2} \Delta_g \varphi({\bf 0})
\left[\langle |\psi({\bf 0},0)|^2 \rangle_0 \right]_\mathrm{av}
- \frac{1}{2} \Delta_g \int d\tau \int d^dx
\varphi({\bf x}) \big[\langle \psi^*
\partial_\tau \psi({\bf x},\tau)
\psi^* \partial_\tau \psi({\bf 0},0) \rangle_0 \big]_{\rm av}
+ O(\Delta_g^2),
\nonumber
\end{eqnarray}
\end{widetext}
where independence of $g({\bf x})$ and $r({\bf x})$ has been used.
Thus
\begin{equation}
-2\left(\frac{\partial f}{\partial \Delta_g} \right)_{\Delta_g=0}
= \varphi({\bf 0}) \varepsilon_0
+ \int d\tau \int d^dx \varphi({\bf x}) {\cal G}_g({\bf x},\tau)
\label{4.27}
\end{equation}
where $\varepsilon_0 = \left[\langle |\psi|^2 \rangle_0
\right]_\mathrm{av}$, and we have defined the correlation function
\begin{equation}
{\cal G}_g({\bf x},\tau) = [\langle \psi^*
\partial_\tau \psi({\bf x},\tau) \psi^*
\partial_\tau \psi({\bf 0},0) \rangle_0 ]_{\rm av}.
\label{4.28}
\end{equation}

Let us define the Fourier transforms
\begin{eqnarray}
\hat \varphi({\bf k}) &=& \int d^dx e^{i{\bf k}\cdot{\bf x}}
\varphi({\bf x})
\nonumber \\
\hat {\cal G}_g({\bf k},\omega) &=& \int d\tau \int d^dx
e^{i({\bf k} \cdot {\bf x} + \omega \tau)}
{\cal G}_g({\bf x},\tau).
\label{4.29}
\end{eqnarray}
Then when $g({\bf x}) \equiv 0$ we have from (\ref{4.6}),
\begin{equation}
\Upsilon_\tau = \varepsilon_0 + \hat{\cal G}_g({\bf 0},0)
\sim |\delta|^{(d-z_0)\nu_0},
\label{4.30}
\end{equation}
where the exponents are appropriate to the random rod problem.  More
generally we expect a scaling form for small $|{\bf k}|$ and
$\omega$:
\begin{equation}
\Upsilon_\tau({\bf k},\omega) \equiv
\varepsilon_0 + \hat{\cal G}_g({\bf k},\omega)
\approx A_1 |\delta|^{(d-z_0)\nu_0}
{\cal Y}(k\xi,\omega \xi_\tau)
\label{4.31}
\end{equation}
where for $\delta > 0$ we have $\lim_{w,s \rightarrow 0}{\cal
Y}(w,s)=0$ while for $\delta < 0$ we have $\lim_{w,s \rightarrow
0}{\cal Y}(w,s)=1$.  Thus,
\begin{eqnarray}
-2\left(\frac{\partial f}{\partial \Delta_g} \right)_{\Delta_g=0}
&=& \varphi({\bf 0}) \varepsilon_0
+ \int_{\bf k} \hat \varphi({\bf k})
\hat {\cal G}_g({\bf k},\omega=0)
\nonumber \\
&=& \int_{\bf k} \hat \varphi({\bf k}) \Upsilon_\tau({\bf k},0),
\label{4.32}
\end{eqnarray}
where we have used the convenient shorthand notation $\int_{\bf k}
\equiv \int \frac{d^dk}{(2\pi)^d}$, $\int_\omega \equiv \int
\frac{d\omega}{2\pi}$, etc. For uncorrelated disorder, $\hat
\varphi({\bf k}) \equiv 1$, while for power law correlated behavior,
$\hat \varphi({\bf k}) \approx c_a k^a + c_0 + \ldots$ as $k
\rightarrow 0$, where $a$ was defined in (\ref{4.22}).  The final
$k$-integral must be treated carefully to extract its
$\delta$-dependence.  Let us rewrite
\begin{equation}
\tilde {\cal Y}(w) = w^{z_0-d} {\cal Y}(w,0),\
\tilde {\cal Y}(w \rightarrow \infty) \rightarrow y_0 > 0,
\label{4.33}
\end{equation}
where the finite limit follows from the requirement that
$\Upsilon({\bf k},0)$ should be a well defined function of ${\bf k}$
alone (in this case a power law $\propto y_0 k^{d-z}$) when $\delta
= 0$.  Clearly one cannot simply scale $\xi$ out of the integral in
(\ref{4.15}) since the resulting integral over $\hat \varphi(w/\xi)
w^{d-z} \tilde {\cal Y}_\pm(w)$ will not converge.  Rather, one must
first subtract out the large $w$ behavior.  Let us write
\begin{equation}
\tilde {\cal Y}(w) \approx y_0 + y_1 w^{-\omega_1}
+ y_2 w^{-\omega_2} + y_3 w^{-\omega_3} + \ldots
\label{4.34}
\end{equation}
where the source of the spectrum of exponents, $0 < \omega_1 <
\omega_2 < \omega_3 < \ldots$, will be discussed below.  Each term
$y_j w^{-\omega_j}$ yields a contribution
\begin{eqnarray}
\int_{\bf k} \hat \varphi({\bf k}) y_j k^{d-z_0}
(k\xi)^{-\omega_j} = b_j|\delta|^{\omega_j \nu}
\nonumber \\
b_j \equiv \xi_0^{-\omega_j}y_j \int_{\bf k}
\hat \varphi({\bf k}) k^{-\omega_j+d-z_0},
\label{4.35}
\end{eqnarray}
so long as $\omega_j < \omega_{\rm max}(a) \equiv
2d-z_0+\min\{a,0\}$ (i.e., the integral converges at small $k$;
convergence at large $k$ being ensured by the implicit lattice
cutoff, $k<k_\Lambda \sim \frac{\pi}{a}$).  Let us define the
subtracted scaling function,
\begin{equation}
\delta \tilde {\cal Y}(w) = \tilde {\cal Y}(w)
- \sum_{\omega_j < \omega_{\rm max}(a)} y_j w^{-\omega_j}.
\label{4.36}
\end{equation}
Then
\begin{eqnarray}
\int_{\bf k} \hat \varphi({\bf k}) \tilde {\cal Y}(k\xi) k^{d-z_0}
&\approx& \sum_{\omega_j < \omega_{\rm max}(a)} b_j
|\delta|^{\omega_j \nu}
\label{4.37}\\
&&+\ \int_{\bf k} k^{d-z_0} \hat \varphi({\bf k})
\delta \tilde {\cal Y}(k\xi),
\nonumber
\end{eqnarray}
where the last term may be evaluated as
\begin{eqnarray}
\int_{\bf k} k^{d-z_0} \hat \varphi({\bf k}) \delta \tilde {\cal Y}(k\xi)
&\approx& b_a |\delta|^{(2d-z_0+a)\nu} + b_0 |\delta|^{(2d-z_0)\nu},
\nonumber \\
b_a &\equiv& c_a \int_{\bf w} w^{d-z_0+a} \delta \tilde {\cal Y}(w)
\nonumber \\
b_0 &\equiv& c_0 \int_{\bf w} w^{d-z_0} \delta \tilde {\cal Y}(w).
\label{4.38}
\end{eqnarray}
In fact $b_0=0$ in our case because, tracing through the
definitions, the integral is dominated by the large $\tau$ limit of
${\cal G}_g({\bf 0},\tau)$, which is proportional to the square of
$\langle \psi^* \partial_\tau \psi \rangle_0$.  The latter
vanishes by particle-hole symmetry of the random rod model.

Now, the origin of the exponents $\omega_j$ is as follows: the
operator
\begin{equation}
P\equiv \int d^dx \int d\tau \int d\tau^\prime
\psi^* \partial_\tau \psi({\bf x},\tau)
\psi^* \partial_\tau \psi({\bf x},\tau^\prime)
\label{4.39}
\end{equation}
will have an operator product expansion in terms of eigenoperators
of a renormalization group transformation near the critical fixed
point of interest: $P = h_1 O_1 + h_2 O_2 + h_3 O_3 + \ldots$, and
$O_i$ is assumed to have renormalization group eigenvalue
$\lambda_i$.  This implies that $\langle P \rangle \sim h_1
|\delta|^{(d+z_0-\lambda_1)\nu} + h_2
|\delta|^{(d+z_0-\lambda_2)\nu} + \ldots$.  But since $\langle P
\rangle = \int_{\bf k} \tilde {\cal Y}(k\xi) k^{d-z_0}$, comparison
with (\ref{4.34}) and (\ref{4.35}) implies that $\omega_j =
d+z_0-\lambda_j$.  The $\omega_j$ therefore reflect the
renormalization group transformation properties of $P$ near the
fixed point.

We can now understand the crossover exponent, $\phi_g$.  Comparing
(\ref{4.24}) and (\ref{4.37}) together with (\ref{4.38}) (with
$b_0=0$), we see that
\begin{eqnarray}
2-\alpha_0-\phi_g &=& \nu\, \min \{\omega_j, 2d-z_0+a\}
\nonumber \\
\Rightarrow\ \lambda_g &\equiv& \frac{\phi_g}{\nu}
= \max\{\lambda_j,2z_0-d-a\}.
\label{4.40}
\end{eqnarray}
The crossover exponent is therefore either $\phi_g = \lambda_1\nu$,
or for small enough $a$, $\phi_g = (2z_0-d-a)\nu$.  Thus
\begin{equation}
\lambda_g = \left\{\begin{array}{ll}
2z_0-d-a, & a < 2z_0-d-\lambda_1
\\
\lambda_1, & a > 2z_0-d-\lambda_1,
\end{array} \right.
\label{4.41}
\end{equation}
implying
\begin{equation}
\phi_g > 0 \Leftrightarrow \left\{\begin{array}{ll}
z_0 > \frac{d+a}{2}, & a < 2z_0-d-\lambda_1
\\
\lambda_1 > 0, & a > 2z_0-d-\lambda_1.
\end{array} \right.
\label{4.42}
\end{equation}
In particular, for short range correlated disorder, where in effect
$a \rightarrow \infty$, we require $\lambda_1 > 0$ in order that
dirty boson disorder destabilize the random rod fixed point.  The
exponent $\lambda_1$ is a nontrivial exponent, and we shall compute
it within the $\epsilon,\epsilon_\tau$-expansion.\cite{DBC}

A naive estimate for $\lambda_1$ is obtained by supposing that the
equality $2z_0-d-a=\lambda_1$ should occur when $a \simeq 0$, yielding
$\lambda_1 \simeq 2z_0-d$ which becomes positive for $z_0 >
\frac{d}{2}$. Note that this same estimate would have been obtained
from the second term in (\ref{4.21}) if we had assumed $b_0 \neq 0$.
As an aside, this estimate is actually exact in the corresponding
derivation of the Harris criterion for classical disordered magnets.
There the correlation function $\langle |\psi({\bf x})|^2 |\psi({\bf
y})|^2 \rangle_0$ appears.  Since $\langle \psi^2 \rangle_0$ does
not vanish, neither does the coefficient analogous to $b_0$.  This
gives rise to a free energy contribution $\langle |\psi|^2
\rangle_0^2 \sim \delta^{2-2\alpha_0}$ which leads immediately to
the Harris criterion, $\phi_g = \alpha_0$.

In fact, we shall find that $\lambda_1 > 2z_0-d$, i.e., $a$ drops
out at some negative value, and $\phi_g$ becomes positive for $z_0$
larger than a value somewhat \emph{less} than $\frac{d}{2}$.  The
random rod result $z_0=1$ in $d=1$ is consistent with this
criterion, although this case is somewhat special because the random
rod fixed point is the \emph{same} as the pure fixed point in $d=1$
(i.e., random rod disorder is irrelevant, though boson disorder is
relevant).  The generalized Harris criterion\cite{DBC} indicates
that rod disorder is irrelevant when $\alpha_\mathrm{pure} +
\nu_\mathrm{pure} < 0$. Using hyperscaling (valid here for $d<3$),
and the fact that $z=1$ at the pure fixed point, this requires
$\nu_\mathrm{pure} > \frac{2}{d}$ (compare the less stringent
requirement, $\nu_\mathrm{pure} > \frac{2}{d_\mathrm{tot}}$ with
$d_\mathrm{tot} = d+1$, for the usual Harris criterion for point
disorder). For $d=1$, $\nu_\mathrm{pure} \rightarrow \infty$, while
for $d=2$, $\nu_\mathrm{pure} \simeq \frac{2}{3}$, so the pure fixed
point becomes unstable to rod disorder somewhere in between $d=1$
and $d=2$.  In all cases where rod disorder is irrelevant, one then
trivially has $z_0 > \frac{d}{2}$, and dirty boson disorder will
certainly be relevant.

We now turn to the question of the relevance  of $g_0 \equiv [g({\bf
x})]_{\rm av}$, i.e., of full breaking of particle-hole symmetry. If
one carries through a naive scaling analysis using ${\cal L}_{g_0} =
\int d^dx \int d\tau \left(\frac{1}{2} g_0^2 |\psi|^2 - g_0 \psi^*
\partial_\tau \psi \right)$ in place of (\ref{4.21}), one
obtains $\partial f/\partial g_0 = 0$, while
\begin{equation}
-\frac{\partial^2 f}{\partial g_0^2}
= \varepsilon_0 + \hat {\cal G}_g({\bf 0},0)
= \Upsilon_\tau \sim |\delta|^{(d-z_0)\nu_0}.
\label{4.43}
\end{equation}
If the $g_0$-dependence of the singular part of the free energy
scales in the form
\begin{equation}
f_s = A' \delta^{(d + z_0)\nu} \Phi_{g_0}
\left(\frac{g_0}{|\delta|^{\phi_{g_0}}} \right),
\label{4.44}
\end{equation}
it is tempting to identify the crossover exponent $\phi_{g_0}$ via
\begin{equation}
2-\alpha_0-2\phi_{g_0} = (d-z_0)\nu\ \Rightarrow\
\phi_{g_0} = z_0 \nu
\label{4.45}
\end{equation}
which is always strongly positive. In the absence of all disorder,
this is the correct exponent (with $z_\mathrm{pure}=1$, and
$\nu_\mathrm{pure} = \nu_{d+1}^\mathrm{XY}$) describing the
crossover from the $(d+1)$-dimensional XY behavior at the tip of the
Mott lobe, to the generic onset of superfluidity in a dilute Bose
gas as the density is increased from zero away from the tip. This
same exponent describes the shape of the pure system Mott lobes near
their tips, Fig.\ \ref{fig:phases}(a): since the transition line is
defined by some critical value $x_c$ of the scaling function
argument $x = g_0/|\delta|^{\phi_{g_0}}$, this leads to $g_{0,c}
\approx x_c |\delta|^{\phi_{g_0}}$, i.e.,\cite{FWGF}
\begin{equation}
\mu_c(J_0) \sim |J_0 - J_c^0|^{\nu_{d+1}^\mathrm{XY}}.
\label{4.46}
\end{equation}
The same argument implies that
\begin{equation}
\mu_c(J_0) \sim |J_0- J_c^\mathrm{RR}|^{z_0 \nu_0}
\label{4.47}
\end{equation}
near the random rod critical points in Fig.\ \ref{fig:phases}(c). It
is likely that $z_0 \nu_0 > 1$ in $d=3$, leading to the pictured
cusps.\cite{DBC}

One must be careful in interpreting the result (\ref{4.45}) in the
presence of random rod disorder. The random rod problem leads to an
incompressible glassy phase [see Fig.\ \ref{fig:phases}(c)].
Perturbing the random rod critical point with either $g_0$ or
$\Delta_g$ therefore not only changes the critical behavior, but
also the nature of the glassy phase. The difference between $\phi_g$
in (\ref{4.40}) and $\phi_{g_0}$ in (\ref{4.45}) therefore reflects
the different rates at which the crossover to the Bose glass phase
occurs under the influence of the two perturbations, with $g_0$
clearly having the stronger effect.

We emphasize that the two crossover exponents describe, in
renormalization group language, the rate at which one is initially
driven away from the random rod fixed point in the two orthogonal
directions described by $g_0$ and $\Delta_g$.  They tell one nothing
about the eventual termination of the flows on some new stable fixed
point.  We have argued physically, supported by the droplet picture
of Sec.\ \ref{sec:excite}, that although initially growing rapidly,
$g_0$ must vanish again as the dirty boson fixed point is
approached, its main role being to induce a finite value of
$\Delta_g$ under renormalization.  Quantifying this expected
\emph{irrelevance} of $g_0$ (i.e., $\phi_{g_0} < 0$) at the
commensurate dirty boson fixed point,\cite{foot:zvsd} would require
an analysis of the right hand side of (\ref{4.43}) in the presence
of a finite value of $\Delta_g$.  We saw in the previous subsection
that in this case $\Upsilon_\tau = \kappa$ is dominated by analytic
terms in the free energy, and hence that $\phi_{g_0}$ is related to
subleading terms in the singular part of the free energy that cannot
be inferred from a simple scaling analysis.  In Sec.\
\ref{sec:epstau} will obtain $\phi_{g_0}$ within the $\epsilon,
\epsilon_\tau$-expansion, and demonstrate explicitly the manner in
which particle-hole asymmetry becomes irrelevant at the commensurate
fixed point for sufficiently large $\epsilon_\tau$.

\section{Calculations in one dimension}
\label{sec:1d}

\subsection{Sine-Gordon model}
\label{sec:sinegordon}

In this section we review and expand upon the analysis of
one-dimensional versions of the dirty boson problem, with emphasis
on the weak disorder limit.\cite{FWGF,GS88} In App.\ \ref{app:b} we
derive various dual representations for the one-dimensional
Lagrangian based on the discrete-time Villain representation,
(\ref{B4}).  We shall analyze the sine-Gordon version, (\ref{B14})
with (\ref{B15}):
\begin{eqnarray}
{\cal L}_{\rm SG} &=& -\frac{1}{2} \sum_{\bf R}
\left[\frac{1}{K_I}(\partial_T S_{\bf R})^2
+ V_0(\partial_I S_{\bf R})^2 \right]
\nonumber \\
&&+\ \sum_{\bf R} \mu_I (\partial_I S_{\bf R})
+ 2 y_0 \sum_{\bf R} \cos(2\pi S_{\bf R}),\ \ \ \
\label{5.1}
\end{eqnarray}
where we have assumed that $V_{IJ} \equiv V_0 \delta_{IJ}$ is
diagonal.  Here ${\bf R} = (I,T)$, with integer $I,T$, are points on
a discrete space-time (dual) lattice, $-\infty < S_{\bf R} < \infty$
are continuous spin variables, the discrete derivatives are defined
by $\partial_I S_{\bf R} = S_{(I+1,T)} - S_{(I,T)}$, $\partial_T
S_{\bf R} = S_{(I,T+1)} - S_{(I,T)}$, and the cosine term represents
an external periodic potential which prefers integer values of
$S_{\bf R}$. This model has the physical interpretation of a
fluctuating interface, represented by the ``height'' variables
$S_{\bf R}$.  The coefficient $K_I$ is proportional to the Josephson
coupling [see equation (\ref{B1})]. In the absence of $\mu_I$, which
has the interpretation of a random tilt potential, the phase
transition in this model is from a flat phase at large $1/K_I$,
where $S_{\bf R}$ has only small fluctuations about some integer
value and exponentially decaying correlations, to a rough phase, at
large $K_I$, in which the interface wanders and has logarithmically
divergent height-height correlations. This rough phase corresponds
to the superfluid phase in the boson model, and the renormalized,
long wavelength value of $y_0$ vanishes.  In the presence of the
random tilting potential, $\mu_I$, the rough phase is qualitatively
unchanged, but the flat phase is no longer necessarily quite so
flat: see below.

Let us decompose $\mu_I = \mu_0 + \delta \mu_I$ into a uniform part
$\mu_0$ and a random part with $[\delta \mu_I]_{\rm av} = 0$.  Thus,
nonzero $\mu_0$ represents the breaking of particle-hole symmetry.
When $\mu_0 = 0$ the interface will be globally flat.  For
sufficiently large $\mu_0 > \mu_{0,c}$, where $\mu_{0,c}$ represents
the Mott gap (which will vanish for some combination of sufficiently
large $\delta \mu_I$, large $K_I$, small $V_0$ and small $y_0$) the
interface will acquire a global tilt, with $[\langle S_{\bf R} -
S_{\bf R'} \rangle]_{\rm av} = Q_0(I-I')$.  The exact form of the
slope function
\begin{equation}
Q_0(\mu_0,K_0,V_0,y_0)
= [\langle \partial_I S_{\bf R} \rangle]_{\rm av}
= -\frac{\partial f_\mathrm{SG}}{\partial \mu_0},
\label{5.2}
\end{equation}
which is proportional to the density difference from the Mott phase,
can be computed perturbatively in powers of $y_0$ (see below).  In
the special case $\mu_I/V_0 = \frac{1}{2}$, the spatial derivative
terms in (\ref{5.1}) may be combined in the form $V_0 (\partial_I
S_{\bf R} - \frac{1}{2})^2$, and there is an exact degeneracy
between $\partial_I S_{\bf R} = 0,1$ (the integer values preferred
by the cosine term). This is the particle-hole symmetric model at
half filling, $Q_0 =
\frac{1}{2}$.\cite{Refael,foot:1dsinglet,foot:1dsingletinterface}

\subsection{Perturbation theory in the superfluid phase}
\label{sec:sfperturb}

Let us define the random walk
\begin{equation}
w_I = \frac{1}{V_0} \sum_{J=0}^I \delta \mu_J,
\label{5.3}
\end{equation}
(defined to be minus the sum from $J=I$ to 0 for $I<0$), and let
\begin{equation}
\tilde S_{\bf R} = S_{\bf R} - Q_0 I - w_I,\ {\bf R}=(I,T).
\label{5.4}
\end{equation}
With this choice, one will have $[\langle \tilde S_{\bf R} - \tilde
S_{\bf R'} \rangle]_{\rm av} \equiv 0$.  The sine-Gordon Lagrangian
takes the form
\begin{eqnarray}
{\cal L}_{SG} &=& \frac{1}{2} \sum_{\bf R}
\left[\frac{1}{K_I}(\partial_T\tilde S_{\bf R})^2
+ V_0(\partial_I \tilde S_{\bf R})^2 \right]
\nonumber \\
&&-\ 2y_0 \sum_{\bf R} \cos[2\pi (\tilde S_{\bf R} + w_I + Q_0 I)]
\nonumber \\
&&+\ E_0 \beta L,
\label{5.5}
\end{eqnarray}
where the constant term is
\begin{eqnarray}
E_0 &=& -\frac{1}{\beta L} \sum_{\bf R}
\frac{(Q_0 V_0 + \delta \mu_I)
(2\mu_0 - Q_0 V_0 + \delta \mu_I)}{2 V_0}
\nonumber \\
&=& -\frac{[\delta \mu_I^2]_{\rm av}}{2 V_0}
- \mu_0 Q_0 + \frac{1}{2} V_0 Q_0^2.
\label{5.6}
\end{eqnarray}
In (\ref{5.4}) we have dropped a sub-extensive boundary term
$-(\mu_0 - Q_0 V_0) \sum_{\bf R} \partial_I \tilde S_{\bf R}$ since
$\sum_{\bf R} \partial_I \tilde S_{\bf R} = \sum_T (\tilde S_{L,T} -
\tilde S_{0,T}) = O(\sqrt{\beta L})$, and therefore yields vanishing
contribution in the thermodynamic limit.  This term may in fact be
made to vanish identically by choosing periodic boundary conditions
for $\tilde S_{\bf R}$.

The explicit dependence on $\mu_0$ is only in the last (constant)
term of (\ref{5.5}), and if follows from (\ref{5.2}) that $Q_0$ may
be determined determined by minimizing the free energy at fixed
$\mu_0$:
\begin{equation}
\left(\frac{\partial f_{\rm SG}}{\partial Q_0}\right)_{\mu_0} = 0.
\label{5.7}
\end{equation}
When $y_0=0$ the condition (\ref{5.7}) yields $Q_0 = \mu_0/V_0$ and
the $\delta \mu_I$ yield only a trivial additive constant to the
free energy. In this limit, for $K_I \equiv K_0$ fixed, the
two-point correlation function is given by
\begin{eqnarray}
G({\bf R}-{\bf R}') &\equiv& \frac{1}{2}
\langle (\tilde S_{\bf R} -
\tilde S_{{\bf R}^\prime})^2 \rangle_0
\label{5.8} \\
&\approx& \frac{1}{2\pi} \sqrt{\frac{K_0}{V_0}}
\ln \left[\frac{\rho({\bf R}-{\bf R}^\prime)}{\rho_0} \right],\
\rho \rightarrow \infty,
\nonumber
\end{eqnarray}
where,
\begin{equation}
\rho({\bf R}-{\bf R}^\prime)
= \left[\frac{1}{K_0 V_0}(I-I^\prime)^2
+ K_0 V_0(T-T^\prime)^2 \right]^{\frac{1}{2}}
\label{5.9}
\end{equation}
is the appropriately rescaled distance, and $\rho_0 = O(1)$ is a
constant scale factor.  When $K_I$ fluctuates, its disorder average
must be included. The result is still (\ref{5.8}), but $K_0$ then
becomes a complicated effective parameter.  The generalized Harris
criterion (Ref.\ \onlinecite{DBC} and Sec.\ \ref{sec:scaling})
implies that disorder in the coefficient $K_0$ is an
\emph{irrelevant} perturbation at the pure (2D XY) critical point at
integer filling in $d=1$, so we will, for the rest of this section,
simply take $K_I \equiv K_0$ when considering the influence of the
$\delta \mu_I$.

Let us then consider the $y_0$ term as a perturbation on the
quadratic term in ${\cal L}_{SG}$.  Deep in the superfluid/rough
phase, where $K_0/V_0$ is large, this is a well defined expansion.
It is also well defined when $Q_0$ is not too small: the cosine term
in (\ref{5.4}) then oscillates very rapidly from site to site, and
effectively averages itself out. This corresponds to the region
between Mott lobes in Fig.\ \ref{fig:phases}.  The condition
(\ref{5.7}) leads to
\begin{equation}
Q_0 = \frac{\mu_0}{V_0} - C_0(\mu_0,K_0,V_0,\Delta) y_0^2
+ O(y_0^4),
\label{5.10}
\end{equation}
where the positive coefficient of the correction term is given by
\begin{equation}
C_0 = \frac{4\pi}{V_0} \sum_{\bf R} I
\left[\langle \sin[2\pi(S_{\bf R}-S_{\bf 0}+w_I+Q_0 I)]
\rangle_0 \right]_{\rm av},
\label{5.11}
\end{equation}
in which the thermodynamic average is with respect to the Gaussian
Lagrangian ${\cal L}_{\rm SG}(y_0=0)$.  To evaluate this further we
assume that the $\delta \mu_I$ are independent, with a symmetric
distribution.  Let us define a measure of the disorder strength,
$\Delta$, via
\begin{equation}
\left[e^{i\frac{2\pi}{V_0} \delta \mu_I} \right]_{\rm av}
\equiv e^{-2\pi^2 \Delta^2}.
\label{5.12}
\end{equation}
Then,
\begin{equation}
C_0 = \frac{4\pi}{V_0} \sum_{\bf R} e^{-4\pi^2 G({\bf R})}
e^{-2\pi^2 \Delta^2 |I|}I \sin(2\pi \mu_0 I/V_0).
\label{5.13}
\end{equation}
The sum over $I$ clearly converges.  Using (\ref{5.8}), it easily
seen that the sum over $T$ converges so long as $\omega_0 \equiv
2\pi \sqrt{K_0/V_0} > 1$.  We shall see below that the superfluid
phase is define by $\omega_0 > 3$, so this condition is indeed met.

The corrugation due to $y_0$ therefore slows the rate of climb
of the interface from its unperturbed rate, $\mu/V_0$.  When
$K_0/V_0$ and $\mu_0$ become small this perturbation theory breaks
down---a signal of the phase transition into the Bose glass phase.

\subsection{Stability of the superfluid phase}
\label{sec:sfstability}

We consider now the stability of the superfluid phase to $y_0$.  In
Ref.\ \onlinecite{FWGF} this analysis was performed using
Kosterlitz-Thouless-type renormalization group methods.  Here we
will take a less sophisticated route and adapt the scaling approach
described in Sec.\ \ref{sec:xover}. This calculation will also allow
us to examine the effects of finite $\mu_0$ on this stability.  The
latter will allow an explicit confirmation of the irrelevance of
full particle-hole symmetry breaking at the commensurate
(statistically particle-hole symmetric) critical point.

We consider the relevance of the cosine term on the fixed line,
characterized by the long-range correlations (\ref{5.8}).  To this
end, define the local operator
\begin{equation}
O_{\bf R} = \cos\left[2\pi (\tilde S_{\bf R} + w_I + Q_0 I)\right],
\label{5.14}
\end{equation}
and introduce a ``temperature'' variable, analogous to $\delta$ in
Sec.\ \ref{sec:scaling}, by adding a mass term
\begin{equation}
\frac{1}{2} t \sum_{\bf R} \tilde S_{\bf R}^2
\label{5.15}
\end{equation}
to ${\cal L}_{SG}$.  By this device we may discuss the relevance of
the $y_0$ term to the critical behavior as $t \rightarrow 0$.  To
this end, we postulate a scaling form for the singular part of the
free energy,
\begin{equation}
f_s(y_0) \approx A t^{2-\alpha}
\Phi\left(\frac{B y_0^2}{t^{\phi_y}}\right)
\label{5.16}
\end{equation}
so that
\begin{equation}
\frac{1}{2} \left(\frac{\partial^2 f_s}{\partial y_0^2}\right)_{y_0=0}
\approx AB t^{2-\alpha-\phi_y} \Phi^\prime(0).
\label{5.17}
\end{equation}
The superfluid phase always occurs at $y_0=0$, so there will be no
shift in the critical value $t=0$.  As usual, the $y_0$ term is
relevant if $\phi_y > 0$.

The derivative in (\ref{5.17}) may be computed in terms of the
average
\begin{eqnarray}
&&\left(\frac{\partial^2 f_\mathrm{SG}}
{\partial y_0^2}\right)_{y_0=0}
= -\frac{4}{\beta L} \left[\left\langle
\left(\sum_{\bf R} O_{\bf R} \right)^2
\right\rangle_0 \right]_{\rm av}
\label{5.18}\\
&&\ \ \ \ \ \ \ \ \ \
=\ -4 \sum_{\bf R} e^{-2\pi^2 \Delta^2 |I|}
e^{-4\pi^2 G({\bf R},t)} \cos(2\pi Q_0 I),
\nonumber
\end{eqnarray}
where (\ref{5.12}) has been used, and where
\begin{eqnarray}
G({\bf R},t) &=& \left[ \left(-\frac{1}{K_0} \partial_T^2
- V_0 \partial_I^2 + t \right)
\delta_{{\bf R R}^\prime} \right]^{-1}_{\bf R,0}
\nonumber \\
&\approx& \int_{k,\omega} \frac{1- e^{i(kI+\omega T)}}
{\omega^2/K_0 + V_0 k^2 + t},~|{\bf R}| \rightarrow \infty.\ \ \ \ \ \
\label{5.19}
\end{eqnarray}
For $t \rightarrow 0$, $G({\bf R},t)$ has the logarithmic form
(\ref{5.7}).  For finite $t$ one may write
\begin{equation}
e^{-4\pi^2 G({\bf R},t)} \approx [\rho({\bf R})/\rho_0]^{-\omega_0}
E[\rho({\bf R})^2 t/\rho_0^2],\ |{\bf R}| \rightarrow \infty,
\label{5.20}
\end{equation}
where $\omega_0 = 2\pi \sqrt{K_0/V_0}$ determines the power law
decay of correlations at criticality (i.e., in the superfluid phase)
and the scaling function $E(w)$ decays exponentially for large $w$
[this can be seen explicitly by writing $G({\bf R},t) = G({\bf R},0)
+ \delta G({\bf R},t)$ and using (\ref{5.19})] and $E(0)=1$.  This
exhibits the scaling of the correlations when $y_0 = 0$, and and since
$\rho$ scales with $\sqrt{t}$ one immediately identifies the
correlation length exponent $\nu = \frac{1}{2}$.

In addition to the subleading singular part we seek, (\ref{5.18})
contains analytic terms in $t$, whose Taylor coefficients may be
evaluated by taking derivatives of (\ref{5.18}) with respect to $t$
at $t=0$.  Let $n$ be the first positive integer such that the $n$th
derivative of (\ref{5.18}) diverges as $t \to 0$.  Since each
derivative of (\ref{5.20}) with respect to $t$ brings a factor of
$\rho({\bf R})^2 \sim |{\bf R}|^2$ out of the scaling function, this
divergence arises from a failure of the integral to converge at
infinity. The leading singularity may therefore be computed exactly
by considering only the large $|{\bf R}|$ asymptotic behavior of the
integrand.  In this limit one may perform the strongly convergent
sum over $I$ by setting $I=0$ inside $G$.  Defining,
\begin{equation}
D_0 = -8\sum_I \cos(2\pi Q_0 I) e^{-2\pi^2 \Delta^2 |I|},
\label{5.21}
\end{equation}
one finds
\begin{eqnarray}
\frac{\partial^n}{\partial t^n}
\left(\frac{\partial^2 f_\mathrm{SG}}{\partial y_0^2}\right)_{y_0=0}
&\approx& 2 D_0 \int_{T_1}^\infty dT
\left(\frac{K_0V_0 T^2}{\rho_0^2} \right)^{n_0-\omega_0/2}
\nonumber \\
&&\ \ \ \ \ \ \times\ E^{(n)}(K_0 V_0 T^2 t/\rho_0^2)
\nonumber \\
&\approx& E_0 D_0 t^{(\omega_0-1)/2-n},
\label{5.22}
\end{eqnarray}
where $E^{(n)}(x)$ is the $n$th derivative of $E(x)$, $T_1 = O(1)$
is a lower cutoff [whose arbitrariness yields only subleading
corrections to the last line of (\ref{5.22})], and with coefficient
\begin{equation}
E_0 = \frac{\rho_0}{\sqrt{K_0 V_0}}
\int_0^\infty u^{2n_0-\omega_0} E^{(n)}(u^2) du.
\label{5.23}
\end{equation}
It is clear at this point that the $n \geq 0$ we seek is the first
integer for which $\omega_0-2n-1 < 0$.  One finally obtains the
singular part
\begin{equation}
\left(\frac{\partial^2 f_\mathrm{SG}}{\partial y_0^2}
\right)_{y_0=0,\mathrm{sing}}
= E_0 D_0 \frac{\Gamma[(\omega_0+1)/2-n]}
{\Gamma[(\omega_0+1)/2]} t^{(\omega_0-1)/2}
\label{5.24}
\end{equation}

From (\ref{5.7}) and (\ref{5.8}) we see that, up to scale factors,
space-time is isotropic.  Thus $z=1$ and hyperscaling yields
$2-\alpha=2\nu$, so that from (\ref{5.17}) we may finally identify
\begin{equation}
\phi_y = \frac{3-\omega_0}{2} = \frac{3}{2} - \pi \sqrt{K_0/V_0}.
\label{5.25}
\end{equation}
Hence $y_0$ becomes relevant when $\sqrt{K_0/V_0} < \frac{3}{2\pi}$.
This should be compared to the analogous result, $\sqrt{K_0/V_0} <
\frac{2}{\pi}$, for the usual Kosterlitz-Thouless transition where
$\mu_I \equiv 0$.  Thus, the interface roughens \emph{earlier}
(i.e., at smaller $K_0$), meaning that superfluidity is \emph{more}
stable, in the presence of disorder.  For $\sqrt{K_0/V_0} >
\frac{3}{2 \pi}$, $y_0$ is irrelevant and may be set to zero to
calculate universal quantities near the phase transition.  At the
critical point one has $\omega \equiv \omega_c = 3$, which should be
compared to the Kosterlitz-Thouless value, $\omega_c = 4$.  One may
then, for example, invert the duality transformation in this limit
to obtain the actual superfluid correlation function.  One finds
that (\ref{B4}), with (\ref{B16}), takes the form
\begin{eqnarray}
\tilde {\cal L}_J(y_0 \rightarrow 0) &=& -\frac{1}{2}
\sum_{\bf r} \bigg[ K_0 (\tilde \phi_{{\bf r}+\hat{\bf x}}
- \tilde \phi_{\bf r})^2
\nonumber \\
&&\ \ \ \ \ \ +\ \frac{1}{V_0} (\tilde \phi_{{\bf r}+\hat \tau}
- \tilde \phi_{\bf r})^2 \bigg],
\label{5.26}
\end{eqnarray}
where ${\bf r} = (i,\tau)$ is the direct lattice integer position
vector and where now $-\infty < \tilde \phi_{\bf r} < \infty$ is a
continuous phase variable [since (\ref{B16}) forces $\nabla \times
{\bf m} \equiv 0$ as $y \rightarrow 0$, we may write ${\bf m} =
\nabla p$, where $p$ is an integer scalar field, then define $\tilde
\phi_{\bf r} = \phi_{\bf r} - 2 \pi p_{\bf r}$].  Thus
\begin{equation}
{\cal G}({\bf r}) \equiv
\langle e^{i(\phi_{\bf r} - \phi_{\bf 0})} \rangle
= \langle e^{i(\tilde\phi_{\bf r} - \tilde\phi_{\bf 0})} \rangle
\sim \tilde \rho({\bf r})^{-\eta},\ \eta = \frac{1}{\omega},
\label{5.27}
\end{equation}
where $\tilde \rho({\bf r})$ is the same as $\rho({\bf r})$ in
(\ref{5.6}), but with $K_0V_0$ replaced by $\frac{1}{K_0 V_0}$. The
exponent $\eta$ is defined in such a way that ${\cal G}(i,\tau=0)
\sim |i|^{-(d+z-2+\eta)}$ at criticality.  Equation (\ref{5.27})
then follows since $d=z=1$ and $\tilde \rho(i,\tau=0) \propto |i|$
for large $|i|$.  At the critical point we have $\eta=\frac{1}{3}$,
which should be compared to the usual Kosterlitz-Thouless value,
$\eta = \frac{1}{4}$.

The above calculation was performed at $y_0=0$.  When $y_0 > 0$, in
the region where it is irrelevant, the parameters $K_0$ and $V_0$ in
the Lagrangians (\ref{5.5}) and (\ref{5.26}) must be renormalized to
values $K_R(y_0)$ and $V_R(y_0)$ before setting $y_0 = 0$ in the
derivation of (\ref{5.27}).  Thus $\omega = 2\pi \sqrt{K_R/V_R}$ and
$\phi_y = \frac{3}{2} - \pi\sqrt{K_R/V_R}$, but the relation $\eta =
\frac{1}{\omega}$ is still exact.  The parameters $K_R$ and $V_R$
are the exact, long wavelength (hydrodynamic) interface stiffness
moduli that a bulk experimental probe would measure, and are
directly analogous to the superfluid density and compressibility in
the superfluid problem---see (\ref{3.29}).  The above analysis shows
that when the ratio $\sqrt{V_R/K_R}$ exceeds the universal value
$\frac{2 \pi}{3}$, $y_0$ becomes relevant, and simple
renormalization of the Gaussian Lagrangian (\ref{5.26}) is invalid.
We then expect $V_R/K_R \rightarrow \infty$, and the interface
becomes localized.  \emph{At} the critical point separating the
localized and delocalized phases, the interface is still
delocalized, with the universal parameter values quoted above. Using
renormalization group techniques, all of these results may be
confirmed by constructing the detailed flows around this fixed
point.\cite{FWGF,GS88}

\subsection{Restoration of particle-hole symmetry in 1D}
\label{subsec:phsym1d}

Recall now the discussion in Secs.\ \ref{sec:phsym_restore} and
\ref{sec:xover} of asymptotic restoration of statistical
particle-hole symmetry---namely the irrelevance of $\mu_0$ in the
presence of nonzero $\delta \mu_I$.  This is seen trivially in the
1D case because the critical fixed point occurs at $y_0 = 0$, at
which point the mapping $S_{\bf R} = \tilde S_{\bf R}$, Eq.\
(\ref{5.4}) with $Q_0 = \mu_0$, entirely eliminates $\mu_0$, as well
as all the $\delta \mu_I$, from (\ref{5.5}), except for the analytic
additive term (\ref{5.6}). One therefore obtains in this case a
rather extreme form of irrelevance, in which the influence of
$\mu_0$ does not decay with a characteristic exponent $\phi_{g_0} <
0$, but actually disappears entirely.

More generally, when both $\mu_0$ and $y_0$ are nonzero, the fact
that $\mu_0$ appears only in the cosine term in (\ref{5.5}) means
that its influence must vanish on large length scales whenever $y_0$
is irrelevant. Examining the scaling analysis
(\ref{5.17})--(\ref{5.25}), used to determine the range of this
irrelevance, one sees that $\mu_0$ (via $Q_0$) appears only in the
cosine factor in (\ref{5.21}). This factor is completely dominated
by the exponential decay due to the fluctuating part of the $\mu_I$,
and is therefore of no real consequence, producing only analytic
corrections multiplying the leading singularity, and therefore
having no influence on the value of the crossover exponent $\phi_y$.
The ultimate origin of this result can be seen in (\ref{5.14}): only
the value of $\theta_I \equiv w_I \mod 2\pi$ is important, and when
the variance measure, $\Delta$, of $\delta \mu_I$ is sufficiently
large this field is basically uniformly distributed over the
interval $[0,2\pi)$, irrespective of the ``mean drift'' $\mu I/V_0$.

Note, finally, that the irrelevance of a uniform $\mu_0$ has no
bearing on the value the dynamical exponent $z$, and it can easily
be verified that the finite piece in the compressibility comes
purely from the analytic part of the free energy, especially the
term (\ref{5.6}). Thus, although $z = d = 1$ in this case, one may
view this as a `coincidence' that has no bearing on the general
mechanism, discussed in detail in Sec.\ \ref{subsec:revisit_zvsd},
by which $\kappa$ remains finite through the Bose glass--superfluid
transition.

\section{The epsilon expansion}
\label{sec:epstau}

In this final section we turn from exact calculations in one
dimension to approximate calculations in higher dimensions,
expanding on, and providing more context for, our previous
work.\cite{MW96} Unlike the classical point disorder problem, the
classical random rod problem [(\ref{2.4}), or (\ref{4.1}) with
$g({\bf x}) \equiv 0$ but $r({\bf x})$ random], does not have a
simple epsilon expansion about $d=4$. Rather, as shown in Refs.\
\onlinecite{DBC}, one must consider also the limit in which the
dimension, $\epsilon_\tau$ of the rods is small, and perform a
double expansion in $\epsilon=4-D$ and $\epsilon_\tau$ (recall that
$D=d+\epsilon_\tau$ is the total dimensionality).  The exponents
take mean-field values, $z=1$, $\nu = \frac{1}{2}$, $\eta=0$, etc.,
at $\epsilon = \epsilon_\tau = 0$, and deviations from these values
may be computed as two-variable power series in $\epsilon$ and
$\epsilon_\tau$.

Our purpose in this section is to extend this technique to the dirty
boson problem.  We saw in Sec.\ \ref{sec:xover} that a certain
nontrivial crossover exponent $\phi_g$ must be positive if, as
expected, particle-hole symmetric disorder is to lead to new
critical behavior, different from that of the classical random rod
problem. This result was confirmed explicitly for $d=1$ in Sec.\
\ref{sec:1d}: there, random rod disorder was found to be an
\emph{irrelevant} perturbation on the pure (Kosterlitz-Thouless, 2D
XY) critical behavior, whereas dirty boson-type disorder was found
to be \emph{relevant}, leading to new critical behavior.  We shall
find that for \emph{small} $\epsilon_\tau$, particle-hole symmetric
disorder is an irrelevant perturbation on the random rod problem,
and therefore that the crossover exponent \emph{changes sign}, from
negative to positive, at a certain value, $\epsilon_\tau =
\epsilon_\tau^c(D)$. To first order in $\epsilon_\tau$ we obtain the
estimate $\epsilon_\tau^c(D=4) = \frac{8}{29}$ ($D=4$ yielding $d=3$
at $\epsilon_\tau = 1$).  For $\epsilon_\tau > \epsilon_\tau^c$
there are then two fixed points, the stable dirty boson fixed, and
the unstable random rod fixed point.  This then establishes the
nonperturbative nature of the dirty boson fixed point.

\subsection{Scaling for general $\epsilon_\tau$}
\label{sec:epstauscale}

Let us begin by extending the scaling arguments to noninteger
$\epsilon_\tau$. We consider the following generalization of
(\ref{4.1}) (or, equivalently, of ${\cal L}_5$ in Table
\ref{table1}):
\begin{eqnarray}
{\cal L}_c &=& -\int d^dx \int d^{\epsilon_\tau}\tau
\left\{\frac{1}{2} |\nabla \psi|^2
- \frac{1}{2} \psi^*[\nabla_\tau - {\bf g}({\bf x})]^2 \psi \right.
\nonumber \\
&&+\ \left. \frac{1}{2} r({\bf x}) |\psi|^2
+ \frac{1}{4} u |\psi|^4 \right\},
\label{6.1}
\end{eqnarray}
where ${\bf g}({\bf x})$ is an $\epsilon_\tau$-dimensional vector.
This form is based on (\ref{2.9}), with the same simplifications
used in (\ref{4.1}).  We write ${\bf g}({\bf x}) = {\bf g}_0 +
\delta {\bf g}({\bf x})$, and assume that $\delta {\bf g}({\bf x})$
is isotropically distributed in ${\bm \tau}$ space. This yields the
correct $\epsilon_\tau = 1$ limit, and ensures that the free energy
depends only on $g_0 \equiv |{\bf g}_0|$. Clearly, ${\bf g}_0 = 0$
is the generalization of statistically particle-hole symmetric
disorder. As before, we also write $r({\bf x}) = r_0 + \delta r({\bf
x})$, with $[\delta r]_\mathrm{av} = 0$, and $r_0$ is the control
parameter.

\subsubsection{Stiffness constants and hydrodynamic action}
\label{subsec:epstau_stiffness}

The evaluation of the stiffness constants, (\ref{3.14}) and
(\ref{4.6}), is slightly more complicated now: although the spatial
stiffnesses, $\Upsilon_\alpha$, are as before, the temporal
stiffness now takes on a tensor character.  Consider a
$\theta$-boundary condition in the ${\bm \tau}$-subspace:
\begin{equation}
\psi({\bf x}, {\bm \tau} + \beta \hat {\bm \tau}_\mu)
= e^{i\theta_\mu}\psi({\bf x},{\bm \tau}),\
\mu = 1,\ldots,\epsilon_\tau.
\label{6.2}
\end{equation}
Defining the periodic field, $\tilde\psi = e^{-i {\bm \theta} \cdot
{\bm \tau}/\beta} \psi$, and substituting into (\ref{6.1}), we find
that
\begin{equation}
{\cal L}_c[\psi;{\bf g}_0]
= {\cal L}_c[\tilde\psi;{\bf g}_0 - i {\bm \theta}/\beta].
\label{6.3}
\end{equation}
The free energy, $f^{\bm \theta} = -\beta^{-\epsilon_\tau} L^{-d}
\ln \left\{\mathrm{tr}\left[e^{{\cal L}_c[\psi]} \right] \right\}$,
is therefore shifted by
\begin{eqnarray}
\delta f^{\bm \theta} &\equiv& f^{\bm \theta} - f^0
= f^0({\bf g}_0-i{\bm \theta}/\beta) - f^0({\bf g}_0)
\nonumber \\
&=& (-i {\bm \theta}/\beta) \cdot
\frac{\partial f^0}{\partial {\bf g}_0}
\label{6.4} \\
&&+\ \frac{1}{2} (-i {\bm \theta}/\beta)
\cdot \frac{\partial^2 f^0}{\partial {\bf g}_0 \partial {\bf g}_0}
\cdot (-i {\bm \theta}/\beta) + O[(|\bm \theta|/\beta)^3].
\nonumber
\end{eqnarray}
Isotropy implies that
\begin{eqnarray}
\frac{\partial f^0}{\partial {\bf g}_0} &=& -\rho_0 \hat{\bf g}_0
\nonumber \\
\frac{\partial^2 f^0}{\partial {\bf g}_0 \partial {\bf g}_0}
&=& -\frac{\rho_0}{g_0} (\openone - \hat{\bf g}_0 \hat {\bf g}_0)
- \kappa_0 \hat{\bf g}_0 \hat {\bf g}_0,
\label{6.5}
\end{eqnarray}
where we have defined the scalar quantities
\begin{equation}
\rho_0 = -\frac{\partial f^0}{\partial g_0},\
\kappa_0 = -\frac{\partial^2 f^0}{\partial g_0^2},
\label{6.6}
\end{equation}
and (\ref{6.4}) reduces to
\begin{eqnarray}
\delta f^{\bm \theta} &=& \frac{i}{\beta}
({\bm \theta} \cdot \hat g_0) \rho_0
+ \frac{1}{2 \beta^2} [|{\bm \theta}|^2 -
({\bm \theta} \cdot \hat{\bf g}_0)^2] \frac{\rho_0}{g_0}
\nonumber \\
&&+\ \frac{|{\bm \theta}|^2}{2 \beta^2} \kappa_0
+ O[(|{\bm \theta}|/\beta)^3].
\label{6.7}
\end{eqnarray}

It is straightforward to write down expressions for $\rho_0$ and
$\kappa_0$ analogous to (\ref{4.6}), but we will require only
\begin{eqnarray}
&&\kappa_0(g_0=0) \equiv \Upsilon_\tau =
\left[\langle |\psi|^2 \rangle \right]_\mathrm{av}
\label{6.8} \\
&&+\ \int d^dx \int d^{\epsilon_\tau}\tau
\left[\langle \psi^* \partial_{\tau_1} \psi({\bf x},{\bm \tau})
\psi^* \partial_{\tau_1} \psi({\bf 0},{\bf 0})
\rangle\right]_\mathrm{av},
\nonumber
\end{eqnarray}
where $\tau_1$ is any given direction in ${\bm \tau}$-space.  The
long wavelength action generalizing (\ref{3.29}) then takes the form
\begin{eqnarray}
S_\mathrm{eff} &=& -\int d^dx \int d^{\epsilon_\tau}\tau
\left[\frac{1}{2}\Upsilon |\nabla \phi|^2 \right.
\nonumber \\
&&+\ \frac{\rho_0}{2 g_0} [|\nabla_\tau \phi|^2
- (\hat{\bf g}_0 \cdot \nabla_\tau \phi)^2]
\nonumber \\
&&+\ \left. \frac{1}{2} \kappa_0
(\hat {\bf g}_0 \cdot \nabla_\tau \phi)^2
+ i\rho_0 (\hat{\bf g}_0 \cdot \nabla_\tau \phi) \right].
\label{6.9}
\end{eqnarray}
Note that when $g_0 \rightarrow 0$ we have $\rho_0/g_0 \rightarrow
\kappa_0(g_0=0) \equiv \Upsilon_\tau$, and $S_\mathrm{eff}$ reduces
to the more familiar form
\begin{equation}
S_\mathrm{eff}(g_0=0) = -\frac{1}{2} \int d^dx
\int d^{\epsilon_\tau}\tau \left(\Upsilon |\nabla \phi|^2
+ \Upsilon_\tau |\nabla_\tau \phi|^2 \right).
\label{6.10}
\end{equation}
Defining the frequency variables $\omega_\parallel = \hat{\bf g}_0
\cdot {\bm \omega}$ and ${\bm \omega}_\perp = {\bm \omega} -
\omega_\parallel \hat{\bf g}_0$, and following the derivation of
(\ref{4.18}), equation (\ref{6.9}) yields a long wavelength, low
frequency Green function
\begin{eqnarray}
\hat G({\bf k},\omega) &\equiv&
\langle |\hat \psi({\bf k},{\bm \omega})|^2
\rangle_{S_\mathrm{eff}}
\nonumber \\
&\approx& \frac{|\psi_0|^2}
{(\rho_0/g_0)|{\bm \omega}_\perp|^2
+ \kappa_0 \omega_\parallel^2
+ \Upsilon |{\bf k}|^2},
\label{6.11}
\end{eqnarray}
where $\psi_0 = \left[\langle \psi({\bf x},{\bm \tau}) \rangle
\right]_\mathrm{av} \sim |\delta|^\beta$ is the order parameter, and
for slow variations, $\psi({\bf x},{\bm \tau}) \approx \psi_0 e^{i
\phi({\bf x},{\bm \tau})}$.  In deriving this form we have neglected
the surface term, $i\rho_0 \hat{\bf g}_0 \cdot \nabla_\tau \phi$,
relative to the others.  This is valid in the superfluid phase where
$\phi$ has only small fluctuations about the long range ordered
value, which we have taken to be $\phi \equiv 0$. As we saw in Sec.\
\ref{subsec:excbf}, in the disordered Bose glass phase this term
does become important (see also below).

\subsubsection{Dynamical exponents and scaling}
\label{subsec:zscale}

Equation (\ref{6.11}) determines the correlations in the
hydrodynamic limit, even near the critical point. One could
propose a scaling function of the form:
\begin{equation}
G({\bf k},{\bm \omega}) \approx G_0 \xi^{2-\eta}
g(k\xi,\omega_\perp \xi_\tau^\perp,
\omega_\parallel \xi_\tau^\parallel),
\label{6.12}
\end{equation}
where ${\bf k}$, ${\bf \omega}_\perp$ and $\omega_\parallel$ appear
scaled by their appropriate correlation lengths, $\xi$,
$\xi_\tau^\parallel \sim \xi^{z_\parallel}$ and $\xi_\tau^\perp \sim
\xi^{z_\perp}$, where, for $g_0 \neq 0$, we have allowed for
different scalings parallel and perpendicular to ${\bf g}_0$.  When
$g_0 = 0$ we expect $\xi_\tau^\parallel = \xi_\tau^\perp \equiv
\xi_\tau$ and $z_\parallel = z_\perp \equiv z$.  To be consistent
with (\ref{6.11}), equation (\ref{6.12}) requires that
\begin{equation}
g(x,y,z) \approx \frac{1}{g_x x^2 + g_\perp y^2 + g_\parallel z^2},\
x,y,z \rightarrow 0,
\label{6.13}
\end{equation}
in which $g_x$, $g_\perp$ and $g_\parallel$ are universal numbers.
Thus
\begin{eqnarray}
\Upsilon &\approx& G_0^{-1} |\psi_0|^2 g_x \xi^\eta
\nonumber \\
\kappa_0 &\approx& G_0^{-1} |\psi_0|^2 g_\parallel
(\xi_\tau^\parallel)^2 \xi^{-2+\eta}
\nonumber \\
\rho_0/g_0 &\approx& G_0^{-1} |\psi_0|^2
g_\perp (\xi_\tau^\perp)^2
\xi^{-2+\eta}
\label{6.14}
\end{eqnarray}
[compare (\ref{4.18}) and (\ref{4.19}]. The hyperscaling relation is
now $2-\alpha = [d+z_\parallel + (\epsilon_\tau - 1)z_\perp] \nu$.
Along with the usual scaling relations, $\alpha+2\beta+\gamma = 2$
and $\gamma = (2-\eta)\nu$, this immediately implies that $\Upsilon
\sim |\delta|^\upsilon$, $\kappa_0 \sim
|\delta|^{\upsilon_\tau^\parallel}$ and $\rho_0/g_0 \sim
|\delta|^{\upsilon_\tau^\perp}$, where $\delta = r_0 - r_{0,c}(g_0)$
is the deviation from criticality, and the implied exponent
relations
\begin{eqnarray}
\upsilon &=& [d + (\epsilon_\tau - 1)z_\perp + z_\parallel - 2] \nu
\nonumber \\
\upsilon_\tau^\parallel &=&
[d + (\epsilon_\tau - 1)z_\perp - z_\parallel] \nu
\nonumber \\
\upsilon_\tau^\perp &=&
[d + (\epsilon_\tau - 3)z_\perp + z_\parallel] \nu,
\label{6.15}
\end{eqnarray}
generalize (\ref{4.11}). If $z_\perp = z_\parallel = z$, then
$\upsilon = (d + \epsilon_\tau -2)\nu$ and $\upsilon_\tau^\parallel
= \upsilon_\tau^\perp \equiv \upsilon_\tau =
[d+(\epsilon_\tau-2)z]\nu$. If $\kappa_0$ remains finite through the
transition even for noninteger $\epsilon_\tau$, then we would
predict
\begin{equation}
z = \frac{d}{2-\epsilon_\tau}.\ (?)
\label{6.16}
\end{equation}
This is the natural generalization of the proposed scaling
relation $z=d$ at $\epsilon_\tau = 1$.\cite{FWGF,FF88}

However, as discussed in Ref.\ \onlinecite{WM07} and Sec.\
\ref{subsec:revisit_zvsd}, this argument actually fails for
$\epsilon_\tau = 1$, and there is no reason to expect it to fare any
better for general $\epsilon_\tau$.  In particular, the absorption
of the twisted boundary condition into the simple shift, (\ref{6.3})
and (\ref{6.4}), of ${\bf g}_0$ again implies that, at least for
$g_0 \neq 0$, the critical behavior of $\rho_0$ and $\kappa_0$ is
not tied, as in (\ref{6.12}), to the correlation times, but only to
derivatives with respect to $g_0$. Thus, although the relation for
$\upsilon$ in (\ref{6.15}) is expected to be correct, the relations
for $\upsilon_\tau^\parallel$ and $\upsilon_\tau^\perp$ are not.
Rather, as discussed above (\ref{4.14}), the leading singular
contribution to $\rho_0$ and $\kappa_0$ will come from the $g_0$
dependence of $r_{0,c}(g_0)$, leading to
\begin{equation}
\rho_{0,s} \sim |\delta|^{1-\alpha},\
\kappa_{0,s} \sim |\delta|^{-\alpha}.
\label{6.17}
\end{equation}
We will see that the renormalization group analysis generates an
independent value for $z$, disagreeing, in particular, with
(\ref{6.16}), confirming again that there is no
$|\delta|^{\upsilon_\tau}$ contribution to the compressibility.

\subsubsection{Relevance of particle-hole symmetry breaking}
\label{subsec:epstau_phsym}

We next calculate the crossover exponent $\phi_g$ associated with
nonzero $\delta {\bf g}$ at $g_0 = 0$, for general $\epsilon_\tau$.
The calculation is essentially identical to that leading to
(\ref{4.40}).  The perturbation, (\ref{4.21}), now becomes
\begin{equation}
{\cal L}_{\bf g} = \int d^dx \int d^{\epsilon_\tau} \tau
\left\{\frac{1}{2} |{\bf g}({\bf x})|^2 |\psi|^2
- \psi^*[{\bf g}({\bf x}) \cdot \nabla_\tau] \psi \right\}.
\label{6.18}
\end{equation}
We assume $\left[g_\mu({\bf x}) g_\nu({\bf x})\right]_\mathrm{av} =
\Delta_g \phi({\bf x} - {\bf x}^\prime) \delta_{\mu\nu}$, where
$\phi$ is a delta function for short range correlated disorder, and
varies as $x^{-(d+a)}$ for large $x$ for power law correlated
disorder. Equation (\ref{4.32}) now becomes
\begin{equation}
-2 \left(\frac{\partial f}{\partial \Delta_g} \right)_{\Delta_g=0}
= \epsilon_\tau \int_{\bf k} \hat \varphi({\bf k})
\Upsilon_\tau({\bf k},{\bf 0})
\label{6.19}
\end{equation}
in which $\Upsilon_\tau({\bf k},{\bm \omega})$ is defined by
(\ref{4.28})--(\ref{4.31}), but, as in (\ref{6.8}), with time
derivatives $\partial_\tau \to \partial_{\tau_1}$ in (\ref{4.28}),
and random rod compressibility exponent
$\upsilon_{\tau,0}(\epsilon_\tau) = [d+(\epsilon_\tau-2)z]\nu_0$
replacing $(d-z_0)\nu_0$ in the scaling form (\ref{4.31}).

As in (\ref{4.33}) and (\ref{4.34}), $\tilde {\cal Y}(w)$ will have
a similar spectrum of exponents, $0 < \omega_1(\epsilon_\tau) <
\omega_2(\epsilon_\tau) < \omega_3(\epsilon_\tau) < \dots$, now
depending on $\epsilon_\tau$. This leads in the same way to
(\ref{4.37}) and (\ref{4.38}), still with $b_0 \equiv 0$, and a term
$b_a |\delta|^{[2d+(\epsilon_\tau-2)z_0]\nu}$.  We then find
\begin{equation}
\phi_g/\nu_0 = \left\{\begin{array}{ll}
2z_0-d-a, &  a < 2z_0-d-\lambda_1(\epsilon_\tau) \\
\lambda_1(\epsilon_\tau), & a > 2z_0-d-\lambda_1(\epsilon_\tau)
\end{array} \right.
\label{6.20}
\end{equation}
essentially as before [see (\ref{4.41})], but now with $\lambda_j =
d+\epsilon_\tau z-\omega_j$ and all exponents evaluated at the
random rod fixed point in $\epsilon_\tau$ time dimensions.

We shall compute $\lambda_1(\epsilon_\tau)$ to $O(\epsilon_\tau)$
below.  From the naive (and, as we shall see, incorrect) estimate,
$\lambda_1 = 2z_0-d$, we again have $\lambda_1>0$ for $z >
\frac{d}{2}$. Since $z=1$ at $\epsilon_\tau=0$, we expect, as stated
earlier, $\lambda_1<0$ for small $\epsilon_\tau$, becoming positive
only for $\epsilon_\tau > \epsilon_\tau^c > 0$.  We shall find that
for $\epsilon_\tau > \epsilon_\tau^c$ a new stable fixed point with
$\Delta_g^*>0$ bifurcates away from the random rod fixed point (with
$\Delta^*=0$).  The exponent $z_0$ is substantially \emph{smaller}
than $\frac{d}{2}$ at this point, violating (\ref{6.16}) for any
$\epsilon_\tau > 0$.  Assuming that this new fixed point may indeed
be identified with the true dirty boson fixed point when
$\epsilon_\tau = 1$, we conclude again that (\ref{6.16}) is
incorrect. Below we shall generalize the analysis of the excitation
spectrum of the Bose glass phase in Sec.\ \ref{sec:excite} to
general $\epsilon_\tau$. For $\epsilon_\tau < 1$, the issue of
whether the Bose glass phase is compressible turns out to be rather
subtle.

\subsection{Phase diagrams for general $\epsilon_\tau$}
\label{sec:epstau_phases}

To understand the phase diagram for general $\epsilon_\tau$ it is
instructive to first consider the mean field approximation, in the
form of the infinite range hopping model ($J_{ij} = J_0/N$ for all
$i,j$, where $N$ is the number of sites). In the absence of
disorder, and with onsite repulsion only, the mean-field
approximation to the Lagrangian (\ref{2.7}) takes the form
\begin{eqnarray}
{\cal L}_\mathrm{MF} &=& \int d^{\epsilon_\tau}\tau
\left\{\frac{J_0}{N} \sum_{i,j}
\cos[\phi_i({\bm \tau}) - \phi_j({\bm \tau})] \right.
\nonumber \\
&&-\ \left. \frac{1}{2U_0} \sum_i
(\nabla_\tau \phi_i - i{\bm \mu})^2 \right\},
\label{6.21}
\end{eqnarray}
with temporal periodic boundary conditions
\begin{equation}
\phi_i({\bm \tau} + \beta {\bf \hat e}_\alpha)
= \phi_i({\bm \tau}) + 2 \pi n_\alpha,
\label{6.22}
\end{equation}
for any set of integers ${\bf n} =
\{n_\alpha\}_{\alpha=1}^{\epsilon_\tau}$.

\subsubsection{Kac-Hubbard-Stratanovich transformation}
\label{subsec:mft_khs}

A complex Kac-Hubbard-Stratanovich (KHS) field $M({\bm \tau})$ may
now be introduced to decouple the hopping term. Using the identity
\begin{equation}
e^{a|z|^2/N} = \frac{aN}{\pi} \int d^2\psi
e^{-a(\psi z^* + \psi^* z + N |\psi|^2)},
\label{6.23}
\end{equation}
where the integral is over the real and imaginary parts of $\psi$,
the partition function can be written in the form
\begin{equation}
Z_\mathrm{MF} = \int DM({\bm \tau}) e^{N S_\mathrm{MF}[M]},
\label{6.24}
\end{equation}
in which the integral is over all $\epsilon_\tau$-dimensional
complex functions $M({\bm \tau})$, weighted by the effective action
\begin{widetext}
\begin{equation}
S_\mathrm{MF}[M] = -\frac{J_0}{2}
\int d^{\epsilon_\tau}\tau |M({\bm \tau})|^2
+ \ln\left[\int D\phi({\bm \tau})
e^{-\int d^{\epsilon_\tau}\tau \left\{\frac{1}{2U_0}
\left[\nabla_\tau \phi({\bm \tau}) - i {\bm \mu} \right]^2
+\frac{J_0}{2} \left[M^*({\bm \tau}) e^{i\phi({\bm \tau})}
+ M({\bm \tau}) e^{-i\phi({\bm \tau})}\right] \right\}}\right]
\label{6.25}
\end{equation}
\end{widetext}
in which the decoupling has reduced the phase variable trace to one
over a single variable $\phi({\bm \tau})$, which includes a sum over
all possible boundary conditions (\ref{6.22}). The constant $a =
\frac{1}{2} J d^{\epsilon_\tau} \tau$ incorporates the
discretization of the ${\bm \tau}$-integral, and the measures
$DM({\bm \tau})$, $D\phi({\bm \tau})$ incorporate any normalization
factors needed to make sense of the continuum limit.

\subsubsection{Saddle point evaluation}
\label{subsec:mft_saddle}

In the thermodynamic limit, $N \to \infty$, a saddle point
evaluation of (\ref{6.24}) becomes exact. Since the superfluid order
parameter is homogeneous and static, the lowest-energy saddle point
must be a time independent field, $M({\bf \tau}) \equiv M_0$, which
can be chosen real. Near the critical point $M_0$ will be small, and
$S_\mathrm{MF}[M_0]$ has a Landau expansion,
\begin{eqnarray}
f_\mathrm{MF}(M_0) &\equiv& -\beta^{-\epsilon_\tau} S_\mathrm{MF}[M_0]
\nonumber \\
&=& f_0({\bm \mu},J_0) + \frac{1}{2} r_0({\bm \mu},J_0)|M_0|^2
\nonumber \\
&&+\ \frac{1}{4} u_0({\bm \mu},J_0)|M_0|^4 + \ldots.
\label{6.26}
\end{eqnarray}
The connection between $M_0$ and and the order parameter $\psi_0 =
\langle \frac{1}{N} \sum_j e^{i\phi_j} \rangle$ is obtained by
adding a term $-\frac{1}{2} \int d^{\epsilon_\tau} \tau \sum_j
\left(h^* e^{i\phi_j} + h  e^{-i\phi_j} \right)$ to ${\cal
L}_\mathrm{MF}$. Since this term is local, the KHS transformation
simply yields the replacement $M \to \tilde M \equiv M + h/J_0$
inside the logarithmic term in (\ref{6.25}). In terms of $\tilde M$,
the first term in (\ref{6.25}) becomes $\frac{1}{2} J_0 |M|^2 =
\frac{1}{2} J_0 |\tilde M - h|^2$, and the free energy is obtained
by minimizing $\tilde f_\mathrm{MF}(M_0,h) = f_\mathrm{MF}(M_0) -
\frac{1}{2} (h^* \tilde M_0 + h \tilde M_0^*) + |h|^2/2J_0$.  One
therefore obtains $\psi_0 = -2(\partial f_\mathrm{MF}/\partial
h)_{h=0} = M_0$.  Therefore $M_0$ is in fact the order parameter in
this theory.

Assuming only that $u_0 > 0$, the transition line occurs at $r_0 =
0$, and the superfluid phase corresponds to $r_0 < 0$.  To obtain
the phase diagram, we therefore need only compute $r_0$, which is
given by
\begin{eqnarray}
r_0 &=& J_0 - J_0^2 \int d^{\epsilon_\tau}\tau
\left\langle \cos[\phi({\bm \tau})] \cos[\phi({\bf 0})]
\right\rangle_0
\nonumber \\
&=& J_0 - \frac{1}{2} J_0^2 \mathrm{Re} \int d^{\epsilon_\tau}\tau
\left\langle e^{i[\phi({\bm \tau}) - \phi({\bf 0})]} \right\rangle_0
\label{6.27}
\end{eqnarray}
in which the average is with respect to the $M_0$-independent
$\frac{1}{2U_0} \left[\nabla_\tau \phi({\bm \tau}) - i {\bm \mu}
\right]^2$ term. To further evaluate $r_0$, we first
account for the boundary conditions by changing to the periodic
variable
\begin{equation}
\tilde \phi({\bm \tau}) = \phi({\bm\tau})
- \frac{2 \pi {\bf n} \cdot {\bm \tau}}{\beta}.
\label{6.28}
\end{equation}
It follows that $\phi({\bm \tau}) - \phi({\bf 0}) = \tilde \phi({\bm
\tau}) - \tilde \phi({\bf 0}) + 2 \pi {\bf n} \cdot {\bm
\tau}/\beta$, $\nabla_\tau \phi({\bm \tau}) = \nabla_\tau \phi({\bm
\tau}) + 2\pi {\bf n}/\beta$, and $\int d^{\epsilon_\tau}\tau
\nabla_\tau \tilde \phi = 0$, and one therefore obtains
\begin{eqnarray}
\left\langle e^{i[\phi({\bm \tau}) - \phi({\bf 0})]}
\right\rangle_0
&=& \frac{\sum_{\bf n} e^{-\beta^{\epsilon_\tau}
(2\pi{\bf n}/\beta - i{\bm \mu})^2/2U_0}
e^{2\pi i {\bf n} \cdot {\bm \tau}/\beta}}
{\sum_{\bf n} e^{-\beta^{\epsilon_\tau}
(2\pi{\bf n}/\beta - i{\bm \mu})^2/2U_0}}
\nonumber \\
&&\times\ e^{-\frac{1}{2}\langle[\phi({\bm \tau})
- \phi({\bf 0})]^2 \rangle_{00}},
\label{6.29}
\end{eqnarray}
in which $\langle \cdot \rangle_{00}$ is an average with respect to
the the Gaussian term $(\nabla_\tau \tilde \phi)^2/2U_0$, and we
have used property $\langle e^{iX} \rangle = e^{-\langle X^2
\rangle/2}$, valid for any Gaussian random variable $X$.  The
remaining average is the inverse Fourier transform of $U_0/|{\bm
\omega}|^2$, namely the Green function for the Laplacian in
$\epsilon_\tau$ dimensions:
\begin{eqnarray}
\frac{1}{2} \langle[\phi({\bm \tau}) - \phi({\bf 0})]^2 \rangle_{00}
&=& U_0 A(\epsilon_\tau) |{\bm \tau}|^{2-\epsilon_\tau}
\nonumber \\
A(\epsilon_\tau) &=& \frac{\Gamma(\epsilon_\tau/2)}
{2(2-\epsilon_\tau) \pi^{\epsilon_\tau/2}}.
\label{6.30}
\end{eqnarray}
Note that $A(1) = \frac{1}{2}$.

For the ${\bf n}$-sums we make use of the identity (\ref{3.37}).
The ratio of ${\bf n}$-sums in
(\ref{6.29}) becomes
\begin{equation}
\frac{\sum_{\bf l} e^{-\frac{1}{2}U_0 \beta^{2-\epsilon_\tau}
({\bf l} + {\bm \mu}/U_0 \beta^{1-\epsilon_\tau}
+ {\bm \tau}/\beta)^2}}
{\sum_{\bf l} e^{-\frac{1}{2}U_0 \beta^{2-\epsilon_\tau}
({\bf l} + {\bm \mu}/U_0 \beta^{1-\epsilon_\tau})^2}}
\to e^{-({\bm \mu} - U_0 \beta^{1-\epsilon_\tau} {\bf l}_0)
\cdot {\bm \tau}}
\label{6.31}
\end{equation}
in which the right hand side is valid for ${\beta \to \infty}$, and
${\bf l}_0({\bm \mu})$ is the value of ${\bf l}$ that minimizes
$({\bm \mu} + U_0 \beta^{1-\epsilon_\tau} {\bf l}_0)^2$.  For
$\epsilon_\tau < 1$ one clearly obtains ${\bf l}_0 = {\bf 0}$, while
in the special case $\epsilon_\tau = 1$ one obtains the periodic
form $l_0(\mu) = [\frac{\mu}{U_0} + \frac{1}{2}]$.  Adopting the
convention $-\frac{1}{2} < \mu/U_0 \mod 1 \leq \frac{1}{2}$, we
finally obtain,
\begin{equation}
\left\langle e^{i[\phi({\bm \tau}) - \phi({\bf 0})]}
\right\rangle_0 = \left\{\begin{array}{ll}
e^{-U_0 A(\epsilon_\tau) |{\bm \tau}|^{2-\epsilon_\tau}
-{\bm \mu} \cdot {\bm \tau}}, & \epsilon_\tau < 1
\\
e^{-\frac{1}{2}U_0|\tau|-U_0(\mu/U_0 \mod 1) \tau},
& \epsilon_\tau = 1,
\end{array}\right.
\label{6.32}
\end{equation}
which should be compared to the droplet model calculation
(\ref{3.42}), (\ref{3.43}).

\begin{figure}

\includegraphics[width=\columnwidth]{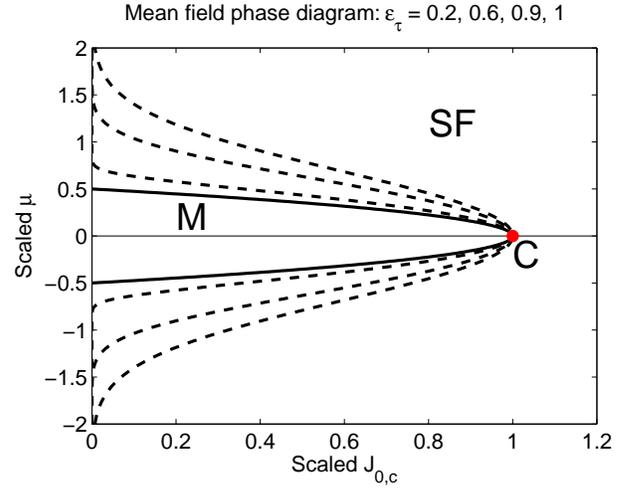}

\caption{(Color online) Mean field phase diagram for the Josephson
junction model, showing Mott (M) and superfluid (SF) phases, with
particle-hole symmetric transition point (C) for $\epsilon_\tau=0.2,
0.6, 0.9, 1$. Larger $\epsilon_\tau$ correspond to smaller Mott
lobes. For $\epsilon_\tau < 1$ (dashed lines) there is a single Mott
lobe persisting for arbitrarily large $\mu$. For $\epsilon_\tau = 1$
the Mott lobe shown (solid line) is actually repeated periodically.
The scaled axis variables are $J_{0,c}(|{\bm \mu}|)/J_{0,c}(0)$ and
$\bar \mu$, as defined in (\ref{6.34}).}

\label{fig:epstau_mott}
\end{figure}

\subsubsection{Superfluid transition line}
\label{subsec:mft_sftrans}

Substituting (\ref{6.32}) into (\ref{6.27}), one obtains the
following equation for the transition line:
\begin{eqnarray}
\frac{2}{J_{0,c}(|{\bm \mu}|)} &=& \int d^{\epsilon_\tau} \tau
e^{-U_0 A(\epsilon_\tau) |{\bm \tau}|^{2-\epsilon_\tau}
-{\bm \mu} \cdot {\bm \tau}}
\nonumber \\
&=& 2\pi^{\epsilon_\tau/2}
\int_0^\infty d\tau \tau^{\epsilon_\tau - 1}
e^{-U_0 A(\epsilon_\tau) \tau^{2-\epsilon_\tau}}
\nonumber \\
&&\times\ \left(\frac{2}{|{\bm \mu}|\tau}
\right)^{\frac{\epsilon_\tau - 2}{2}}
I_{\frac{\epsilon_\tau-2}{2}}(|{\bm \mu}|\tau),
\label{6.33}
\end{eqnarray}
where $I_\nu(z)$ is the modified Bessel function. Note that
$(z/2)^{-\nu} I_\nu(z)$ is analytic at the origin, taking the value
$1/\Gamma(\nu+1)$ there, and that $I_{-\frac{1}{2}}(z) = \sqrt{2/\pi
z} \cosh(z)$ reduces (\ref{6.33}) to the correct result at
$\epsilon_\tau = 1$.  Note that for $\epsilon_\tau > 1$ the integral
(\ref{6.33}) diverges for ${\bm \mu} \neq {\bf 0}$, indicating a
divergent compressibility. Thus, only $\epsilon \leq 1$ displays the
physics of interest to us.

Using the Taylor expansion of the Bessel function,\cite{GR} one obtains,
\begin{eqnarray}
\frac{J_{0,c}(0)}{J_{0,c}(|{\bm \mu}|)}
&=& \sum_{k=0}^\infty
\frac{\Gamma\left(\frac{\epsilon_\tau + 2k}{2-\epsilon_\tau}\right)}
{\Gamma\left(\frac{\epsilon_\tau}{2-\epsilon_\tau}\right)}
\frac{\Gamma\left(\frac{\epsilon_\tau}{2}\right)}
{k! \Gamma\left(k + \frac{\epsilon_\tau}{2}\right)} \bar \mu^{2k}
\nonumber \\
\bar \mu &\equiv& \frac{|{\bm \mu}|}
{2[U_0 A(\epsilon_\tau)]^{\frac{1}{2-\epsilon_\tau}}}
\nonumber \\
\frac{J_{0,c}(0)}{[U_0 A(\epsilon_\tau)
]^{\frac{\epsilon_\tau}{2-\epsilon_\tau}}}
&=& \frac{2\Gamma\left(\frac{2+\epsilon_\tau}{2}\right)}
{\pi^{\frac{\epsilon_\tau}{2}}
\Gamma\left(\frac{2}{2-\epsilon_\tau} \right)}.
\label{6.34}
\end{eqnarray}
It follows that $J_{0,c}(0) \to 2$ as $\epsilon_\tau \to 0$, and
$J_{0,c}(0) \to U_0/2$ as $\epsilon_\tau \to 1$.

For $\epsilon_\tau = 1$ the result (\ref{6.33}) is valid only for
the central Mott lobe. The remainder are constructed by periodic
repetition.  For this case, the integral can be computed
analytically for arbitrary $\mu$, and one obtains,\cite{FWGF}
\begin{equation}
\frac{J_{0,c}(\mu)}{U_0} =
\frac{1}{2}\left[1 - \left(\frac{2\mu}{U_0}\right)^2\right].
\label{6.35}
\end{equation}
For $\epsilon_\tau < 1$ the $|{\bm \tau}|^{2-\epsilon_\tau}$ term
dominates the exponent in (\ref{6.33}) for large ${\bm \tau}$, and
the integral converges for arbitrarily large $|{\bm \mu}|$. A saddle
point estimate yields
\begin{equation}
J_{0,c} \sim \exp\left[-(1 - \epsilon_\tau)
\left(\frac{2 \bar \mu}{2 - \epsilon_\tau}
\right)^{\frac{2-\epsilon_\tau}{1-\epsilon_\tau}}
\right],\ \bar \mu \gg 1,
\label{6.36}
\end{equation}
where non-exponential prefactors have been dropped. In Fig.\
\ref{fig:epstau_mott} we plot numerical solutions of (\ref{6.34})
for the phase boundary for different values of $\epsilon_\tau$.

Since in mean field theory, every site gets effectively decoupled,
it is very easy to incorporate the effect of site disorder within
this formalism.  The generalization of the mean field action is
\begin{equation}
S^\mathrm{dis}_\mathrm{MF}[M] = \int d^{\epsilon_\tau}
\varepsilon p(|{\bm \varepsilon}|)
S_\mathrm{MF}[M;{\bm \mu} - {\bm \varepsilon}],
\label{6.37}
\end{equation}
where $p(|{\bm \varepsilon}|)$ is the (assumed isotropic) single
site distribution for the site disorder ${\bm \varepsilon}$, and
$S_\mathrm{MF}[M;{\bm \mu} - {\bm \varepsilon}]$ is the pure action
(\ref{6.25}) with ${\bm \mu} \to {\bm \mu} - {\bm \varepsilon}$. It
follows that the  phase boundary is now defined by:
\begin{equation}
\frac{2}{J^\mathrm{dis}_{0,c}(|{\bm \mu}|)}
= \int d^{\epsilon_\tau}\tau
\int d^{\epsilon_\tau}\varepsilon p(|{\bm \varepsilon}|)
e^{-A(\epsilon_\tau) U_0 |{\bf \tau}|^{2-\epsilon_\tau}
- ({\bm \mu} - {\bm \varepsilon}) \cdot {\bm \tau}}.
\label{6.38}
\end{equation}
For $\epsilon_\tau = 1$, it is assumed here that $|\mu -
\varepsilon| < U_0/2$ for all allowed values of $\epsilon$,
otherwise the mod factor in (\ref{6.33}) enters.

The result (\ref{6.38}), even for $\epsilon_\tau = 1$, does not
reproduce the result $\mu_c(J_0;\Delta) = \mu_c(J_0;0)-\Delta$
(except at $J_0=0$), where $\Delta$ is the maximum value of
$|\varepsilon|$ supported by $p$, obtained from the rare region
arguments in Sec.\ \ref{sec:ranmudrops}.  In fact, the integral
(\ref{6.38}) is smaller than it would be if all the weight of $p$
were to lie at ${\bm \varepsilon} = -\Delta \hat {\bm \mu}$ (whence
$|{\bm \mu}| \to |{\bm \mu}|+\Delta$), implying that
$J^\mathrm{dis}_{0,c}(|{\bm \mu}|)$ is larger, leading to enlarged
Mott lobes: $\mu_c(J_0;\Delta) \geq \mu_c(J_0;0)-\Delta$.

In the presence of disorder, (\ref{6.38}) implies that the Mott
lobes shrink, but in a nontrivial way that depends on the detailed
shape of $p$, not just its support. This is an artifact of mean
field theory, which couples all sites equally, whereas the rare
region argument result relied on local interactions in spatially
separated, noninteracting droplets. Moreover, the coupling constant
$J_0/N$ in (\ref{6.21}) is scaled by the total number of sites.
Therefore, even if one were to view all sites with values of ${\bm
\varepsilon}_i$ in some small region of size $(\Delta
\varepsilon)^{\epsilon_\tau}$ as a droplet, its critical point would
have to be scaled by the relative number of sites in the droplet,
i.e., by $1/p(|{\bm \varepsilon}|) (\Delta
\varepsilon)^{\epsilon_\tau}$, in order to obtain a consistent
result. This roughly explains the form of (\ref{6.38}).

For $\epsilon_\tau = 1$, the Mott lobes still disappear entirely for
$\Delta > U_0/2$.  However, for $\epsilon_\tau < 1$, the fact that
the Mott lobe extends to arbitrarily large $|{\bm \mu}|$ means that
it will survive for arbitrarily large $\Delta$ as well, including
unbounded disorder.

\subsubsection{Mott phase compressibility}
\label{subsec:mft_mott}

Computation of the density and compressibility within the Mott lobe
requires the $M$-independent part of the free energy $f_0$ in
(\ref{6.26}).  Using (\ref{6.25}) and (\ref{6.28}) one obtains $f_0
= f_{00} + \Delta f_0$, where
\begin{eqnarray}
f_{00} &=& -\frac{1}{\beta^{\epsilon_\tau}}
\ln\left[\int D\tilde \phi(\tau) e^{-\frac{1}{2U_0}
\int d^{\epsilon\tau}\tau (\nabla_\tau \tilde \phi)^2} \right]
\nonumber \\
&=& -\frac{1}{2} \int \frac{d^{\epsilon_\tau} \omega}
{(2\pi)^{\epsilon_\tau}}
\ln\left(\frac{2\pi U_0}{|{\bm \omega}|^2}\right)
\label{6.39}
\end{eqnarray}
is the ${\bm \mu}$-independent Gaussian contribution, and
\begin{eqnarray}
\Delta f_0 &=& -\frac{1}{\beta^{\epsilon_\tau}}
\ln\left[\sum_{\bf n} e^{-\beta^{\epsilon_\tau}
(2\pi {\bf n}/\beta - i {\bm \mu})^2} \right]
\nonumber \\
&=& -\frac{|{\bm \mu}|^2}{2U_0}
-\frac{\epsilon_\tau}{2\beta^{\epsilon_\tau}}
\ln\left(\frac{\beta^{2-\epsilon_\tau} U_0}{2\pi}\right)
\nonumber \\
&&-\ \frac{1}{\beta^{\epsilon_\tau}} \ln\left[
\sum_{\bf l} e^{-U_0 \beta^{2-\epsilon_\tau}
\left({\bf l} - \beta^{1-\epsilon_\tau}
{\bm \mu}/U_0 \right)^2/2}\right]
\nonumber \\
&\to& \left\{\begin{array}{ll}
\frac{1}{2} U_0 l_0(\mu)^2 - l_0(\mu) \mu, & \epsilon_\tau = 1 \\
0, & \epsilon_\tau < 1
\end{array} \right.
\label{6.40}
\end{eqnarray}
in which, once again, $l_0(\mu) = \left[\frac{\mu}{U_0} +
\frac{1}{2} \right]$ (nonzero only for $\epsilon_\tau = 1$) is the
integer that minimizes $(l-\mu/U_0)^2$, and the last line follows
for $\beta \to \infty$. The derivative with respect to ${\bm \mu}$
therefore recovers the Mott lobe integer density $\rho = l_0(\mu)$
for $\epsilon_\tau = 1$ and ${\bm \rho} = 0$ for $\epsilon_\tau <
1$. The vanishing compressibility follows immediately.

In the presence of disorder, $l_0({\bm \mu}-{\bm \varepsilon})$ is
unchanged within a (smaller) Mott lobe, so $\Delta
f^\mathrm{dis}_0({\bm \mu}) = \int d^{\epsilon_\tau} \epsilon
p(|{\bm \varepsilon}|) \Delta f_0({\bm \mu} - {\bm \varepsilon}) =
\Delta f_0({\bm \mu})$ is also unchanged, and identical results
follow for the density and compressibility.

\subsubsection{Phase diagram for finite range hopping}
\label{subsec:epstau_phases}

In both pure and disordered cases, the fact that $M_0 \approx
\sqrt{-r_0/u_0}$ is nonzero immediately outside of the Mott phase
boundary, with unchanged mean field critical behavior, implies that
there is no glassy phase in mean field theory. Only for
finite-ranged hopping will a finite width Bose glass phase appear
between the Mott and superfluid phases, as in Fig.\
\ref{fig:phases}(b).

In considering the interpolation between small $\epsilon_\tau$ and
$\epsilon_\tau = 1$ it is clear that it makes sense only to consider
the zero density Mott lobe.  The renormalization group treatment in
Sec.\ \ref{sec:rg_epstau} will therefore focus on the vicinity of
the commensurate transition at $|{\bm \mu}| = 0$.  The mean field
calculations give one some confidence that the critical behavior
near this point will vary continuously with $\epsilon_\tau$, and
that extrapolations from small values to the physical value
$\epsilon_\tau = 1$ will be at least qualitatively valid.

\subsection{Droplet model for general $\epsilon_\tau$}
\label{sec:droplet_epstau}

In order to gain insight into the nature of the Bose glass phase for
general $\epsilon_\tau$, we next generalize the droplet model of
Sec.\ \ref{sec:excite}. The identical rare region argument as for
$\epsilon_\tau = 1$ implies that the Mott phase boundary occurs at
$\mu_\pm(J_0,\Delta) = \mu_\pm(J_0,0) \mp \Delta$, where $\Delta$ is
the maximum supported value of ${\bm \varepsilon}_i$. For ${\bm
\mu}| > |\mu_\pm(J_0,\Delta)|$ there will . We consider then the
effective action (\ref{6.9}) in a finite superfluid droplet, with
spatial volume $V$, in which $\phi$ is assumed constant in space
[compare (\ref{3.30})]:
\begin{equation}
S_\mathrm{eff}^{(1)} = V \int d^{\epsilon_\tau}\tau
\left[\frac{1}{2} \kappa (\nabla_\tau \phi)^2
+ i {\bm \rho} \cdot \nabla_\tau \phi \right]
\label{6.41}
\end{equation}
in which, as before, ${\bm \rho}$, $\kappa$ are the bulk density and
compressibility if the droplet were extended into an infinite
homogeneous medium. The usual $2\pi$-periodic boundary conditions on
$\phi$ apply.

\subsubsection{Droplet density and compressibility}
\label{subsec:dropdensity_epstau}

Let us first generalize the free energy calculation, (\ref{3.32}).
By performing a calculation identical to (\ref{6.39}) and
(\ref{6.40}), the generalization of (\ref{3.36}) in the vicinity of
chemical potential ${\bm \mu}_0$ then reads
\begin{eqnarray}
f &=& f_0(|{\bm \mu}|) + f_{00}(\kappa^0)
\label{6.42} \\
&&-\ \frac{1}{\beta^{\epsilon_\tau}V}
\ln \left[\sum_{\bf l} e^{-\beta^{2-\epsilon_\tau}
(\beta^{\epsilon_\tau-1}
{\bm \rho} V - {\bf l})^2/2\kappa^0 V} \right],
\nonumber
\end{eqnarray}
in which the functions $f_0$ and $f_{00}$ are defined analogously to
(\ref{3.34}) and (\ref{3.36}):
\begin{equation}
f_{00}(\kappa_0) =  -\frac{1}{2} \int
\frac{d^{\epsilon_\tau} \omega}{(2\pi)^{\epsilon_\tau}}
\ln\left(\frac{2\pi}{\kappa^0 V |{\bm \omega}|^2}\right)
\label{6.43}
\end{equation}
is the free energy associated with the $\frac{1}{2} \kappa^0
(\nabla_\tau \tilde \phi)^2$ term [compare (\ref{6.39})], and
\begin{equation}
f_0(|{\bm \mu}|) = f_0(|{\bm \mu}_0|)
- |{\bm \rho}^0| (|{\bm \mu}|-|{\bm \mu}_0|)
- \frac{1}{2}\kappa^0 (|{\bm \mu}|-|{\bm \mu}_0|)^2,
\label{6.44}
\end{equation}
is the analytic background free energy in the neighborhood of ${\bm
\mu}_0$. In the limit $\beta \to \infty$, for $\epsilon_\tau = 1$,
one recovers the discrete particle addition result (\ref{3.39}) and
(\ref{3.40}).  For $\epsilon_\tau < 1$ only the ${\bf l}={\bf 0}$
term in (\ref{6.42}) survives, and one obtains
\begin{equation}
f = f_0 + f_{00} + \frac{|{\bm \rho}|^2}{2\kappa^0}.
\label{6.45}
\end{equation}
Noting that $\frac{\partial {\bm\rho}}{\partial {\bm \mu}} = \kappa
\openone$, one obtains
\begin{eqnarray}
{\bm \rho}_\mathrm{drop}({\bm \mu}_0)
&=& -\left(\frac{\partial f}{\partial {\bm \mu}}
\right)_{{\bm \mu} = {\bm \mu}_0}
\nonumber \\
&=& {\bm \rho}^0 - \frac{1}{\kappa^0} {\bm \rho}^0 \cdot
\left(\frac{\partial {\bm \rho}}{\partial {\bm \mu}}
\right)_{{\bm \mu} = {\bm \mu}_0}
\nonumber \\
&=& {\bm \rho}^0 - {\bm \rho}^0 = {\bf 0}.
\label{6.46}
\end{eqnarray}
The superfluid droplet therefore has vanishing density, and hence,
like the Mott phase, is \emph{incompressible,} for arbitrary ${\bm
\mu}$ and $V$. We conclude that for $\epsilon_\tau < 1$ both the
Mott phase and the Bose glass phase are incompressible, and $\kappa$
(and ${\bm \rho}$) will become nonzero only as one crosses into the
superfluid phase.

\subsubsection{Droplet temporal correlations and superfluid
susceptibility}
\label{subsec:dropcorr_epstau}

One must therefore seek a different measure to distinguish the Mott
and Bose glass phases for $\epsilon_\tau < 1$.  We examine,
therefore, the temporal correlation function and superfluid
susceptibility.

By following through the derivation (\ref{3.41})--(\ref{3.43}) for
$\epsilon_\tau < 1$, analogous to the survival of only the ${\bf l}
= 0$ term in the free energy series (\ref{6.42}), the $\beta \to
\infty$ limit for the temporal correlation function leads to the
simple result
\begin{eqnarray}
G^{(0)}_\rho({\bm \tau} - {\bm \tau}^\prime)
&=& \left\langle e^{i[\phi({\bm \tau})
-\phi({\bm \tau}')]} \right \rangle_{S^{(1)}_\mathrm{eff}}
\nonumber \\
&=& G_0({\bm \tau} -{\bm \tau}^\prime)
e^{-{\bm \rho} \cdot ({\bm \tau}-{\bm \tau}')/\kappa}
\label{6.47}
\end{eqnarray}
in which
\begin{equation}
G_0({\bm \tau}) = e^{-A(\epsilon_\tau)
|{\bm \tau}|^{2-\epsilon_\tau}/\kappa V}
\label{6.48}
\end{equation}
is the Gaussian correlation at ${\bm \rho} = {\bf 0}$, with
coefficient $A(\epsilon_\tau)$ defined by (\ref{6.30})---compare
(\ref{6.32}).

Clearly the $G_0$ factor, which decays faster than exponentially,
dominates the asymptotic behavior at large $|{\bm \tau} - {\bm
\tau}'|$, but for large $\kappa V$ this may not occur until $|{\bm
\tau} - {\bm \tau}'|$ is extremely large.  To quantify this, we
compute the droplet averaged correlation function
\begin{eqnarray}
G_\rho({\bm \tau}) &=& \int dV d\kappa
p(V,\kappa) G^{(0)}_\rho({\bm \tau})
\nonumber \\
&\approx& e^{-\Delta {\bm \mu} \cdot {\bm \tau}}
\int dV d\kappa p(V,\kappa) G_0({\bm \tau})
\label{6.49}
\end{eqnarray}
where $p(V,\kappa)$ is the probability density for a droplet of size
$V$ and bulk compressibility $\kappa$. The last line holds in the
vicinity of the Mott lobe boundary, where ${\bm \rho} \approx
\kappa_0 \Delta {\bm \mu}$, where $\Delta {\bm \mu} = {\bm \mu}-{\bm
\mu}_c(J_0,\Delta)$ is the deviation from the phase boundary, and
$\kappa_0(J_0)$ is the compressibility just inside the background
superfluid phase---compare (\ref{3.31}). Therefore ${\bm \rho} \cdot
{\bm \tau}/\kappa \approx \Delta {\bm \mu} \cdot {\bm
\tau}$ is approximately independent of $\kappa$.

For large $|{\bm \tau}|$, the integral (\ref{6.49}) is dominated by
large volumes for which $p(V,\kappa) \sim e^{-V/V_0(\kappa)}$, where
$V_0(\kappa)$ is a characteristic scale for droplets with background
compressibility $\kappa$. A steepest decent evaluation of
(\ref{6.49}) then leads to
\begin{equation}
G_\rho({\bm \tau}) \sim e^{-\Delta {\bm \mu} \cdot {\bm \tau}}
e^{-(|{\bm \tau}|/\tau_0)^{1-\epsilon_\tau/2}},
\label{6.50}
\end{equation}
where $\tau_0(J_0) = [\kappa_0 V_0(\kappa_0)/4
A(\epsilon_\tau)]^{1/(1-\epsilon_\tau)}$, and various
non-exponential prefactors have been dropped. The $|{\bm
\tau}|/\tau_0)^{1-\epsilon_\tau/2}$ exponent reproduces (\ref{3.20})
at $\Delta {\bm \mu} = 0$ and $\epsilon_\tau = 1$. This is quite a
remarkable result: Although the temporal correlations decay at
infinity within any given droplet, the large droplet tail of the
distribution causes the droplet averaged correlation function to
\emph{diverge} in directions such that $\Delta {\bm \mu} \cdot {\bm
\tau} \to -\infty$.  This strongly distinguishes the Bose glass
phase from the Mott phase.

A related quantity is the superfluid susceptibility,
\begin{equation}
\chi_s = \left(\frac{\partial \psi_0}{\partial h}\right)_{h=0}
= \int d^dx \int d^{\epsilon_\tau}\tau G({\bf x},{\bm \tau}).
\label{6.51}
\end{equation}
The saddle point estimate of the contribution from a single droplet
is given by,
\begin{eqnarray}
\chi_\mathrm{drop}(V) &\approx& V \int
d^{\epsilon_\tau}\tau G^{(0)}_\rho({\bm \tau})
\nonumber \\
&\sim& e^{(1-\epsilon_\tau)
\kappa_0^{\frac{3-\epsilon_\tau}{1-\epsilon_\tau}}
\left[\frac{|\Delta {\bm \mu}|}
{(2-\epsilon_\tau)}\right]^{\frac{2-\epsilon_\tau}
{1-\epsilon_\tau}} \left[\frac{V}{A(\epsilon_\tau)}
\right]^{\frac{1}{1-\epsilon_\tau}}},\ \ \ \ \ \ \ \
\label{6.52}
\end{eqnarray}
which grows faster than exponentially in $V$, and is therefore
dominated by large droplets. The total susceptibility
\begin{equation}
\chi_s \sim \int dV e^{-V/V_0(\kappa_0)} \chi_\mathrm{drop}(V)
\to \infty,
\label{6.53}
\end{equation}
therefore diverges, providing a more direct signature of the Bose
glass phase.

\subsubsection{Droplet model conclusions: $\epsilon_\tau < 1$
vs.\ $\epsilon_\tau = 1$}
\label{subsec:epstau_dropsummary}

The droplet model exhibits some strong differences in the physics of
the Bose glass phase, all related to the physics of superfluid
droplets, between $\epsilon_\tau = 1$ and $\epsilon_\tau < 1$ that
may raise questions about the smoothness of the limit $\epsilon_\tau
\to 1$.

The vanishing density and compressibility in the Bose glass phase
implies that there should be no analytic background contribution to
$\kappa$ for $\epsilon_\tau < 1$.  Thus (\ref{6.17}), with some
exponent $\alpha(\epsilon_\tau) < 0$, is expected to describe the
critical behavior of the full compressibility, not just its singular
part. This means that the analytic contribution to the density and
compressibility must appear \emph{discontinuously} at $\epsilon_\tau
= 1$.

Due to droplet incompressibility, the mechanism leading to a
divergent susceptibility for $\epsilon_\tau < 1$ is also rather
different than that for $\epsilon_\tau = 1$. As discussed in Sec.\
\ref{subsec:excbf}, for the latter, the temporal correlations decay
as a slow $1/\tau$ power law---see (\ref{3.24}) and
(\ref{3.48})---due to the spectrum of \emph{finite sized} droplets
that are very close to adding or giving up an extra particle.  For
$\epsilon_\tau < 1$, the long-time correlations are dominated by
large droplets, and it is the resulting stretched exponential
behavior that generates the divergent susceptibility.

These strong differences do not require that critical exponents,
such as $\alpha$, be discontinuous, but one could certainly imagine
some kind of nonanalytic behavior, as when one approaches an upper
or lower critical dimension of a transition. We shall see below
that, at least, the qualitative features of the critical behavior do
behave as functions of $\epsilon_\tau$ in the expected fashion.
Since the expansion is around $\epsilon_\tau = 0$, it has absolutely
nothing to say about possible singularities at $\epsilon_\tau = 1$.
However, the results do lend some support to it as an analytic tool
for exploring some features of the Bose glass--superfluid transition
in higher dimensions where numerical results are not yet available.

\subsection{Renormalization group calculations}
\label{sec:rg_epstau}

\subsubsection{Replicated Lagrangian}
\label{subsec:replicas}

In this final section, we revisit the renormalization group
calculation carried out in Ref.\ \onlinecite{WK89}, but now with
explicit attention to issues of particle-hole symmetry.  To this
end, we use the standard replica trick on ${\cal L}_c$ in
(\ref{6.1}) to average over the disorder,\cite{A75} and obtain the
replicated Lagrangian, ${\cal L}_c^{(p)} = {\cal L}_1^{(p)} + {\cal
L}_2^{(p)}$, where $p \rightarrow 0$ is the number of replicas, and
\begin{widetext}
\begin{eqnarray}
{\cal L}_1^{(p)} &=& -\sum_{\alpha=1}^p \int d^dx
\int d^{\epsilon_\tau}\tau \left[\frac{1}{2} e_\tau
|\nabla_\tau \psi_\alpha|^2
+ {\bf g}_0 \cdot \psi^* \nabla_\tau \psi_\alpha
+ \frac{1}{2} e_x |\nabla \psi_\alpha|^2
+ \frac{1}{2} r_0 |\psi_\alpha|^2
+ \frac{1}{4} u |\psi_\alpha|^4 \right]
\label{6.54} \\
{\cal L}_2^{(p)} &=& \frac{1}{2} \sum_{\alpha,\beta=1}^p
\int d^dx \int d^{\epsilon_\tau}\tau
\int d^{\epsilon_\tau}\tau^\prime
[\tilde \Delta_r |\psi_\alpha({\bf x},{\bm \tau})|^2
|\psi_\beta({\bf x},{\bm \tau}')|^2
\nonumber \\
&&\ \ \ \ \ \ \ \ \
+\ \Delta_g [\psi^* \nabla_\tau \psi
- {\bf g}_0|\psi|^2]({\bf x},{\bm \tau})
\cdot [\psi^* \nabla_\tau \psi - {\bf g}_0
|\psi|^2]({\bf x},{\bm \tau}')].
\label{6.55}
\end{eqnarray}
\end{widetext}
In the end we will take $e_\tau = e_x = 1$, but in setting up the RG
calculation it is useful to leave them as free parameters. For
simplicity we have taken $\delta {\bf g}({\bf x})$ and $\delta
\tilde r({\bf x}) \equiv \delta r({\bf x}) - \delta {\bf g}({\bf
x})^2$ to be independent Gaussian random fields with
\begin{eqnarray}
\left[\delta {\bf g}({\bf x}) \right]_\mathrm{av} &=& {\bf 0},\
\left[\delta {\bf g}_\mu({\bf x})
\delta {\bf g}_\nu({\bf x}') \right]_\mathrm{av}
= \Delta_g \delta({\bf x}-{\bf x}') \delta_{\mu\nu}
\nonumber \\
\left[\delta \tilde r({\bf x}) \right]_\mathrm{av} &=& 0,\
\left[\delta \tilde r({\bf x}) \delta \tilde r({\bf x}')
\right]_\mathrm{av}
= \tilde \Delta_r \delta({\bf x}-{\bf x}').
\label{6.56}
\end{eqnarray}
The analysis in Ref.\ \onlinecite{WK89} was carried out with ${\bf
g}_0$ finite and $\Delta_g = 0$, but $\tilde\Delta_r > 0$.  However,
considerations of particle-hole symmetry now lead us to seek fixed
points with ${\bf g}_0 = 0$ and $\Delta_g > 0$.

In Fourier space we have
\begin{eqnarray}
{\cal L}_2^{(p)} &=& \frac{1}{2} \sum_{\alpha,\beta=1}^p
\int_{{\bf k}_1} \int_{{\bf k}_2} \int_{{\bf k}_3} \int_{{\bf k}_4}
\delta({\bf k}_1-{\bf k}_2+{\bf k}_3-{\bf k}_4)
\nonumber \\
&&\times\ \int_{\bm \omega} \int_{{\bm \omega}'}
[\Delta_r - 2i \Delta_g {\bf g}_0 \cdot {\bm \omega}
- \Delta_g {\bm \omega} \cdot {\bm \omega}']
\nonumber \\
&&\ \ \ \ \ \times\ \psi_\alpha^*({\bf k}_1,{\bm \omega})
\psi_\alpha({\bf k}_2,{\bm \omega})
\psi_\beta^*({\bf k}_3,{\bm \omega}')
\psi_\beta({\bf k}_4,{\bm \omega}'),
\nonumber \\
\label{6.57}
\end{eqnarray}
where $\Delta_r \equiv \tilde \Delta_r + {\bf g}_0^2 \Delta_g$.
Nominally, by naive power counting, it would appear that the leading
term at low frequencies should be the $\Delta_r$ term, and one might
expect the other two frequency dependent terms to be strongly
irrelevant. This will turn out to be true for small $\epsilon_\tau$,
where naive power counting is almost valid.  However, because ${\bm
\omega}$ and ${\bm \omega}'$ refer to different replicas, $\alpha$
and $\beta$, these terms break particle-hole symmetry, and are not
as strongly irrelevant as one might expect (as compared to, for
example, $\omega^2$ corrections to the $|\psi|^4$ coefficient, $u$),
and, in fact, will be seen to become relevant beyond a critical
value of $\epsilon_\tau$.

We shall also begin by setting ${\bf g}_0 = 0$.  By power counting,
the ${\bf g}_0 \cdot \psi^* \nabla_\tau \psi$ term (focused on in
Ref.\ \onlinecite{WK89}) again appears strongly relevant compared to
the $e_\tau |\nabla_\tau \psi|^2$, and the $\Delta_g {\bf g}_0
(\psi^* \nabla \tau \psi) |\psi|^2$ appears strongly relevant
compared to the $\Delta_g (\psi^* \nabla \tau \psi) \cdot (\psi^*
\nabla \tau \psi)$ term. Again, although relevant at small
$\epsilon_\tau$, there is a (different) critical value beyond which
these terms become irrelevant.

\begin{figure}

\includegraphics[width=\columnwidth]{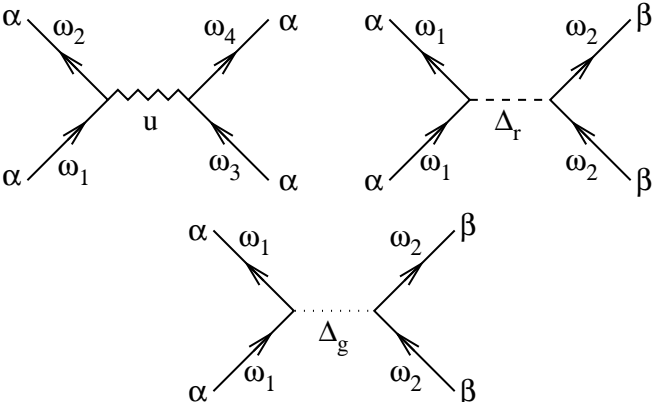}

\caption{Vertices corresponding to $u$, $\Delta_r$ and $\Delta_g$,
where $\alpha,\beta$ are replica indices, and the $u$ vertex also
enforces energy conservation, $\omega_1 - \omega_2 + \omega_3 -
\omega_4 = 0$.}

\label{fig:RG1}
\end{figure}

\begin{figure}

\includegraphics[width=0.8\columnwidth]{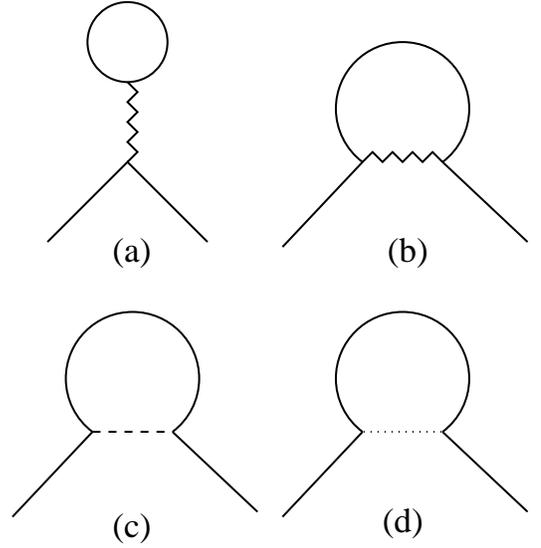}

\caption{Diagrams that contribute to the propagator renormalization.}

\label{fig:RG2}
\end{figure}

\begin{figure}

\includegraphics[width=\columnwidth]{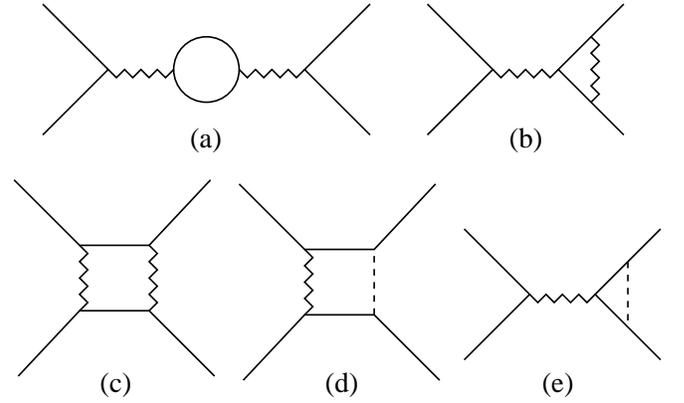}

\caption{Diagrams that contribute to the renormalization of $u$.}

\label{fig:RG3}
\end{figure}

\begin{figure}

\includegraphics[width=\columnwidth]{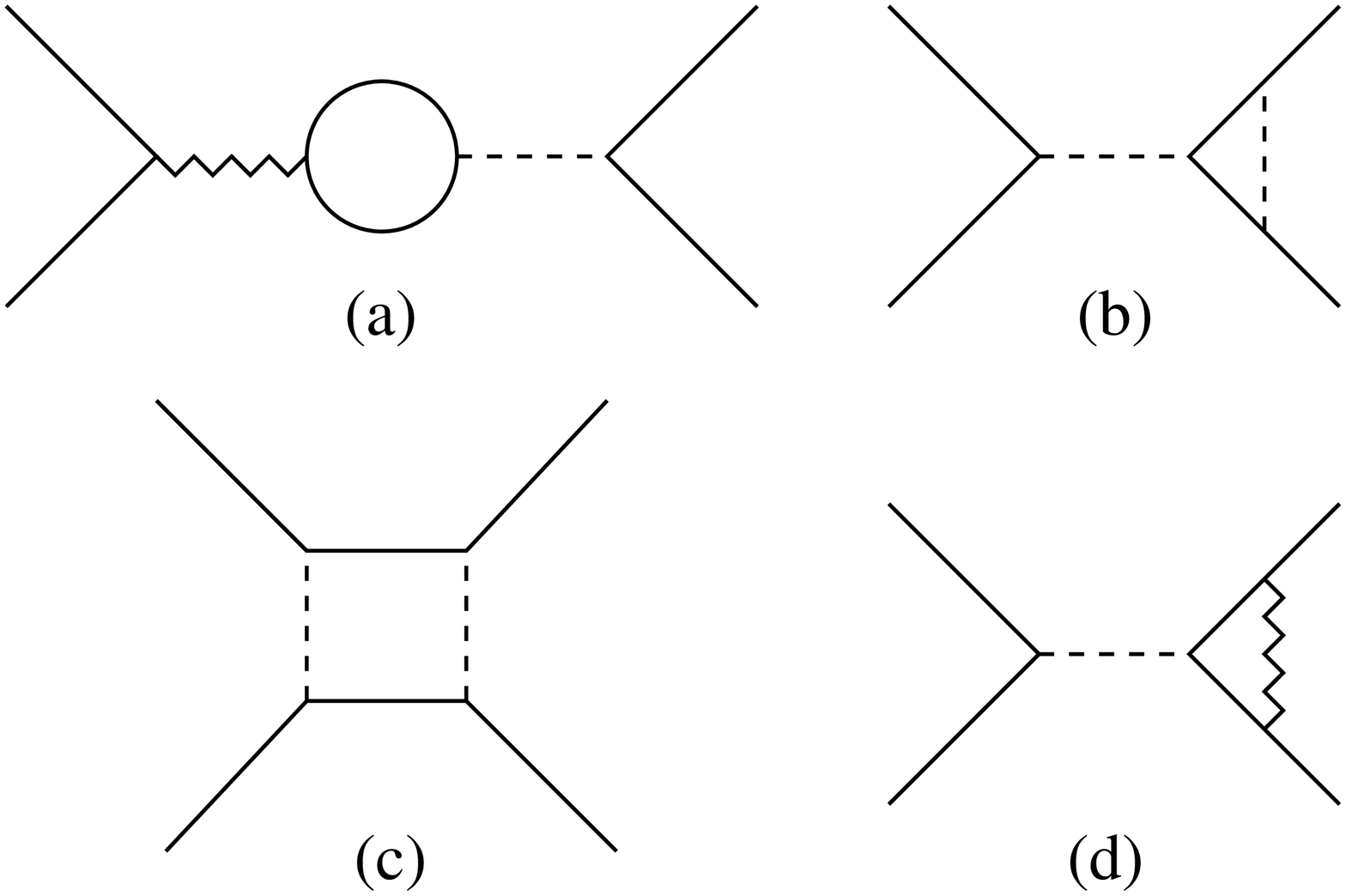}

\caption{Diagrams that contribute to the renormalization of $\Delta_r$.}

\label{fig:RG4}
\end{figure}

\begin{figure}

\includegraphics[width=0.8\columnwidth]{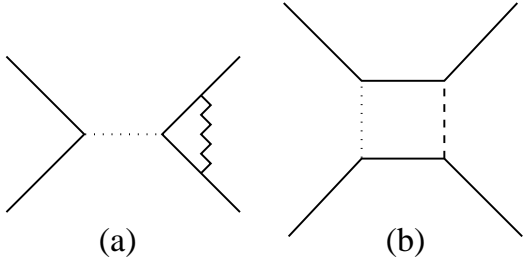}

\caption{Diagrams that contribute to the renormalization of $\Delta_g$.}

\label{fig:RG5}
\end{figure}

\subsubsection{Recursion relations}
\label{subsec:recur}

We shall perform a standard Wilson momentum shell renormalization
group transformation\cite{WK74} in which successive shells in
$k$-space are integrated out.  For each such ${\bf k}$ all
frequencies, ${\bm \omega}$, are integrated out.  Since the
frequency is unbounded, the Brillouin zone is really a
hypercylinder.  After each integration, we rescale ${\bf k}$ and
${\bm \omega}$ in order to maintain the same wavevector cuttoff,
$k_\Lambda$.  The spin rescaling factor and the dynamical exponent,
$z$, are determined in the usual way by setting $e_x$ and $e_\tau$
equal to unity.

In Fig.\ \ref{fig:RG1} are shown the four basic diagrammatic
vertices corresponding to $u$, $\Delta_r$, and $\Delta_g$, while in
Figs.\ \ref{fig:RG2}--\ref{fig:RG5} are shown the lowest order
diagrams that then contribution to the renormalization of the
propagator and of the vertices themselves.  Note that there are a
number of ``missing'' diagrams not included because they do not
contribute in the replica limit, $p \to 0$.  We obtain, then, in a
straightforward way the recursion relations:
\begin{eqnarray}
\frac{d\bar r}{dl} &=& 2 \bar r + \frac{2(m+1) \bar u}{1 + \bar r}
- \frac{2 \bar \Delta_r}{1 + \bar r}
+ O(\bar u^2,\bar\Delta_r^2,\bar\Delta_g^2)
\nonumber \\
\label{6.58} \\
\frac{d\bar u}{dl} &=& \epsilon \bar u - 2(m+4) \bar u^2
+ 12 \bar u \bar \Delta_r + O(\bar u^3,\ldots)
\nonumber \\
\label{6.59} \\
\frac{d \bar \Delta_r}{dl} &=& (\epsilon + \epsilon_\tau)\bar \Delta_r
+ 8 \bar \Delta_r^2 - 4(m+1) \bar u \bar \Delta_r
+ O(\bar u^3,\ldots)
\nonumber \\
\label{6.60} \\
\frac{d \bar \Delta_g}{dl} &=& \bar \Delta_g (\epsilon +
\epsilon_\tau + 10 \bar \Delta_r - 2 \bar\Delta_g - 2)
\nonumber \\
&\equiv& \lambda_g \bar\Delta_g - 2\bar\Delta_g^2,
\label{6.61}
\end{eqnarray}
where $m$ is the number of boson species ($m=1$ physically), $\bar r
= r_0/k_\Lambda^2$, $\bar u = K_d u/4$, $\bar \Delta_r = K_d
\Delta_r$, $\bar\Delta_g = k_\Lambda^2 K_d \Delta_g$ are
appropriately rescaled by the cutoff, and $K_d = 2/(4\pi)^{d/2}
\Gamma(d/2)$ is $(2\pi)^{-d}$ times the area of the unit sphere in
$d$-dimensions. The conditions $e_x = e_\tau = 1$ lead to the
identifications
\begin{equation}
z = 1 + \bar \Delta_r + \bar\Delta_g,\ \ \eta=0.
\label{6.62}
\end{equation}

\subsubsection{Dirty bosons fixed point}
\label{subsec:fixedpts}

Note that $\bar\Delta_g$ does not enter any recursion relations
except its own at this order.  If one sets $\bar\Delta_g=0$ one
obtains the usual Boyanovsky-Cardy lowest order recursion
relations,\cite{DBC} with fixed point
\begin{eqnarray}
\bar r^* &=& -\frac{3m\epsilon + (5m+2)\epsilon_\tau}{8(2m-1)},\
\bar u^* = \frac{\epsilon + 3 \epsilon_\tau}{4(2m-1)}
\nonumber \\
\bar\Delta_g^* &=& 0,\ \bar\Delta_r^*
= \frac{(2-m)\epsilon + (m+4)\epsilon_\tau}{8(2m-1)},
\label{6.63}
\end{eqnarray}
correct to linear order in $\epsilon$ and $\epsilon_\tau$. For
sufficiently small $\epsilon, \epsilon_\tau$ we see that
$\lambda_g^* < 0$ and, consistent with the previous power counting
estimates, this fixed point is stable against the perturbation $\bar
\Delta_g$.  However, for
\begin{equation}
\epsilon_\tau > \epsilon_\tau^c \equiv
\frac{8(2m-1) - 3(m+2)\epsilon}{13m + 16}
\label{6.64}
\end{equation}
this fixed point becomes unstable.  Setting $m=1$ and $\epsilon=0$
(so that $\epsilon_\tau = 1$ corresponds to $d=3$) we find
\begin{equation}
\epsilon_\tau^c = \frac{8}{29},
\label{6.65}
\end{equation}
which is actually quite small, and is therefore a potentially
reasonable estimate.  Using (\ref{6.61}) and (\ref{6.62}) at $\bar
\Delta_g = 0$, one may write $\lambda_g = 2z - d + 8\bar\Delta_r$.
The last term shows the rather large deviation from the naive result
$\lambda_g = 2z - d$, discussed in Sec.\ \ref{subsec:epstau_phsym}.
The latter would have led to the estimate $\epsilon_\tau^c =
\frac{8}{9}$, which is uncomfortably close to unity, considering how
poorly controlled this expansion is for larger
$\epsilon_\tau$.\cite{DBC}

For $\epsilon_\tau > \epsilon_\tau^c$, there is now a new stable
finite $\bar \Delta_g$ fixed point,
\begin{equation}
\bar \Delta_g^* = \frac{1}{2}(\epsilon + \epsilon_\tau
+ 10 \bar \Delta_r^* - 2) = \frac{13m + 16}{4(2m-1)}
(\epsilon_\tau - \epsilon_\tau^c),
\label{6.66}
\end{equation}
which bifurcates continuously away from the random rod fixed point.
The corresponding dynamical exponent,
\begin{equation}
z = 1 + \bar\Delta_r^* + \bar\Delta_g^* =
\frac{(m+4)\epsilon + (7m+10)\epsilon_\tau}{4(2m-1)},\
\epsilon_\tau > \epsilon_\tau^c,
\label{6.67}
\end{equation}
is substantially \emph{larger} than the random rod value. For $m=1$,
$\epsilon=0$, and $\epsilon_\tau=1$ one obtains $z = \frac{17}{4}$,
which should be considered a very crude extrapolation. The
``thermal'' eigenvalue, determining $\nu$, is
\begin{eqnarray}
\frac{1}{\nu} &=& 2 - 2(m+1)\bar u^* + 2 \bar \Delta_r^*
\nonumber \\
&=& 2 - \frac{(m+4)\epsilon + (7m+10)\epsilon_\tau}{8(2m-1)},
\label{6.68}
\end{eqnarray}
while, as stated above, $\eta=0$.  These results are both unchanged
from their random rod values at this order.

\begin{figure}

\includegraphics[width=0.9\columnwidth]{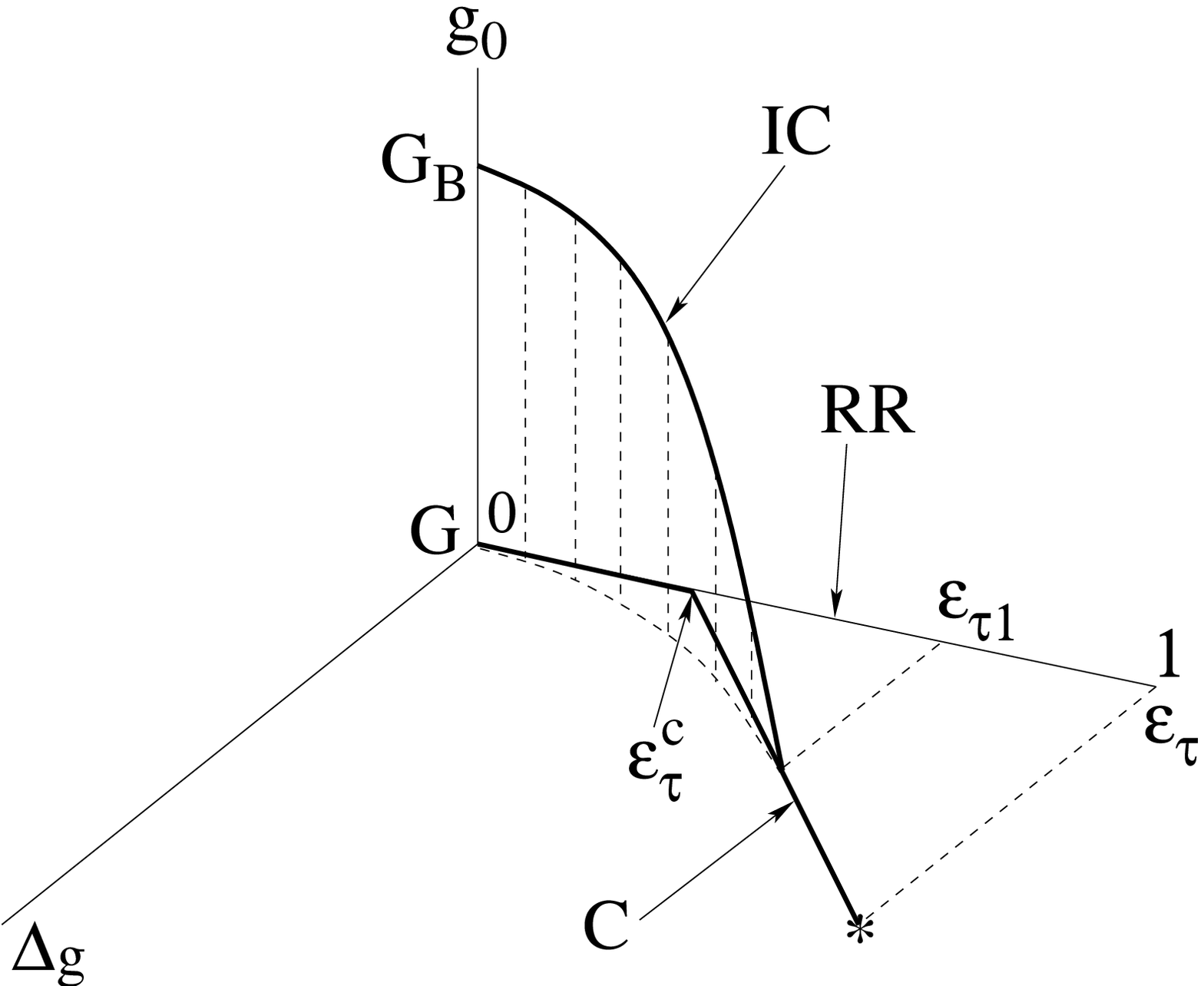}

\caption{Proposed behavior of the random rod (RR), commensurate
(statistically particle-hole symmetric) dirty boson (C), and
incommensurate (fully particle-hole asymmetric) dirty boson (IC)
fixed points as functions of $\epsilon_\tau$.  The $\Delta_r$ axis
has been suppressed. Here $G$ and $G_B$ are commensurate and
incommensurate Gaussian fixed points. The commensurate fixed point
bifurcates away from the random rod fixed point at $\epsilon_\tau^c
\simeq \frac{8}{29}$. At $\epsilon_\tau = \epsilon_{\tau 1} \simeq
\frac{2}{3}$, C and IC merge and at $\epsilon_\tau = 1$, C is the
stable fixed point that describes the physical dirty boson problem.}

\label{fig:fixedpoints}
\end{figure}

\subsubsection{Relevance of particle-hole symmetry breaking}
\label{subsec:phsym_relevance}

Finally, let us include the ${\bf g}_0$ term.  To linear order the
flow equation for $g_0 = |{\bf g}_0|$ is found to be
\begin{equation}
\frac{dg_0}{dl} = g_0 [1 + \bar \Delta_r - \bar \Delta_g].
\label{6.69}
\end{equation}
Consistent with the power counting estimates, $g_0$ is strongly
relevant for small $\epsilon_\tau$, and is always relevant at the
random rod fixed point, $\bar \Delta_g = 0$. However, $g_0$ becomes
\emph{irrelevant} at the dirty boson fixed point for
\begin{equation}
\epsilon_\tau \geq \epsilon_{\tau 1}
\equiv \frac{4(2m-1)}{3(1+m-\epsilon)}.
\label{6.70}
\end{equation}
For $m=1$ and $\epsilon = 0$ one obtains $\epsilon_{\tau 1} =
\frac{2}{3}$. This is again an uncontrolled estimate, but does
indeed indicate that the statistical particle-hole symmetry is
restored prior to $\epsilon_\tau = 1$.

For $\epsilon_\tau < \epsilon_{\tau 1}$ there is a new
incommensurate fixed point at nonzero $g_0$.  In order to locate it
one should choose $z$ to keep $g_0$, rather than $e_\tau$, fixed
during the RG flow.  The flow equation for $\Delta_g$ then becomes
\begin{equation}
\frac{d\bar \Delta_g}{dl} = \bar \Delta_g
[\epsilon + \epsilon_\tau - 4 + 8 \bar\Delta_r]
+ 8 \bar\Delta_r^2,
\label{6.71}
\end{equation}
while the remaining flow equations are identical to those in Ref.\
\onlinecite{WK89}. The fixed point found in that work did not
account for $\Delta_g$, but we see that if $\Delta_r^* =
O(\epsilon,\epsilon_\tau)$ then $\Delta_g^* = O(\epsilon^2,\epsilon
\epsilon_\tau,\epsilon_\tau^2)$, so the results given there are
indeed correct to $O(\epsilon,\epsilon_\tau)$.  However, as
$\epsilon_\tau$ grows, so does $\Delta_g^*$, eventually leading to
the merging with the dirty boson fixed point.

The technical problems encountered in Ref.\ \onlinecite{WK89} are
also overcome in our analysis.  Specifically, the $|\nabla_\tau
\psi|^2$ term, which was also previously ignored, has the flow
equation
\begin{equation}
\frac{de_\tau}{dl} = -(1 + \bar\Delta_r) e_\tau + \bar{\Delta}_g,
\label{6.72}
\end{equation}
implying that it is of the same order as $\bar \Delta_g^*$ at the
fixed point. A nonzero $e_\tau$ fixes the convergence problems, and
allows one to remove an unphysical frequency cutoff
$\omega_\Lambda$.

Finally, to see how the two fixed points merge, we write down flow
equations in the intermediate region, $\epsilon_{\tau 1} >
\epsilon_\tau > \epsilon_\tau^c$, where one must consider both
$e_\tau$ and $\bar g_0 \equiv k_\Lambda g_0$. We choose $z$ so that
$e_\tau + \bar g_0 \equiv 1$ remains fixed. The flow equation for
$g_0$ is then
\begin{equation}
\frac{dg_0}{dl} = (2 + 2 \bar\Delta_r - z) g_0
\label{6.73}
\end{equation}
with
\begin{equation}
z = (1 + \bar\Delta_r \bar \Delta_g)
+ (1 +\bar \Delta_r)\bar g_0.
\label{6.74}
\end{equation}
At the fixed point we therefore find
\begin{equation}
\bar g_0^* = \frac{1 + \bar \Delta_r^* - \bar \Delta_g^*}
{1 + \bar \Delta_r^*},
\label{6.75}
\end{equation}
which vanishes precisely when $g_0$ becomes irrelevant at the dirty
boson fixed point.

The entire proposed fixed point structure as a function of
$\epsilon_\tau$ is summarized in Fig.\ \ref{fig:fixedpoints}). For
small $\epsilon_\tau$ the unstable random fixed point and stable
filly particle-hole asymmetric fixed points exist.  For
$\epsilon_\tau^c < \epsilon_\tau < \epsilon_{\tau 1}$ there are
three fixed points, with the new commensurate dirty boson fixed
point being more stable than the random rod fixed point, but less
stable than the asymmetric fixed point. Finally, for $\epsilon_\tau
> \epsilon_{\tau 1}$ the incommensurate fixed point merges with the
dirty boson fixed point, which is then completely stable.  This
provides a detailed scenario by which statistical particle-hole
symmetry is restored . We caution, however, that due both to the
uncontrolled nature of the double $\epsilon$-expansion at the dirty
boson fixed point, and the special nature of $\epsilon_\tau = 1$,
extrapolation of these results to $\epsilon_\tau = 1$ should be
treated as, at best, qualitative estimates. The general scenario we
propose, however, seems very natural and illuminating.

\appendix

\section{Functional integrals based on the Trotter decomposition}
\label{app:a}

Given a Hamiltonian, ${\cal H}$, thermodynamics is obtained from the
partition function $Z = \mathrm{tr}\left[e^{-\beta{\cal H}}\right]$.
The Trotter decomposition involves identifying a convenient complete
set of states $\{|\alpha\rangle\}$, writing $e^{-\beta{\cal H}} =
\left[e^{-\Delta\tau{\cal H}} \right]^M$, where $M =
\beta/\Delta\tau$, and inserting the states $|\alpha\rangle$ between
each element of the product:
\begin{eqnarray}
Z &=& \sum_{\alpha_0} \sum_{\alpha_1} \ldots \sum_{\alpha_{M-1}}
\langle \alpha_0|e^{-\Delta\tau{\cal H}} |\alpha_1 \rangle
\langle \alpha_1|e^{-\Delta\tau{\cal H}} |\alpha_2 \rangle
\nonumber \\
&&\times \ldots \times \langle \alpha_{M-1}|e^{-\Delta\tau{\cal H}}
|\alpha_0 \rangle.
\label{A1}
\end{eqnarray}

Often one can decompose ${\cal H} = {\cal H}_0 + {\cal H}_1$,
and choose the $|\alpha \rangle$'s to be eigenstates of ${\cal
H}_0$.  This is especially convenient when ${\cal H} = {\cal
H}[\{\hat q_i\},\{\hat p_i\}]$ is written in terms of a set of
$N$ canonically conjugate positions, $\hat q_i$, and momenta, $\hat
p_i$, and takes the special form
\begin{equation}
{\cal H} = {\cal H}_0[\{\hat q_i\}] + {\cal H}_1[\{\hat p_i\}]
\label{A2}
\end{equation}
in which ${\cal H}_1$ is a quadratic polynomial in the $\hat p_i$:
\begin{equation}
{\cal H}_1 = \frac{1}{2} \sum_{i,j}
S_{ij} \hat p_i \hat p_j + \sum_i t_i \hat p_i.
\label{A3}
\end{equation}
The $|\alpha \rangle$'s are then taken to be product eigenstates of
the positions $\hat q_i$:  $|q \rangle \equiv |q_1, \ldots, q_N
\rangle = \otimes_{i=1}^N |q_i \rangle$ with $\hat q_i |q \rangle =
q_i |q \rangle$. We also define product eigenstates of the momenta,
$|p\rangle \equiv |p_1,\ldots,p_N \rangle = \otimes_{i=1}^N
|p_i\rangle$, with $\hat p_i |p\rangle = p_i |p\rangle$.  From the
canonical commutation relations, $[\hat q_i, \hat p_i] =
i\delta_{ij}$, which in turn imply
\begin{eqnarray}
\left[\hat q_i,e^{-i\lambda \hat p_j}\right]
&=& \lambda e^{-i\lambda \hat p_j} \delta_{ij}
\nonumber \\
\left[e^{i\lambda \hat q_i}, \hat p_j \right] &=&
-\lambda e^{i\lambda \hat q_i} \delta_{ij},
\label{A4}
\end{eqnarray}
we immediately infer that
\begin{eqnarray}
e^{-i\lambda \hat p_j} |q_j \rangle &=& |q_j+\lambda \rangle
\nonumber \\
e^{i\lambda \hat q_j} |p_j \rangle &=& |p_j + \lambda \rangle.
\label{A5}
\end{eqnarray}
From this we obtain the \emph{wavefunctions}
\begin{eqnarray}
\langle q_i|p_i \rangle &=& \langle q_i=0|
e^{iq_i\hat p_i} |p_i \rangle
= e^{iq_ip_i} \langle q_i=0 | e^{ip_i \hat q_i} |p_i=0 \rangle
\nonumber \\
&=& \frac{1}{\cal N} e^{i q_i p_i},
\label{A6}
\end{eqnarray}
where $\frac{1}{\cal N} = \langle q_i=0 | p_i=0 \rangle$ normalizes
the wavefunction.  Note that if the $q_i$ are periodic variables,
defined only modulo $2\pi$, say, then $|q_i\rangle \equiv |q_i+2\pi
\rangle$ are identified.  This requires the same periodicity of the
wavefunctions, and determines the allowed eigenvalues, $p_i$---which
therefore must be integers in this case, and one obtains ${\cal N} =
\sqrt{2\pi}$ as well.

With the shorthands $\hat q = \{\hat q_i\}$, $\hat p = \{\hat
p_i\}$, etc., we may now compute, for small $\Delta\tau$,
\begin{eqnarray}
\langle q| e^{-\Delta\tau {\cal H}} |q^\prime \rangle
&\approx& \langle q| e^{-\Delta\tau {\cal H}_0(\hat q)}
e^{-\Delta\tau {\cal H}_1(\hat p)} |q^\prime\rangle
\nonumber \\
&=& e^{-\Delta\tau {\cal H}_0(q)} \sum_p \langle q|p \rangle
\langle p| e^{-\Delta\tau {\cal H}_1(\hat p)} |q^\prime \rangle
\nonumber \\
&=& e^{-\Delta\tau {\cal H}_0(q)} \sum_p e^{-\Delta\tau {\cal
H}_1(p)} \langle q|p \rangle \langle p|q^\prime \rangle
\nonumber \\
&=& \frac{e^{-\Delta\tau {\cal H}_0(q)}}{{\cal N}^N}
\sum_p e^{-\Delta\tau {\cal H}_1(p)}
e^{i \sum_i p_i(q_i-q_i^\prime)},
\nonumber \\
\label{A7}
\end{eqnarray}
and we are now left only with computing the inverse Fourier
transform of the Gaussian function $e^{-\Delta\tau {\cal H}_1(p)}$.
To do this, we first complete the square:
\begin{eqnarray}
{\cal H}_1[\hat p] &=& \frac{1}{2} \sum_{i,j} S_{ij} (\hat p_i +
\nu_i)(\hat p_j + \nu_j) - \frac{1}{2} \sum_{i,j} \nu_i \nu_j
\nonumber \\
\nu_i &\equiv& \sum_j (S^{-1})_{ij} t_j;\ t_i = \sum_j S_{ij}
\nu_j.
\label{A8}
\end{eqnarray}
We specialize now to the case of integer $p_i$. Using the formula
\begin{eqnarray}
\sum_{p_i=-\infty}^\infty &=& \int_{-\infty}^\infty dp_i
\sum_{n_i=-\infty}^\infty \delta(p_i-n_i)
\nonumber \\
&=& \int_{-\infty}^\infty dp_i \sum_{m_i=-\infty}^\infty
e^{i2\pi m_i p_i},
\label{A9}
\end{eqnarray}
one obtains
\begin{eqnarray}
&&\frac{1}{{\cal N}^N}
\sum_p e ^{i\sum_i p_i(q_i-q_i^\prime)}
e^{-\Delta\tau {\cal H}_1(p)}
\nonumber \\
&&\ \ =\ \sum_m \int \frac{dp}{{\cal N}^N}
e^{i \sum_i p_i(q_i-q_i^\prime + 2\pi m_i)}
e^{-\Delta\tau {\cal H}_1(p)}
\label{A10} \\
&&\ \ =\ \sum_m e^{\frac{1}{2}\Delta\tau
\sum_{i,j}S_{ij} \nu_i\nu_j
- i\sum_i \nu_i(q_i-q_i^\prime+2\pi m_i)}
\nonumber \\
&&\ \ \ \ \ \ \times\ \int \frac{d \bar p}{{\cal N}^N}
e^{-\frac{1}{2}\Delta \tau \sum_{i,j} S_{ij}
\bar p_i \bar p_j + i\sum_i \bar p_i (q_i-q_i^\prime+2\pi m_i)}
\nonumber \\
&&\ \ =\ \frac{1}{{\cal N}(\Delta\tau)}
\sum_m e^{\frac{1}{2}\Delta \tau \sum_{i,j} S_{ij} \nu_i \nu_j
- \sum_i \nu_i(q_i-q_i^\prime + 2\pi m_i)}
\nonumber \\
&&\ \ \ \ \ \ \times\ e^{-\frac{1}{2 \Delta\tau} \sum_{i,j}
(S^{-1})_{ij} (q_i-q_i'+2\pi m_i) (q_j-q_j'+2\pi m_i)}
\nonumber
\end{eqnarray}
where, in the second equality, we have changed variables to $\bar
p_i = p_i+\nu_i$, and ${\cal N}(\Delta\tau) = \sqrt{{\rm
det}(\Delta\tau S)}$.

Now consider the limit $\Delta\tau \rightarrow 0$.  For given
$\{q_i\}$ only a {\it single} term in the $m$-sum will survive,
namely that which minimizes the exponent.  Furthermore, only if
$q_i-q_i^\prime+2\pi m_i = O(\Delta\tau^{\frac{1}{2}})$ will the
term contribute to the path integral, (\ref{A1}).  Therefore, modulo
$2\pi$, $q_i(\tau)$ becomes a continuous function in the limit
$\Delta\tau \rightarrow 0$, and $q_i-q_i^\prime+2\pi m_i \rightarrow
-\Delta\tau \dot q_i$.  Clearly we will nearly always have $m_i=0$,
with $m_i=\pm 1$ every so often when the periodic boundary
conditions are enforced (for example, when $q_i = 2\pi^-$ and
$q_i^\prime = 0^+$, or vice versa).  When $\Delta\tau \rightarrow 0$
we may equivalently define a continuous function $q_i(\tau)$, taking
arbitrary real values, and then sum over all boundary conditions
$q_i(\tau)=q_i(0)+2\pi m_i$, with $m_i$ running over all integers.
Thus, we finally have, as $\Delta\tau \rightarrow 0$:
\begin{eqnarray}
&&\langle q| e^{-\Delta\tau {\cal H}_1(\hat p)}
|q^\prime \rangle
\nonumber \\
&&\ \ \ \ =\ \frac{1}{{\cal N}(\Delta\tau)}
e^{-\frac{1}{2} \Delta\tau \sum_{i,j}
(S^{-1})_{ij} \dot q_i \dot q_j}
\label{A11} \\
&&\ \ \ \ \ \ \ \ \ \
\times\ e^{-i\Delta\tau \sum_i \nu_i \dot q_i + \frac{1}{2}
\Delta\tau \sum_{i,j} S_{ij} \nu_i\nu_j}
\nonumber \\
&&\ \ \ \ =\ \frac{1}{{\cal N}(\Delta\tau)}
e^{-\frac{1}{2}\Delta\tau \sum_{i,j} (S^{-1})_{ij}
(\dot q_i + it_i) (\dot q_j + it_j)},
\nonumber
\end{eqnarray}
and
\begin{eqnarray}
Z &=& \frac{1}{\sqrt{{\rm det}(S)}} \int {\cal D}q(\tau)
\exp\bigg\{-\int_0^\beta d\tau \Big[{\cal H}_0[q(\tau)]
\nonumber \\
&+& \frac{1}{2} \sum_{i,j} (S^{-1})_{ij}(\dot q_i + it_i)
(\dot q_j + it_j) \Big] \bigg\},
\label{A12}
\end{eqnarray}
where $\int{\cal D}q(\tau)$ is a functional integral over all
paths with a uniform (Wiener) measure.

Equation (\ref{A12}) now leads directly to the path integral
representation for the Josephson Hamiltonian (\ref{2.3}), with
$\hat q$ replaced by $\hat \phi$ and $\hat p$ replaced by $\hat n$.
The term ${\cal H}(\hat q)$ is just the Josephson cosine
coupling term.

If, instead of canonical coordinates, ${\cal H} = {\cal
H}[a^\dagger,a]$ is written instead in terms of raising and lowering
operators, another convenient complete set of states is that of the
coherent states,
\begin{eqnarray}
|\alpha \rangle &=& \sum_{n=0}^\infty
\frac{\alpha^n}{\sqrt{n!}} |n \rangle,~a|\alpha \rangle
= \alpha |\alpha \rangle
\nonumber \\
\langle \alpha | \alpha^\prime \rangle &=& \sum_{n=0}^\infty
\frac{(\alpha^* \alpha)^n}{n!} e^{-\frac{1}{2}|\alpha|^2
- \frac{1}{2} |\alpha^\prime|^2}
\nonumber \\
&=& e^{\alpha^*\alpha^\prime - \frac{1}{2}|\alpha|^2
- \frac{1}{2}|\alpha^\prime|^2}.
\label{A13}
\end{eqnarray}
Thus, for any normally ordered operator, $O(a^\dagger,a)$, we have
\begin{equation}
\langle \alpha | O(a^\dagger,a) |\alpha^\prime \rangle =
O(\alpha^*,\alpha^\prime) \langle \alpha | \alpha^\prime \rangle,
\label{A14}
\end{equation}
and hence for $\Delta \tau \rightarrow 0$
\begin{eqnarray}
\langle \alpha | e^{-\Delta\tau {\cal H}[a^\dagger,a]}
|\alpha^\prime \rangle &\approx& \langle \alpha |(1-\Delta\tau {\cal
H}[a^\dagger,a]) |\alpha^\prime \rangle
\nonumber \\
~&\approx& \langle \alpha |\alpha^\prime \rangle e^{-\Delta \tau
{\cal H}[\alpha^*,\alpha^\prime]}.
\label{A15}
\end{eqnarray}
Recognizing that
\begin{eqnarray}
\sum_n \bigg[\alpha_n^* \alpha_{n+1} &-& \frac{1}{2} |\alpha_n|^2
- \frac{1}{2} |\alpha_{n+1}|^2 \bigg]
\nonumber \\
&=& \Delta\tau \sum_n \alpha_n^*
\left(\frac{\alpha_{n+1}-\alpha_n}{\Delta\tau}\right),
\label{A16}
\end{eqnarray}
where the periodic boundary conditions in imaginary time have been
used, we arrive at (\ref{2.1}) (with $\psi$ replacing $\alpha$) when
$\Delta\tau \rightarrow 0$.

\section{Duality transformations}
\label{app:b}

In order to obtain a model amenable to analysis in one dimension one
must derive a \emph{dual representation} for the Josephson
Lagrangian, (\ref{2.3}).  In Ref.\ \onlinecite{FWGF} this was done
in a somewhat \emph{ad hoc} fashion using the Haldane representation
for one-dimensional bosons.  Here we perform the duality
transformation directly, on a variant of the Josephson Lagrangian
(\ref{2.3}), in a much more transparent manner, following closely
the analogous derivation for the two-dimensional $XY$-model and its
associated Kosterlitz-Thouless transition.\cite{JKKN}

The transformation is best carried out in discrete time.  The
continuous time limit is mathematically well defined but, as we
shall see, physically less transparent: one runs into
logarithmically divergent coupling constants, just as in the Trotter
decomposition of the quantum Ising model in a transverse
field.\cite{S76}  These are consequence of the usual exponential
weighting times in continuous time, discrete state Markov processes.
In order to avoid introducing the probabilistic formalism necessary
for dealing with the continuum limit we shall maintain a discrete
time variable.

We begin by introducing the Villain, or periodic Gaussian form, of
the $XY$-coupling:\cite{JKKN}
\begin{equation}
e^{-K_0(1-\cos(\phi))} \rightarrow
\sum_{m=-\infty}^\infty e^{-\frac{1}{2}K(\phi - 2\pi m)^2}
\equiv e^{V_0(\phi,K)},
\label{B1}
\end{equation}
which allows the duality transformation to be carried out
\emph{exactly}. We will consider only the case where $J_{ij}$ in
(\ref{2.3}) is nearest neighbor, but possibly random.  In (\ref{B1})
$K_0 = J_{ij} \Delta \tau$ for the bond $\langle ij \rangle$, and
therefore already includes the effect of discretizing the
$\tau$-integral in (\ref{2.3}).  There are two limits in which $K$
and $K_0$ may be quantitatively compared.  For large $K_0$ the
variable $\phi$ will have only small fluctuations about zero, and
only the curvature, $K_0$, near the minimum of the cosine potential
at $\phi=0$ is important.  In this limit only the $m=0$ term
contributes to the Villain form, and the two potentials therefore
match when $K \approx K_0$.  Conversely, when $K_0$ is small, $\phi$
fluctuates strongly, and many $m$ contribute to the Villain sum.  It
is then convenient to use the Fourier representation\cite{JKKN}
\begin{equation}
e^{V_0(\phi,K)} = \sum_{l=-\infty}^\infty
\frac{e^{-l^2/2K}}{\sqrt{2\pi K}} e^{il\phi},
\label{B2}
\end{equation}
where now, for small $K$, only the $l=0,\pm 1$ terms are important.
This yields
\begin{equation}
V_0(\phi,K) \approx -\frac{1}{2} \ln(2\pi K)
+ 2 e^{-\frac{1}{2K}} \cos(\phi),\ K \rightarrow 0,
\label{B3}
\end{equation}
and hence the correspondence $K \approx 1/2\ln(2/K_0)$ in this limit.
Now, the continuum limit, $\Delta\tau \rightarrow 0$, corresponds to
$K_0\rightarrow 0$, and therefore (\ref{B3}) is appropriate.  This
yields $K \approx -1/2 \ln(\tilde J_{ij}\Delta\tau)$ as $\Delta \tau
\rightarrow 0$---the logarithmic behavior alluded to above.
We shall keep $\Delta\tau$ small but finite, setting aside the
question of its optimal value, which would have to be addressed for
\emph{quantitative} comparison of the Josephson and Villain forms of
our model. For our purposes it is important only that the discrete
and continuous time versions lie in the same universality class.

Using (\ref{B1}) we then define the Villain form of the Josephson
Lagrangian, (\ref{2.3}):
\begin{eqnarray}
e^{\tilde{\cal L}_J[\phi]} &\equiv& \sum_{\bf m}
e^{\bar{\cal L}_J[\phi,{\bf m}]}
\nonumber \\
\bar{\cal L}_J &\equiv& -\frac{1}{2} \sum_{{\bf r},\alpha\neq 0}
K_i^\alpha (\partial_\alpha \phi_{\bf r} - 2 \pi m_{\bf r}^\alpha)^2
\nonumber \\
&&-\ \frac{1}{2} \sum_{i,j,\tau} ({\bf V}^{-1})_{ij}
(\partial_\tau \phi_{i\tau}-i\nu_i-2\pi m^0_{i\tau})
\nonumber \\
&&\ \ \ \ \ \ \ \ \ \
\times\ (\partial_\tau \phi_{j\tau}-i\nu_j-2\pi m^0_{j\tau}),
\label{B4}
\end{eqnarray}
where $\partial_\alpha \phi_{\bf r} \equiv \phi_{{\bf r}+\hat\alpha}
- \phi_{\bf r}$ and ${\bf m}_{\bf r} = (m_{\bf r}^0,m_{\bf
r}^1,\ldots,m_{\bf r}^d)$ is a $(d+1)$-dimensional integer vector
field defined at each space-time lattice site ${\bf r} \equiv
(i,\tau)$, and the index $\alpha=0,1,\ldots,d$ runs over the
neighboring sites in the $\hat{\bm \alpha} \equiv \hat {\bf
x}_\alpha$ direction, with $\hat{\bf x}_0 = \hat {\bm \tau}$.  Until
further notice there is no restriction on the dimensionality, $d$.
The parameters $K_i^\alpha$, $V_{ij}$ and $\nu_i$ are, respectively,
the Villain analogues of $J_{ij}\Delta\tau$, $U_{ij}\Delta\tau$ and
$\mu_i \Delta \tau$ in (\ref{2.3}).

Since $e^{\tilde {\cal L}_J}$ is separately periodic in all the
differences $\partial_\alpha \phi_{\bf r}$ we may write it as a
Fourier series,
\begin{equation}
e^{\tilde{\cal L}_J[\phi]} = \frac{1}{\cal M}
\sum_{\bf n} e^{i\sum_{{\bf r},\alpha} n_{\bf r}^\alpha
(\phi_{{\bf r}+\hat \alpha} - \phi_{\bf r})}
e^{\hat {\cal L}_J[{\bf n}]},
\label{B5}
\end{equation}
where
\begin{equation}
\frac{1}{\cal M} e^{\hat{\cal L}_J[{\bf n}]} =
\prod_{{\bf r},\alpha} \int_0^{2\pi}
\frac{d\theta^\alpha_{\bf r}}{2\pi}
e^{-i \sum_{{\bf r},\alpha} n_{\bf r}^\alpha
\theta_{\bf r}^\alpha}
e^{-\tilde{\cal L}_J[\phi_{{\bf r}+\hat \alpha}-\phi_{\bf r}
\rightarrow \theta_{\bf r}^\alpha]}
\label{B6}
\end{equation}
so that
\begin{equation}
\hat {\cal L}_J[{\bf n}] = -\frac{1}{2}
\sum_{i,\tau,\alpha\neq 0}
\frac{(n_{i\tau}^\alpha)^2}{K_i^\alpha}
- \frac{1}{2} \sum_{i,j} V_{ij} n_{i\tau}^0 n_{j\tau}^0
+ \sum_{i\tau} \nu_i n_{i\tau}^0,
\label{B7}
\end{equation}
and the normalization is
\begin{equation}
{\cal M} = \sqrt{\det(2\pi{\bf V})
\prod_{i,\tau,\alpha\neq 0} \frac{2\pi} {K_i^\alpha}}.
\label{B8}
\end{equation}
To derive (\ref{B7}) we have used the identity
\begin{equation}
\int_{-\infty}^\infty \frac{du}{2\pi} e^{-imu}
e^{-\frac{1}{2} K (u+iv)^2} = \frac{1}{\sqrt{2\pi K}}
e^{-mv} e^{-m^2/2K}
\label{B9}
\end{equation}
and its higher-dimensional generalizations.  The partition function
now becomes
\begin{equation}
Z = \mathrm{tr}^\phi e^{\tilde {\cal L}_J[\phi]} = \frac{1}{\cal M}
\sum_{\bf n} \prod_{\bf r} \delta_{\nabla\cdot{\bf n}_{\bf r},0}
e^{-\hat {\cal L}_J[{\bf n}]},
\label{B10}
\end{equation}
where $\nabla \cdot {\bf n}$ is the discrete space-time divergence,
\begin{equation}
\nabla \cdot {\bf n}_{\bf r} = \sum_\alpha
(n_{\bf r}^\alpha - n_{{\bf r}-\hat\alpha}^\alpha).
\label{B11}
\end{equation}
This formulation is entirely real, and is therefore a convenient
basis for Monte Carlo simulations of the dirty boson
problem.\cite{SWGY}

Let us now restrict attention to $d=1$.  One may then solve the
constraint $\nabla \cdot {\bf n} = 0$ by introducing a \emph{dual
lattice} integer field, $S_{\bf R}$, such that
\begin{equation}
{\bf n}_{\bf r} = (\nabla \times S)_{\bf r} \equiv
(S_{{\bf R}-\hat{\bf x}}-S_{\bf R}, S_{\bf R}-S_{{\bf R}-\hat\tau}),
\label{B12}
\end{equation}
where the dual lattice bond connecting ${\bf R}-\hat{\bm \tau}$ to
${\bf R}$ is the one that cuts the real lattice bond connecting
${\bf r}$ to ${\bf r}+\hat{\bf x}$, while that connecting ${\bf
R}-\hat{\bf x}$ to ${\bf R}$ cuts the one connecting ${\bf r}$ to
${\bf r}+\hat {\bm \tau}$, i.e., ${\bf R} \equiv (I,T) = {\bf r} +
\frac{1}{2}(\hat{\bf x}+\hat{\bm \tau})$.  Thus the
$\alpha$-component of the discrete curl of a scalar is the
difference between its values on the two dual sites that border the
bond from ${\bf r}$ to ${\bf r}+\hat {\bm \alpha}$.  The field
$S_{\bf R}$ is defined uniquely up to an overall additive constant,
and the constrained trace over the ${\bf n}_{\bf r}$ is precisely
equivalent to the free trace over the $S_{\bf R}$.  One therefore
obtains
\begin{equation}
Z= \frac{1}{\cal M} \sum_S e^{\hat{\cal L}_J[\nabla \times S]}
\label{B13}
\end{equation}
with
\begin{eqnarray}
\hat{\cal L}_J[\nabla \times S] &=& -\frac{1}{2}
\sum_{\bf R} \frac{1}{K_I} (\partial_\tau S_{\bf R})^2
\nonumber \\
&&-\ \frac{1}{2} \sum_{I,J,T} V_{IJ}
(\partial_x S_{IT}) (\partial_x S_{JT})
\nonumber \\
&&+\ \sum_{\bf R} \nu_I (\partial_x S_{\bf R}).
\label{B14}
\end{eqnarray}
Here $\partial_\alpha S_{\bf R} = S_{\bf R} - S_{{\bf R}-\hat {\bm
\alpha}}$, $K_I$ is the Villain coupling on the real lattice bond
that cuts $({\bf R}-\hat {\bm \tau},{\bf R})$, and similarly for
$\nu_I$ and $V_{IJ}$.  Note that in this representation the
Lagrangian is purely real and has a very natural classical
interpretation, namely that of a three-dimensional interface model.
The field $S_{\bf R}$ represents the height of a surface over a
two-dimensional plane. The first two terms in $\hat {\cal L}_J$
determine the energy cost for steps in the $\tau$ and $x$
directions, respectively.  In this case the energy associated with
steps in the $\tau$ direction is random, but only in the spatial
index. The last term represents a random tilting potential which
{\it favors} steps in the $x$ direction with the same sign as
$\nu_I$.  It is precisely this breaking of the symmetry between left
and right steps that reflects the broken particle-hole symmetry in
the original quantum Hamiltonian.  Note that in this dual model the
symmetry being broken is associated with parity ($x \rightarrow -x$)
rather than time reversal ($\tau \rightarrow -\tau$).

The more common sine-Gordon representation is obtained from
(\ref{B5}) by softening the integer constraint on the $S_{\bf R}$.
Thus $\sum_{S_{\bf R}=-\infty}^\infty = \int_{-\infty}^\infty
dS_{\bf R} \sum_{h_{\bf R}=-\infty}^\infty \delta(S_{\bf R}-h_{\bf
R})$ is replaced by $\int dS_{\bf R} q(S_{\bf R})$, where $q(t)$ is
periodic with period one, and is peaked around $t=0$.  The
sine-Gordon model (\ref{5.1}) results from the choice
\begin{equation}
q(t) = e^{2y_0 \cos(2\pi t)}
\label{B15}
\end{equation}
where $y_0$ is called the fugacity.  The integer constraint is
recovered in the limit $y_0 \rightarrow \infty$.  For small $y_0$
(see below) this term may be obtained directly by including a term
\begin{equation}
\ln(y_0) \sum_{\bf R} (\nabla \times {\bf m})_{\bf R}^2
\label{B16}
\end{equation}
in the original Lagrangian, (\ref{B4}).  The discrete curl of a
vector field is a scalar field on the dual lattice obtained by
summing the vector field around the dual lattice plaquette,
\begin{equation}
(\nabla \times {\bf m})_{\bf R} = m_{\bf r}^1
+ m_{{\bf r}+\hat{\bf x}}^0 - m_{{\bf r}+\hat {\bm \tau}}^1
- m_{\bf r}^0,
\label{B17}
\end{equation}
and is precisely the \emph{vorticity} on that plaquette.

Finally, the Coulomb gas representation is obtained either from the
sine-Gordon representation by expanding the exponential in
(\ref{B15}) and integrating out the $S_{\bf R}$, or from the
discrete version (\ref{B14}), by substituting $\sum_{l_{\bf
R}=-\infty}^\infty e^{i2\pi l_{\bf R} S_{\bf R}}$ for $\sum_{h_{\bf
R}=-\infty}^\infty \delta(S_{\bf R}-h_{\bf R})$:
\begin{eqnarray}
Z &=& \mathrm{tr}^l \mathrm{tr}^S
\left[e^{2\pi i\sum_{\bf R} l_{\bf R}S_{\bf R}}
e^{\hat {\cal L}_J[\nabla\times S]}\right]
\nonumber \\
&=& \mathrm{tr}^l \left[e^{{\cal L}_C[l]}\right],
\label{B18}
\end{eqnarray}
where [we include the term (\ref{B15}) for completeness],
\begin{eqnarray}
{\cal L}_C[l] &=& \frac{1}{2} \sum_{{\bf R},{\bf
R}^\prime} {\cal G}_{{\bf RR}^\prime}
(2\pi l_{\bf R} + i \partial_x \nu_I)
(2\pi l_{{\bf R}^\prime} + i\partial_x \nu_{I^\prime})
\nonumber \\
&+& \ln(y_0) \sum_{\bf R} l_{\bf R}^2,
\label{B19}
\end{eqnarray}
and ${\cal G}_{{\bf RR}^\prime}$ is the inverse of the quadratic
form:
\begin{equation}
({\cal G}^{-1})_{{\bf RR}^\prime} = \frac{1}{K_I}
(\partial_T \partial_{T^\prime} \delta_{{\bf RR}^\prime})
+ (\partial_I \partial_{I^\prime}
V_{II^\prime}) \delta_{TT^\prime}.
\label{B20}
\end{equation}
For diagonal $V_{IJ} = V_0 \delta_{IJ}$ and fixed $K_I \equiv K_0$,
${\cal G}_{{\bf RR}^\prime}$ is, modulo a trivial rescaling, the
inverse of the two-dimensional lattice Laplacian, and yields the
usual logarithmic Coulomb interaction at large distances.  So long
as $V_{IJ}$ is short ranged and $K_I = K_0 + \delta K_I$ with
$\delta K_I/K_I \ll 1$, ${\cal G}_{{\bf RR}^\prime}$ will remain
Coulomb-like at large distances.  Note that ${\cal L}_C$ is once
again complex, with $\partial_x \nu$ playing the role of complex
offset charges.

The sine-Gordon form yields the same Coulomb gas form (\ref{B19}),
except that the values of $l_{\bf R}$ are restricted to $0,\pm 1$
only. When $y_0$ is small, large values of $l_{\bf R}$ are
suppressed anyway, and the difference between (\ref{B15}) and
(\ref{B16}) is negligible.


\begin{thebibliography}{}
\bibitem{foot:fractau} There are, however, also not so obvious
fractional time dimension generalizations of (\ref{2.1}): see Ref.\
\onlinecite{WK89}.
\bibitem{foot:Mottgap} The random site energies may be thought of as
locally reducing the local chemical potential $\mu - \varepsilon_i =
\mu + \Delta - (\Delta + \varepsilon_i)$ from a uniform value
$\mu+\Delta$.  A free energy convexity argument shows that that such
a reduction can only reduce, or at best maintain, the particle
density. Since, by assumption, $\mu + \Delta < \mu_+(J_0)$ lies
within the Mott phase, the density will remain fixed at the Mott
integer value.
\bibitem{foot:signdelta} There are actually two distinct sets of
scaling functions and amplitudes in (\ref{4.7}), one for $\delta >
0$ and one for $\delta < 0$.  For notational simplicity we shall not
make this distinction explicit since $\rho_s$ (and $\kappa$ in the
random rod problem case) are nonzero only in the superfluid phase.
\bibitem{foot:zvsd} Irrelevance of $g_0$ means that it drops out
completely sufficiently close to the dirty boson critical point,
ruling out in yet a different way, any scaling (\ref{4.7}) of shifts
$g_0 \to g_0 - i\omega_0$ with $\xi_\tau$.
\bibitem{foot:1dsingletinterface} In this interface picture, the
singlet glass phase at half filling\cite{foot:1dsinglet} corresponds
to sufficiently small $K_0 = [K_I]_\mathrm{av}$ in which the
hierarchy of different values of $K_I$ (organized from smallest to
largest) generates an unusual hierarchy of long-range spatial
correlations where all sites divide up into (possibly well
separated) pairs. Although the individual tilts $\partial_I
S_{(I,T)}$, $\partial_I S_{(I',T)}$ at two such sites $I,I'$
fluctuate strongly, as a function of $T$, between 0 and 1, the sum
$\partial_I S_{(I,T)} + \partial_I S_{(I',T)}$ remains close to
unity---the path integral analogue of the quantum state
$\frac{1}{\sqrt{2}}(|0,1\rangle + |1,0 \rangle)$.\cite{Refael} This
pairing reduces the interface tilt fluctuations at large distances,
reducing the roughness below that of the superfluid phase.  Since
our main concern in this paper is with the neighborhood of integer
filling, we shall not pursue the physics of this singlet glass any
further here.
\end{thebibliography}
\end{document}